\documentclass[acmtog]{acmart}
\acmSubmissionID{411}

\usepackage{booktabs}
\usepackage{graphicx}
\usepackage{enumitem}
\usepackage{soul}
\usepackage{dsfont}
\usepackage{gensymb}
\usepackage{microtype}
\usepackage{pifont}
\usepackage{threeparttable}

\usepackage[normalem]{ulem}

\usepackage{tikz,libertine,xcolor,colortbl}
\tikzstyle{closeup} = [
  opacity=1.0,          
  height=1.58cm,         
  width=1.58cm,          
  connect spies, green  
]
\tikzstyle{largewindow} = [red, line width=0.35mm]
\tikzstyle{smallwindow} = [red,line width=0.20mm]
\tikzstyle{largewindow2} = [blue, line width=0.35mm]
\tikzstyle{smallwindow2} = [blue,line width=0.20mm]
\usetikzlibrary{calc,spy}

\newcommand{\eg}{e.g.,\ }
\newcommand{\ie}{i.e.,\ }

\newcommand{\specpoint}{reflection point}
\newcommand{\specpoints}{reflection points}
\newcommand{\diffpoint}{primary point}
\newcommand{\diffpoints}{primary points}
\newcommand{\diff}{primary}

\newcommand{\diffpc}{primary point cloud}

\newcommand{\specpc}{reflection point cloud}

\newcommand{\WarpField}{Neural Warp Field}
\newcommand{\warpfield}{neural warp field}
\newcommand{\Polytope}{Reflection Volume}
\newcommand{\polytope}{reflection volume}

\newcommand{\RB}[1]{#1}

\setcopyright{licensedothergov}
\acmJournal{TOG}
\acmYear{2022}
\acmVolume{41}
\acmNumber{6}
\acmArticle{201}
\acmMonth{12} 
\acmDOI{10.1145/3550454.3555497}

\begin{document}
\title{Neural Point Catacaustics for Novel-View Synthesis of Reflections}

\author{Georgios Kopanas}
\affiliation{%
  \institution{Inria \& Université Côte d’Azur}
  \country{France}}
\email{georgios.kopanas@inria.fr}

\author{Thomas Leimkühler}
\affiliation{%
  \institution{Max-Planck-Institut für Informatik}
  \country{Germany}}
\email{thomas.leimkuehler@mpi-inf.mpg.de}

\author{Gilles Rainer}
\affiliation{%
  \institution{Inria \& Université Côte d’Azur}
  \country{France}}
\email{gilles.rainer.enst@gmail.com}

\author{Clément Jambon}
\affiliation{%
  \institution{Inria \& Université Côte d’Azur and Ecole Polytechnique}
  \country{France}}
\email{clement.jambon@polytechnique.edu}

\author{George Drettakis}
\affiliation{%
  \institution{Inria \& Université Côte d’Azur}
  \country{France}}
\email{george.drettakis@inria.fr}

\begin{abstract}

View-dependent effects such as reflections pose a substantial challenge for image-based and neural rendering algorithms. Above all, curved reflectors are particularly hard, as they lead to highly non-linear reflection flows as the camera moves. We introduce a new point-based representation to compute Neural Point Catacaustics allowing novel-view synthesis of scenes with curved reflectors, from a set of casually-captured input photos. At the core of our method is a neural warp field that models catacaustic trajectories of reflections, so complex specular effects can be rendered using efficient point splatting in conjunction with a neural renderer. One of our key contributions is the explicit representation of reflections with a \specpc~ which is displaced by the \warpfield, and a \diffpc~ which is optimized to represent the rest of the scene. \RB{After a short manual annotation step,} our approach allows interactive high-quality renderings of novel views with accurate reflection flow. Additionally, the explicit representation of reflection flow \RB{supports} several forms of scene manipulation in captured scenes, such as reflection editing, cloning of specular objects, reflection tracking across views, and comfortable stereo viewing. We provide the source code and other supplemental material on \textcolor{blue}{\url{https://repo-sam.inria.fr/fungraph/neural_catacaustics/}}

\end{abstract}



\begin{teaserfigure}
	\includegraphics[width=\textwidth]{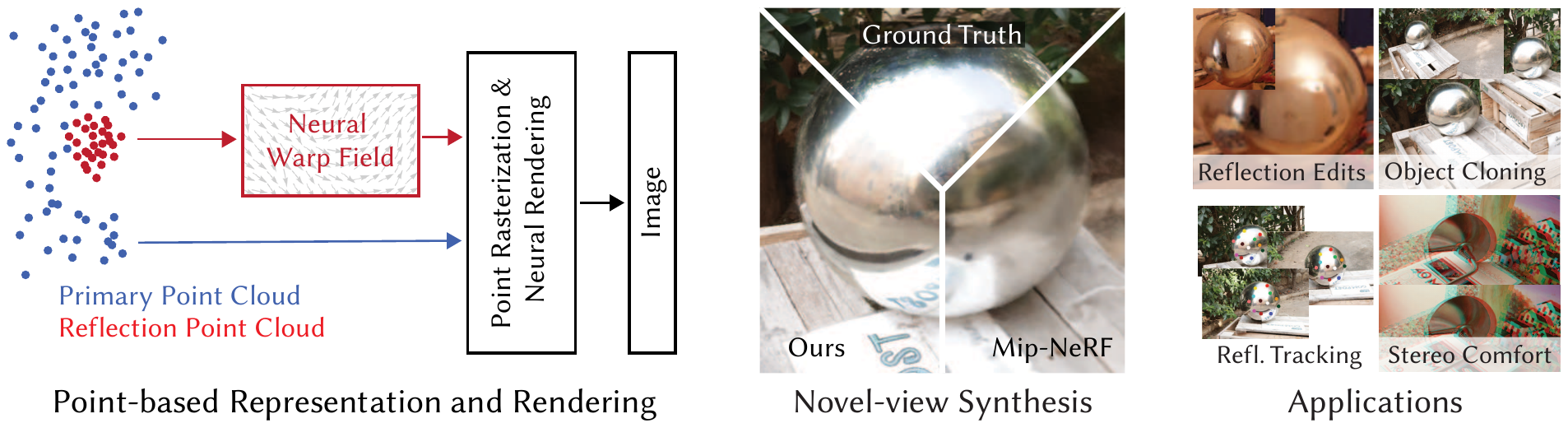}
	\caption{We propose a method to perform novel-view synthesis of curved reflectors. We employ a dynamic point-based scene representation that allows to model catacaustic trajectories of reflections for accurate reflection flow estimation. Our approach outperforms the state of the art in terms of image quality and \RB{supports} a range of additional applications.}
	\label{fig:teaser}
	\Description[TeaserFigure]{TeaserFigure}
\end{teaserfigure}

\maketitle

\section{Introduction}

\begin{figure}[!ht]
	\includegraphics[width=0.99\linewidth]{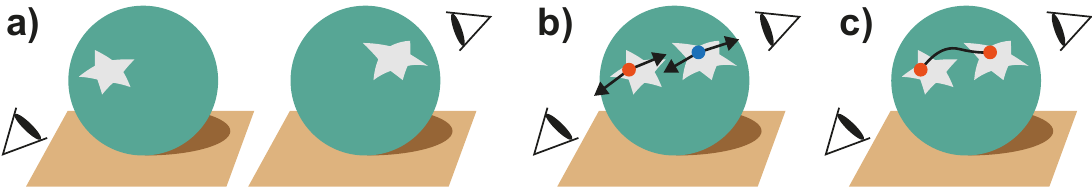}
	\caption{
	Eulerian vs. Lagrangian modeling of view-dependent effects.
	(\emph{a}) A curved reflector is seen from two viewpoints. The reflection of a star appears at different positions with different distortions.
	(\emph{b}) In the Eulerian approach, each surface point evaluates appearance as a function of view direction (black arrows). The red point is on in the left view, but is off in the right view. Reflections and their distortions must be learned for each surface point.
	(\emph{c}) In a Lagrangian approach, reflected objects are represented once and their motion is modeled explicitly. Distortions naturally emerge from the motion of many reflection points.
	}
	\label{fig:EulerLagrange}
\end{figure}

Recent neural rendering methods \cite{tewari2020state,tewari2021advances} provide impressive visual quality for free-viewpoint rendering of captured scenes. 
Such scenes often contain important high-frequency view dependent effects, such as reflections from shiny objects, which can be modelled in two fundamentally different ways (Fig.~\ref{fig:EulerLagrange}): with an \emph{Eulerian} approach where we consider a fixed representation of reflections and model directional variation in appearance, or via a \emph{Lagrangian} solution where we track the \emph{flow} of reflections as the observer moves. Most previous methods adopt the former by representing color on static points as a function of position and view direction using either expensive volumetric \cite{mildenhall2020nerf,Wizadwongsa2021NeX,tewari2021advances}, or mesh-based~\cite{riegler2021stable,HPPFDB18} rendering.  Instead, our solution directly learns reflection \emph{flow} as a function of viewpoint via a \emph{Neural Warp Field}, in effect adopting a \emph{Lagrangian} approach \cite{Bemana2020xfields}. 
Our \emph{point-based} neural rendering method naturally allows \specpoints~ to be warped via the neural field, enabling interactive rendering.

Previous solutions often involve an inherent tradeoff between quality and speed, since to model (somewhat) high-frequency reflections they often use slow volumetric ray-marching with view-dependent queries. Fast approximate alternatives \cite{hedman2021snerg, yu2021plenoctrees} sacrifice angular resolution and thus reflection quality/sharpness.
Overall, such methods use a multi-layer perceptron (MLP) to model density and view-dependent color parameterized by view direction, creating reflected geometry behind the reflector. 
Combined with volumetric ray-marching this often results in ``foggy'' appearance, lacking sharp detail in reflections. 
A recent solution~\cite{verbin2021ref} improves the quality of such methods but still suffers from slow volumetric rendering. In addition, manipulating scenes with reflections is hard with such solutions.

Our Lagrangian, point-based approach avoids bias towards low frequencies inherent in implicit MLP-based neural radiance fields, which persists even when using different encodings and parameterizations. Our approach has two additional advantages: The overhead is lower during inference, allowing interactive rendering, and the direct representation makes scene manipulation easy.

We extract an initial point cloud using standard 3D reconstruction stereo from a multi-view dataset; after a minimal manual step to define a reflector mask on 3-4 images, we optimize two separate point clouds with additional high-dimensional features. During rendering, the \emph{\diffpc}~is static, and represents the mostly diffuse scene component, while the second \emph{\specpc}~represents highly view-dependent reflection effects; these latter points are displaced by the learned \warpfield~ (see Fig.~\ref{fig:teaser}). Points also carry footprint and opacity parameters that are optimized with their position during training. The learned features of the two point clouds are then rasterized and interpreted by a neural renderer to produce the final image.

We are motivated by the theoretical foundation of geometric optics of curved reflectors, which shows that reflections from a curved object move on \emph{catacaustic surfaces} \cite{lawrence2013catalog,hamilton1828theory}; These often result in highly irregular, fast-moving reflection flows (Fig.~\ref{fig:CatacausticGeometry}). 
We train a flow field to learn these trajectories, which we call Neural Point Catacaustics, allowing interactive free-viewpoint neural rendering. Importantly, the explicit nature of our point-based representation facilitates manipulation of scenes with reflections, e.g., reflection editing, cloning of reflective objects, etc. 

We first present the geometric background of complex reflection flow for curved reflectors that guides our algorithm, and then present the following contributions:
\begin{itemize}[leftmargin=5.5mm,topsep=1pt]
\item A novel direct scene representation for neural rendering, with a separate \specpc~ displaced by a reflection \warpfield~ that learns to compute Neural Point Catacaustics, 
and a \diffpc~ with optimized parameters to represent the rest of the scene content.
\item A Neural Warp Field that learns the \emph{displacement} of reflection points as a function of viewpoint. Stable training of our end-to-end method -- including this field -- requires careful parameterization and initialization, progressive training and point densification.
\item
Finally, we present a general, \emph{interactive} neural rendering algorithm, that achieves high quality for both the diffuse and view-dependent radiance in a scene, allowing free-viewpoint navigation in captured scenes and interactive rendering.
\end{itemize}
We illustrate our method on several captured scenes, and show that it is quantitatively and qualitatively superior to previous neural rendering methods for reflections from curved objects, while allowing fast rendering and manipulation of such scenes: e.g., editing reflections, cloning reflective objects, or finding reflection correspondences in input images.

\section{Related Work}
\label{sec:related}

We review only closely related literature in traditional and image-based rendering of reflections -- especially warping methods -- and discuss neural rendering methods that use directional representations of radiance and also point-based (neural) rendering.

\subsection{Rendering Reflections}

Modeling and rendering reflections has been a major goal of Computer Graphics (CG) from its outset. Ray-tracing ~\cite{whitted1979improved} directly simulated optics to render reflections, greatly enhancing realism of CG rendering. Initially, only simplified solutions such as environment or reflection mapping \cite{blinn1976texture} were practical.

\paragraph{Traditional Reflection Rendering.}
Several solutions were initially proposed to render reflections with rasterization. For planar mirrors, a new virtual camera \cite{foley1996computer} and multi-pass rendering with virtual objects \cite{diefenbach1997multi} were used to render reflections, by rendering the reflected scene a second time.
This idea was extended to curved reflectors~\cite{ofek1998interactive,estalella2005accurate} by deforming a virtual mesh to render reflected shapes, using different acceleration structures. Several solutions were subsequently proposed exploiting GPU shaders; see survey~\cite{szirmay2009specular}.
Similarly, we use points to represent reflection geometry, but we train a neural network to learn the \emph{flow} of reflected points with viewpoint motion.

Glossy surfaces with varying levels of roughness can be treated by pre-filtering environment maps \cite{greene1986environment}; this approach was extended by fast image-based warping of reflection probes ~\cite{cabral1999reflection}. Path pertubation theory \cite{chen2000theory} has been used to estimate displacement of specular paths; the same theory guided  computation of probe-based reflection flow for ray-tracing~\cite{RLPWSD20}. Lochmann et al. \shortcite{lochmann2014real} use an optimization to estimate reflection flow for distributed rendering. 
Our method has similar motivation to these methods, since we use reflection flow, but in the context of captured real scenes, which implies many additional challenges, e.g., inaccurate geometry.

\paragraph{Reflection Reconstruction and Image-Based Rendering with Reflections}

3D reconstruction \cite{goesele2007multi} and image-based rendering (IBR) of scenes with reflections is notoriously difficult, since photoconsistency is violated.  Several methods capture and render reflections using image-based approaches, typically by separating the input images into reflected and transmitted layers~\cite{szeliski2000layer}. Layer separation for reflections is a long-standing topic in computer vision with user-assisted solutions~\cite{levin2007user}, or recent deep learning methods based on polarization (e.g., ~\cite{wieschollek2018separating}).
More sophisticated algorithms estimate reflected depth~\cite{sinha2012image} and allow efficient IBR~\cite{kopf2013image}, but have difficulty with curved reflectors. A recent specific solution estimates curved reflector geometry for car windows, and performs ad-hoc layer separation to estimate reflection flow~\cite{rodriguez2020image}.
In contrast, we present a solution that learns reflections on arbitrary curved reflectors.

Theoretically, geometry and material optimization using differentiable rendering \cite{nimier2019mitsuba} could be used, 
but such solutions have not yet been demonstrated to work for captured scenes with incomplete geometry, especially in our case of shiny reflectors that have very incomplete reconstruction. 

\subsection{Neural and Point-Based Rendering}

Neural Rendering is a vast and very fast-moving field; we thus only discuss research directly relevant to our work, i.e., reflection rendering/flow and point-based rendering; recent surveys \cite{tewari2020state,tewari2021advances,xie2021neural} cover the field in detail.

Several neural rendering methods propose specific treatment for reflections/specularities. Thies et al.~\shortcite{thies2019deferred} use feature vectors in texture space to encode specular effects, but require dense angular sampling of input cameras.
The X-fields method~\cite{Bemana2020xfields} learns flow of various parameters -- including specular reflections -- but focuses on small-baseline ``light-field like'' camera motion. Our focus is on wide-baseline casual capture, and unconstrained free-viewpoint camera motion in full scenes.

NeX~\cite{Wizadwongsa2021NeX} improves multi-plane image methods by learning a directional color representation as a linear combination of basis functions.
Neural radiance fields (NeRF)~\cite{mildenhall2020nerf,barron2021mipnerf} learn density and view-dependent color using an MLP.
NeRFs tend to create density corresponding to reflected objects behind the reflector, and use the view-dependent color term to simulate the desired appearance.
This can represent reflection flow to a certain extent, but often fails to reconstruct sharp features and requires expensive volumetric rendering. 
The latter problem has been addressed in several variants \cite{garbin2021fastnerf, mueller2022instant}.
In contrast to these representations, our explicit point-based method generally gives sharper results, and \RB{supports} scene manipulation.

Other methods use the mesh provided by multi-view stereo reconstruction as a ``scaffold''. 
Deep Blending~\cite{HPPFDB18} uses view-dependent meshes for input image projection subsequently blended with learned weights; reflections blend in and out with simple interpolation, often creating severe temporal artifacts. 
Also, the per-input view data storage does not scale to datasets with large numbers of images. Recent work explicitly reconstructs the reflected image for planar reflections, using a superresolution neural network for rendering~\cite{xu2021scalable}.
Neural methods build view dependent deep features from input images merged onto this common space~\cite{riegler2021stable}, while Philip et al.~\shortcite{PMGD21} handle non-diffuse reflections explicitly, by providing mirror images to the neural renderer.
All the above methods propose an \emph{Eulerian} approach~\cite{wu2012eulerian}, since they 
store directional information in a fixed spatially-guided structure (explicit or implicit) as opposed to our \emph{Lagrangian} estimation of reflection \emph{flow}.

\RB{Recently published} NeRF-based methods improve results for reflections, either by reparameterizing view-dependent color by the \emph{direction of reflection} instead of view~\cite{verbin2021ref}, improving reflections for directional incoming lighting, or by learning a separate neural representation for planar reflections~\cite{guo2021nerfren}. Our method treats the general case of reflections on curved reflectors, and allows interactive rendering.

Finally, recent methods learn deformations using neural warp fields (e.g.,~\cite{park2021nerfies,tretschk2021nonrigid}), but for the different goal of handling dynamic content, while Wang et al.~\shortcite{MirrorNeRF} use a related idea to perform one-shot capture from an array of spherical mirrors. These methods can be considered Lagragian, but in a different context, i.e., for motion, or using reflections to reconstruct the non-specular scene.

\paragraph{Point-Based Neural Rendering}
Point-based rendering is a flexible approach to rendering geometry~\cite{gross2007point}, since it does not require mesh connectivity. Points are typically \emph{splatted} to the screen; choosing the size and shape of the splats must be done carefully~\cite{zwicker2001surface}.
Interest in point-based rendering has been revived with the introduction of differentiable solutions \cite{wiles2020synsin,yifan2019differentiable}, including fast approximations~\cite{lassner2021pulsar}. Several neural rendering methods have been presented, directly rendering a point-based representation of the scene \cite{aliev2020neural}\cite{meshry2019neural}, by projecting view-dependent features into the novel view~\cite{kopanas2021point}, or in a fast rendering approach that optimizes camera parameters to improve quality~\cite{adop}.
In recent work, points are used to represent an implicit light field~\cite{ost2021neural}. 
Other methods use points as neural basis functions ~\cite{pointnerf} or local neural fields~\cite{feng2022np}; both have mechanisms to upsample or grow the point cloud when necessary. 
Our method follows this line of work, exploiting the flexibility of points that are naturally adapted to the use of estimated reflection flow in our context, but also allow flexible representation for the rest of the scene, e.g., easy point densification when needed. 

\begin{figure*}[ht]
	\includegraphics[width=0.99\linewidth]{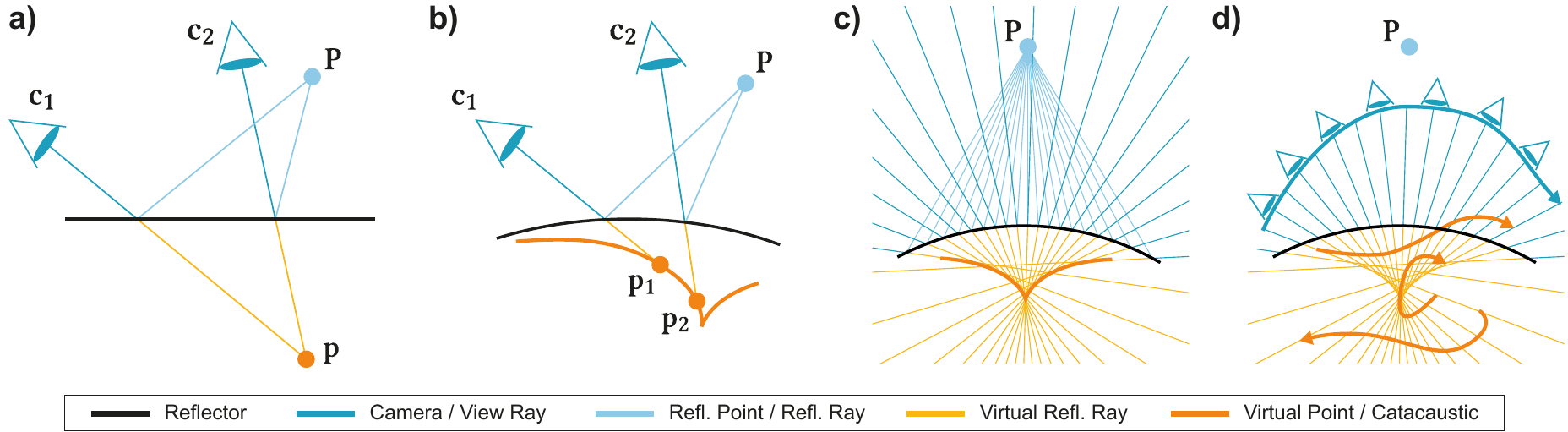}
	\caption{
	The geometry of catacaustics:
	\emph{(a)} In the case of a planar reflector, a reflected point $\mathbf{P}$ results in a static virtual point $\mathbf{p}$, independent of camera position $\mathbf{c}$.					\emph{(b)} For curved reflectors (here a convex example), camera motion leads to $\mathbf{p}$ tracing a surface, called \emph{catacaustic}.
	\emph{(c)} The catacaustic for a single point $\mathbf{P}$ is defined by the envelope of virtual reflected rays, \RB{depicted as the bold orange curve, which at each point is tangent to one of the virtual reflected rays}. 
Virtual images of $\mathbf{P}$ are formed only on the catacaustic trajectory.
	\emph{(d)} In a point-based rendering setup, \RB{where the tangent constraint is not satisfied,} the trajectory of virtual points is not unique. An infinite number of trajectories (three examples are shown) lead to the same apparent reflection. Reflection rays are omitted for clarity.
	}
	\label{fig:CatacausticGeometry}
\end{figure*}

\section{Background \& Overview}
\label{sec:Background}

\noindent
Our goal is to develop a neural rendering method for scenes with reflectors, allowing interactive free-viewpoint rendering and scene manipulation.
We next review the background on reflection geometry in this context, and present an overview of our method.

\subsection{Geometry of Reflections from Curved Objects}
For planar reflectors (Fig.~\ref{fig:CatacausticGeometry}a) the geometry is simple.
Consider point $\mathbf{P}$; its reflection is a static \emph{virtual} point $\mathbf{p}$ on the other side of the reflector, which does not move when the viewpoint changes. To obtain proper reflection flow during rendering, $\mathbf{p}$ just needs to be reprojected to the novel view \cite{foley1996computer,xu2021scalable,guo2021nerfren}. 

For curved reflectors (Fig.~\ref{fig:CatacausticGeometry}b) things are more complex. Here, the position of the virtual point depends on the viewpoint, i.e., the point follows a \emph{trajectory in space} as the view changes. The virtual points $\mathbf{p}$ trace a surface, which is called a \emph{catacaustic} \cite{lawrence2013catalog,hamilton1828theory}.
The shape of the surface depends both on the position of the reflected point $\mathbf{P}$ and the shape of the reflector. The catacaustic surface is usually highly non-linear and has significantly higher complexity than the reflector geometry \cite{josse2014degree}. Closed-form solutions exist for analytic reflector geometries. For example, a circular reflector results in a cardioid \cite{glaeser1999reflections} and can be found with optimization for more complex shapes when exact geometry is available~\cite{mitchell1992illumination}. The catacaustic surface can lie behind or in front of the reflector surface, depending on $\mathbf{P}$ and whether the reflector is convex or concave, potentially resulting in large minifications and magnifications and other irregular motion.

In physics/geometric optics, given all geometric information, the position and shape of the catacaustic are uniquely defined: Only on the catacaustic surface, (virtual) optical images of $\mathbf{P}$ appear, since (virtual) reflected rays intersect only on these surfaces (Fig.~\ref{fig:CatacausticGeometry}c). 
\RB{For set of curves, the envelope is defined as a curve tangent to each one of the curves in the set. Specifically, the envelope is defined by these points of tangency~\cite{bruce1992curves}.}
As catacaustics are the envelope of the virtual reflected rays \cite{hamilton1828theory} \RB{(Fig.~\ref{fig:CatacausticGeometry}(c))}, the virtual points are only visible along the tangent ray of the catacaustic (also see Fig.~\ref{fig:CatacausticGeometry}b).

\subsection{Neural Rendering of Reflection from Curved Objects}

Free-viewpoint neural rendering of these moving reflections is complex. 
Our \emph{Lagrangian} methodology estimates the \emph{catacaustic trajectories} of the points reflected on curved reflectors, therefore storing reflections once and reusing them.

We learn reflection motion by training a Neural Warp Field \cite{sitzmann2019srns, xie2021neural, tewari2021advances} from multi-view images. This warp field is used to deform a virtual point cloud to match the reflections captured in the input views and, together with a neural rendering network, allow interactive free-view navigation in a scene with curved reflector shapes.

However, the use of points results in a deviation from physics: Since in point-based rendering a point emits light in all directions, it is a (virtual) optical image by construction, independent of its position. This means the trajectory of reflections is no longer unique (Fig.~\ref{fig:CatacausticGeometry}d): Any trajectory crossing virtual reflected rays in the correct order at the appropriate time results in the same apparent reflection as seen by a moving camera (modulo occlusions), resulting in a depth ambiguity. 
Recovering the physically correct catacaustic surface is not necessary for rendering since all the different solutions in Fig. \ref{fig:CatacausticGeometry}d generate the same image.

Our method builds on the theory of catacaustic surfaces to learn the trajectories of reflection points, which we refer to as \emph{Neural Point Catacaustics}. We perform an qualitative analysis comparing physical to neural point catacaustics in Sec.~\ref{sec:gt-cata}.

\begin{figure}[!ht]
	\includegraphics[width=0.99\linewidth]{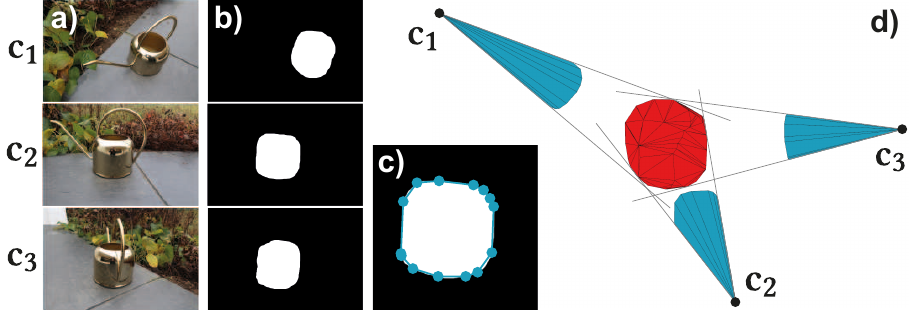}
	\caption{Bounding reflector volume: 
	\emph{(a, b)} The user is asked to paint rough masks marking the reflector in a small set \RB{(here 3)} of images. 
	\emph{(c)} From these masks, we compute simple 2D bounding polylines (shown for $\mathbf{c_2}$ only).
	\emph{(d)} Finally, we solve for the convex 3D polyhedron that satisfies all mask constraints.
	}
	\label{fig:Polyhedron}
\end{figure} 

\begin{figure*}[ht]
	\includegraphics[width=0.99\linewidth]{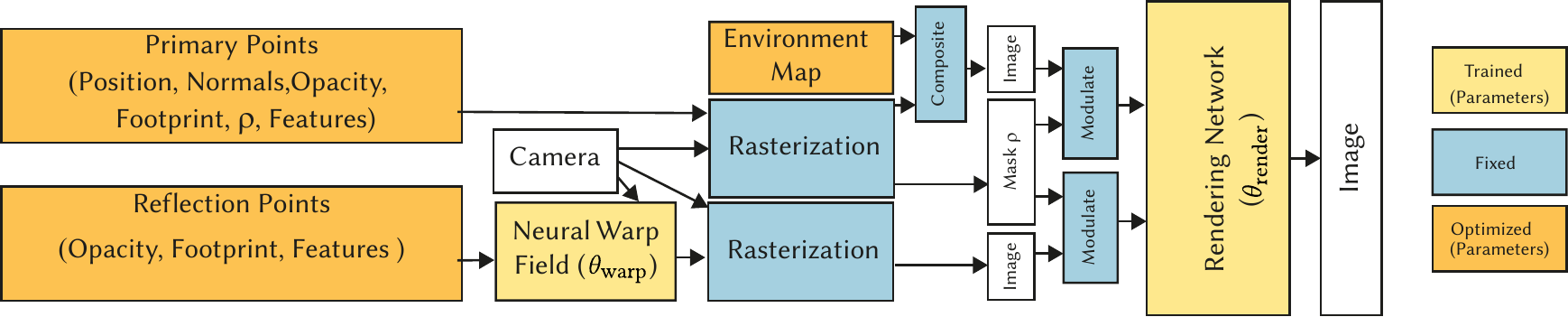}
	\caption{Overview of our method.}
	\label{fig:Overview}
\end{figure*}
\subsection{Method Overview}
\label{sec:overview}

Our method takes multiple wide-baseline photos of a scene as input, typically 200-300 
for a room or an outdoors scene, containing curved reflectors. We run standard structure-from-motion (SfM) to calibrate the cameras, and a standard multi-view stereo (MVS) method to extract a dense point cloud. 
A minimal manual step is then performed, where the user creates a coarse mask of the reflector(s) in 3-4 views in the dataset (this step typically takes less than 1 minute); these masks are used to extract a \Polytope~ that encapsulates the reflectors.

Our method is illustrated in Fig.~\ref{fig:Overview}.
The MVS reconstruction is used to initialize a \emph{\diffpc}, representing the diffuse and low-frequency view-dependent content of the scene. We also initialize a \emph{\specpc}~ using the \polytope~created earlier.
During training, we optimize the attributes of these two point clouds jointly.

To render an image given a free-viewpoint camera, we first displace the \specpc~ in 3D using a \warpfield, moving reflections to the correct position for a given camera.
Then, both the \diff~ and the \specpc~ are rasterized separately using EWA splatting~\cite{zwicker2001surface}.
The rasterized point clouds are composited together with an optimized environment map used to compensate for missing distant geometry.
Finally, the resulting high-dimensional features are fed to a decoder neural renderer to synthesize the final image.
Our architecture is trained end-to-end and results in an interactive free-viewpoint renderer.

\section{Method}

We now present the main components of our method, following Fig.~\ref{fig:Overview} (left to right). We first discuss our dual point cloud representations, explain how the \specpoints~ are displaced by our \WarpField, then discuss how we rasterize the points and their parameters, and finally discuss how we perform neural rendering.


\subsection{Point Clouds}

We use 3D points to represent the scene. In addition to their natural parameters i.e., positions and normal, we augment the points with additional parameters, all of which are optimized by our method. These properties allow the \diffpc~ to recover from incorrect and incomplete geometry from MVS reconstruction \RB{(see Fig.~\ref{fig:geometry-fix})}, and are central in learning the \warpfield~ so that the \specpoints~ can be displaced to follow reflection flow.

The parameters we optimize for all points in both clouds are: opacity, footprint and high-dimensional features.
For \diffpoints~ we also optimize position, normals and parameter $\rho$ that will be used to modulate rasterized point clouds.

\subsection{\WarpField}

We define a \warpfield~ which is responsible for displacing \specpoints~ with camera motion.
Formally, given an initial \specpoint~ position $\mathbf{p} \in \mathds{R}^3$ and the current camera position $\mathbf{c} \in \mathds{R}^3$, we seek to compute the reflection position $\mathbf{p'} \in \mathds{R}^3$ on the Neural Point Catacaustic, using the field 
$
\mathcal{F} \in \left( \mathds{R}^3 \times \mathds{R}^3 \right) \rightarrow \mathds{R}^3
$
via:
\begin{equation*}
\mathbf{p'} 
=
\mathbf{p}
+
\mathcal{F} (\mathbf{p}, \mathbf{c}).
\end{equation*}
We realize $\mathcal{F}$ using an MLP with trainable parameters $\theta_\text{warp}$.
The initial point positions $\mathbf{p}$ are fixed, not optimized (see Sec.~\ref{sec:polyhedron} for our initialization procedure), but each point is displaced by the \warpfield.
Notice that the warp field takes the initial point position $\mathbf{p}$ as input, so that we can use the same field for all \specpoints, resulting in a compact representation of the entire reflection flow.

\subsection{Rasterization}
\label{sec:rasterize}

We use differentiable point splatting~\cite{yifan2019differentiable,kopanas2021point} to rasterize our point clouds. Each point can be seen as an oriented disk with position $\mathbf{p}$ and a normal, projected onto the screen as a soft ellipsoid. Each point also has a 3D footprint, defined as the squared radius of the disk, a 6-channel neural feature, and an opacity $o$.

The \diffpc~ is tasked to represent the geometry of the scene. 
We learn an additional parameter $\rho$ for each \diffpoint;
This scalar value will later control how much of the \specpc~ is blended in the final rendering. 

To rasterize a point, we need to determine its 2D footprint, respecting the distance to the camera and its slant as defined by the point normal. To this end, we compute the covariance matrix for each point as follows~\cite{zwicker2001surface,ren2002object}:
\begin{equation}
\label{eq:sigma}
\Sigma ~=~ h^2 \mathbf{J}  \mathbf{V} {\mathbf{J} }^T +~ \nu \mathbf{I}
\end{equation}

\noindent
where $\mathbf{J} \in \mathds{R}^{2 \times 2}$ is the Jacobian of the transformation from world space to view space, $h$ is a scaling factor accounting for pixel resolution~\cite{zwicker2001surface}, and $\mathbf{V}$ is the identity matrix scaled by the 3D footprint of the point.
$\nu \mathbf{I}$ is a low-pass filter; we use $\nu~=~0.3$ for all our tests, striking a balance between gradient instability arising from aliasing, and recovering high-frequency features.

We perform front-to-back compositing of all the splats to compute a pixel feature value $c$. 
We follow previous work~\cite{yifan2019differentiable,kopanas2021point} and compute for all points $\mathcal{N}$ that fall on a pixel:
\begin{equation}
\label{eq:front-to-back}
	c = \sum_{i \in \mathcal{N}}
	c_{i}\alpha_{i}
	\prod_{j=1}^{i-1}(1-\alpha_{j}).
\end{equation}
Our $\alpha_i$ is given by evaluating a 2D Gaussian with covariance $\Sigma$ (Eq.~\ref{eq:sigma})~\cite{yifan2019differentiable} and multiplying it with the learned per-point opacity parameter $o$. The opacity value $o$ permits the optimization to make points ``disappear'' which allows to correct for erroneous overreconstruction from MVS.
We call the full product:
\begin{equation}
\label{eq:accopacity}
	\bar{o}~=~\prod_{j \in \mathcal{N}}(1-\alpha_{j})
\end{equation}
accumulated opacity, to be used in a loss described in Sec.~\ref{sec:training} and for the environment map described below.

We use an environment map for the cases where MVS did not manage to compute geometry; this happens for objects far away, viewed from only a few views in large scenes.
We parameterize an environment map with polar coordinates and we initialize with 6 features with a constant value of zero.
During rendering, we project the environment map to the view and blend it using accumulated opacity $\bar{o}$.

\begin{figure}[!t]
	\centering
	\begin{tabular}{cc}
	Final&Primary Points\\
	\includegraphics[width=3.96cm]{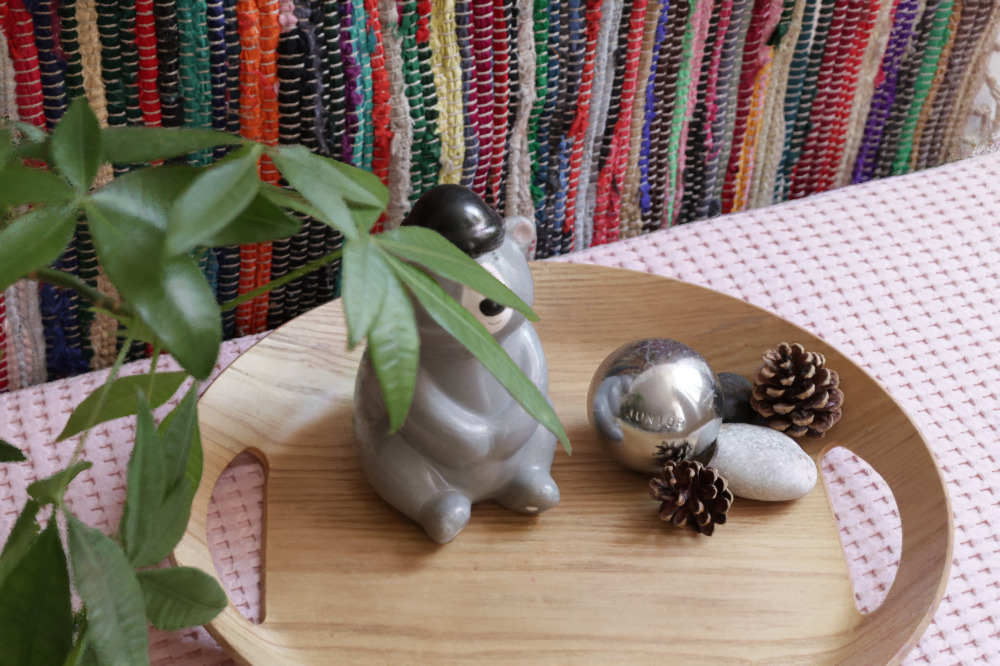}&
	\includegraphics[width=3.96cm]{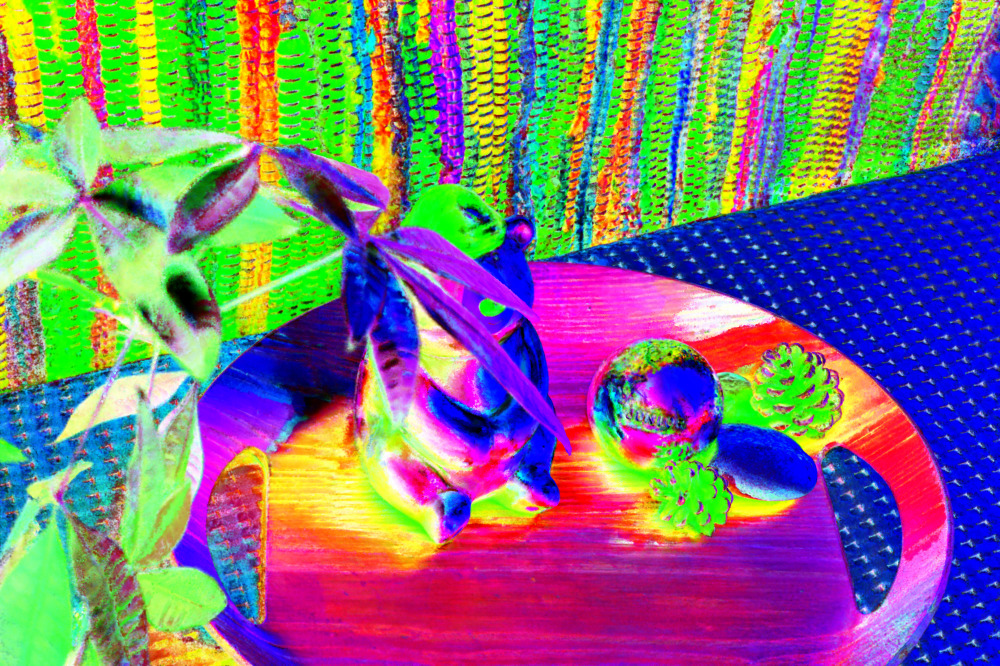}\\
	Reflection Points&$\rho$ Mask\\
	\includegraphics[width=3.96cm]{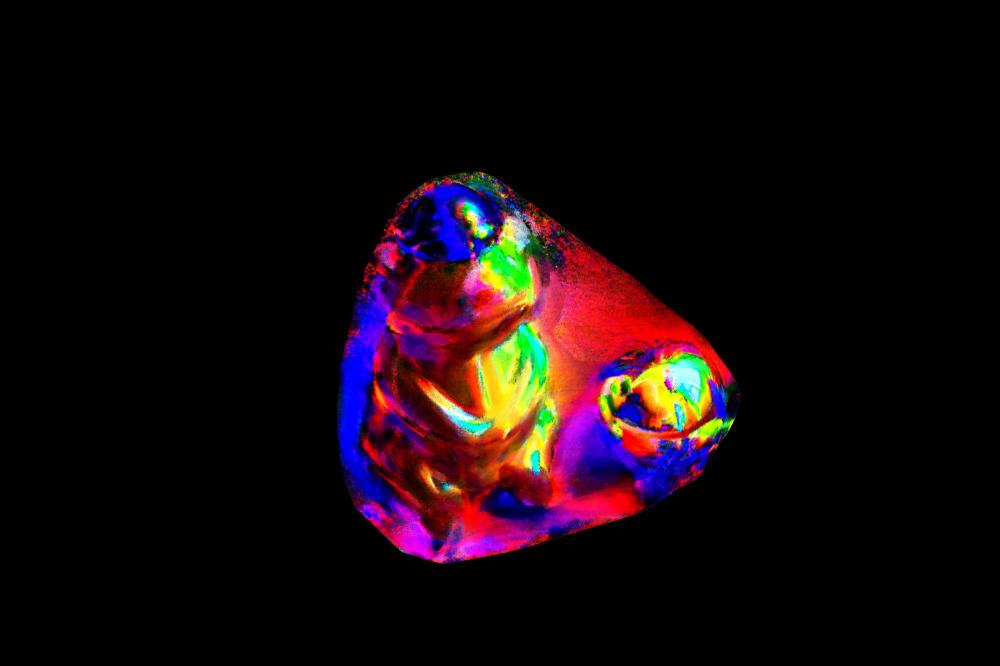}&
	\includegraphics[width=3.96cm]{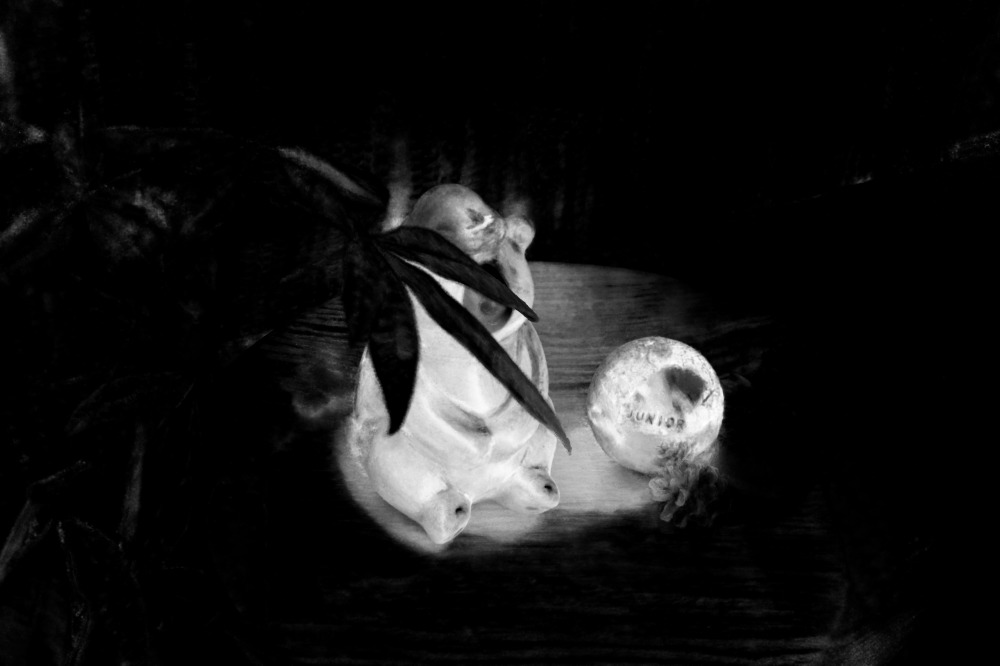}\\
	
	\end{tabular}
	
	\caption{
	\label{fig:intermediate_buffers}
	\RB{Intermediate buffers provided to the neural rendering network. For the primary and reflection point clouds we show the first three channels as RGB. Notice how the primary point cloud captures the non-specular appearence of the objects \ie the texture on the porcelain bear, while the reflection point cloud captures all the specular details. The mask $\rho$ handles occlusions between specular and non-specular objects.}
	}
\end{figure}

\subsection{Neural Rendering}

Once we have rasterized and composited all points and their corresponding parameters, i.e., 6-channel features and $\rho$, we feed this information into a neural renderer MLP with two heads.
The first head encodes rasterized features of the \diffpoints~ plus the view direction into a latent feature. Feeding the view direction allows the \diffpc~to model low-frequency view-dependent effects (Eulerian-style), which almost all real-world materials exhibit to some extent.

We use \RB{a} mask $\rho$ to modulate these primary rasterized features.
The second head does the same for the \specpoint~ features independently, and is modulated by $1-\rho$.  We concatenate the outputs of the two branches and feed them into a decoder network to obtain the final image.
\RB{All the layers used by the renderer are illustrated in Fig.~\ref{fig:intermediate_buffers}.}
We denote the trainable parameters of this module $\theta_\text{render}$.

\section{Optimization and Training}

Recall we have two point clouds, one for reflections and one for the remaining scene content; we optimize the parameters of these two point clouds and the two networks jointly.
This joint optimization process poses several significant challenges, especially for the reflection flow that 
models complex motion and thus is prone to instabilities which
we address with our method. 
We discuss how to initialize and regularize \specpoints~ and their motion in space using a \emph{\polytope}, created with minimal user intervention. We then discuss our loss function and our optimization using a multi-resolution solver and point densification.

\subsection{Initializing and Regularizing Reflection Points}
\label{sec:polyhedron}

In many scenes, reflective surfaces tend to be restricted in space;
Our test scenes each contain one main curved reflector, and the input photos include views around the object to ensure good angular coverage of reflections and view of the entire scene for free-viewpoint rendering.
We consider the 3D volume that contains the reflective object(s).
This \polytope~ represents the approximate region in space where we expect our Neural Point Catacaustics to move.
We initialize the \specpc~ by randomly placing points on the bounding surface of the \polytope. 
We use 400K points in all our experiments.
The initialization on a 2D surface is a natural choice, since the scene elements we see in the reflections mostly consist of (or can be approximated by) surfaces.
Additionally, we will use the \polytope~ to regularize the warped position of the \specpc~ during the optimization (Sec.~\ref{sec:training}).

We define the reflection volume by identifying the regions corresponding to reflection objects in the input images.
Methods for automatic detection of reflections exist \cite{whelan2018reconstructing}, but have not yet been shown to robustly handle curved reflectors.
Therefore, we opt for an approach with minimal user intervention: The user is asked to paint rough masks in a small number of input images, marking the objects with reflections (Fig.~\ref{fig:Polyhedron}a and b). \RB{We use three or four images in all our examples; Painting takes less than a minute for a given dataset}.

From these masks, and the 3D calibration of the corresponding cameras, we compute a convex polyhedron bounding the 3D space occupied by the reflector in three steps.
First, we compute the convex hull of each 2D mask and simplify it using the 
Douglas-Peucker \shortcite{douglas1973algorithms} algorithm. The result is a closed polyline $L_i$ per image $i$  (Fig.~\ref{fig:Polyhedron}c).
We then lift everything to 3D: Using information from the calibrated cameras $\mathbf{c}_i$, we can associate each vertex of $L_i$ with a corresponding 3D point on $\mathbf{c}_i$'s image plane. Together with the camera's center of projection, each two adjacent vertices define a plane (blue triangle fans in Fig.~\ref{fig:Polyhedron}d), each separating 3D space into two half spaces; one contains the reflector, the other does not.
Finally, we solve for the convex polyhedron which bounds the 3D space where all constraints are fulfilled \cite{preparata1985convex} (red shape in Fig.~\ref{fig:Polyhedron}d).

\subsection{Loss}

\label{sec:training}
Our architecture is trained end-to-end for each specific scene -- which is standard practice for recent neural rendering methods ~\cite{mildenhall2020nerf,tewari2021advances,kopanas2021point}. 

Our loss consists of five terms:
\begin{equation}
\mathcal{L} 
= 
\lambda_{\ell_1} \mathcal{L}_{\ell_1}
+
\lambda_\textrm{DSSIM} \mathcal{L}_\textrm{DSSIM}
+
\lambda_\mathbf{p} \mathcal{L}_\mathbf{p}
+ 
\lambda_m \mathcal{L}_m
+
\lambda_{m_{TV}} \mathcal{L}_{m_{TV}}.
\end{equation}

$\mathcal{L}_{\ell_1}$ and $\mathcal{L}_\textrm{DSSIM}$ penalize differences between the output of the neural renderer and the corresponding input view in terms of $\ell_1$-norm and the DSSIM metric~\cite{loza2006structural}, respectively.
$\mathcal{L}_\mathbf{p} $ encourages the \specpoints~ to stay within the \polytope.
For this, we use a binary mask $m_{\mathrm{RV}}$ created by projecting the \polytope~ into the current view.
We take the (soft) accumulated opacity $\bar{o}$ (Eq.~\ref{eq:accopacity}) of the \specpoints~ and compute
\begin{equation}
\mathcal{L}_\mathbf{p} ~=~ \|(\bar{o} ~-~m_{\mathrm{RV}}) m_{\mathrm{RV}} \|_1.
\end{equation}
For points that lie outside the projection of the reflection volume, this term has no effect, allowing some freedom for points to move outside the volume, but encourages points to stay inside to ``fill'' the reflector surface.
$\mathcal{L}_m$ and $\mathcal{L}_{m_{TV}}$ regularize the rasterized mask $\rho$.
$\mathcal{L}_m$ compares it to the binary mask $m_{\mathrm{RV}}$ in terms of the $\ell_1$-norm, encouraging the mask to light up for the reflector.
Finally, $\mathcal{L}_{m_{TV}}$ encourages smoothness of $\rho$ by penalizing its total variation. \RB{An example of the $\rho$ mask is shown in Fig.~\ref{fig:intermediate_buffers}.}

We use an ADAM optimizer and weight decay for $\theta_\text{warp}$ for all our scenes, and
we employ learning rate scheduling. 
The values of the parameters are $\lambda_{\ell_1}=0.05,~
\lambda_\text{DSSIM}=0.2,
\lambda_\mathbf{p}=\lambda_m=0.01,~
~\lambda_{m_{TV}}=1e-5$.

\subsection{Progressive Optimization}

We optimize point parameters and the neural networks end-to-end by rendering patches of the input views.
In each iteration, we render a random 150x150 pixel patch from a random input view and compute the loss given a ground truth image. We back-propagate the gradients every 20 iterations, in a manner equivalent to batches to avoid memory saturation. 

If done naively, the early stages of the optimization can be unstable; the process may collapse to a degenerate case i.e., moving the \specpc~ outside the camera frustum. To avoid this, we use multi-resolution optimization. 
We also progressively densify the \diffpc~ to recover from large errors of MVS while maintaining sufficient point density.

\begin{figure}[!t]
	\centering
	\begin{tabular}{cc}
	\includegraphics[trim={200 100 450 300 }, clip, width=3.96cm]{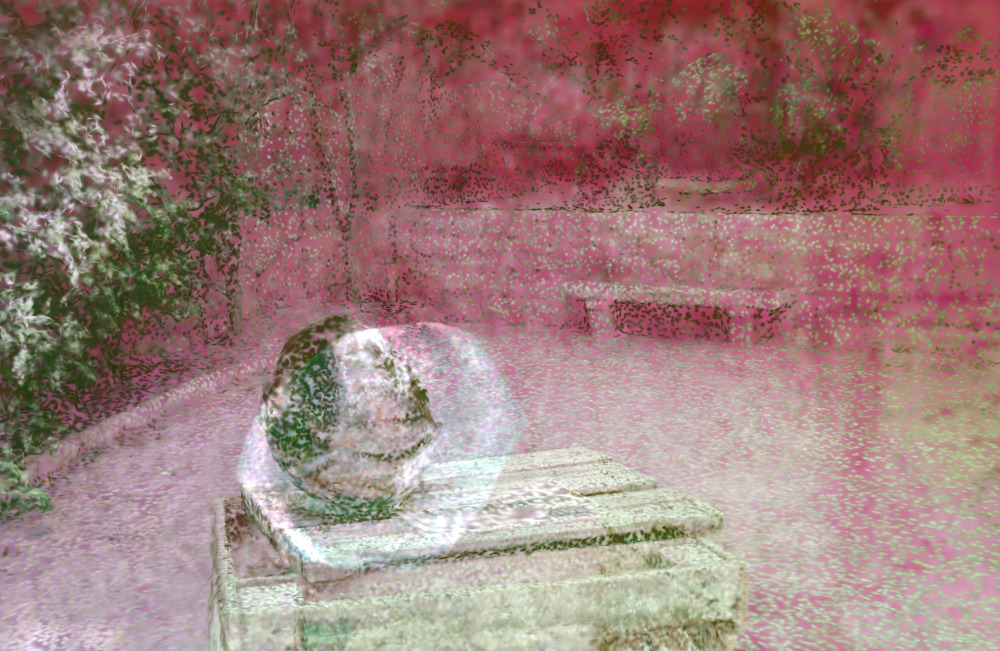}&
	\includegraphics[trim={200 100 450 300 }, clip, width=3.96cm]{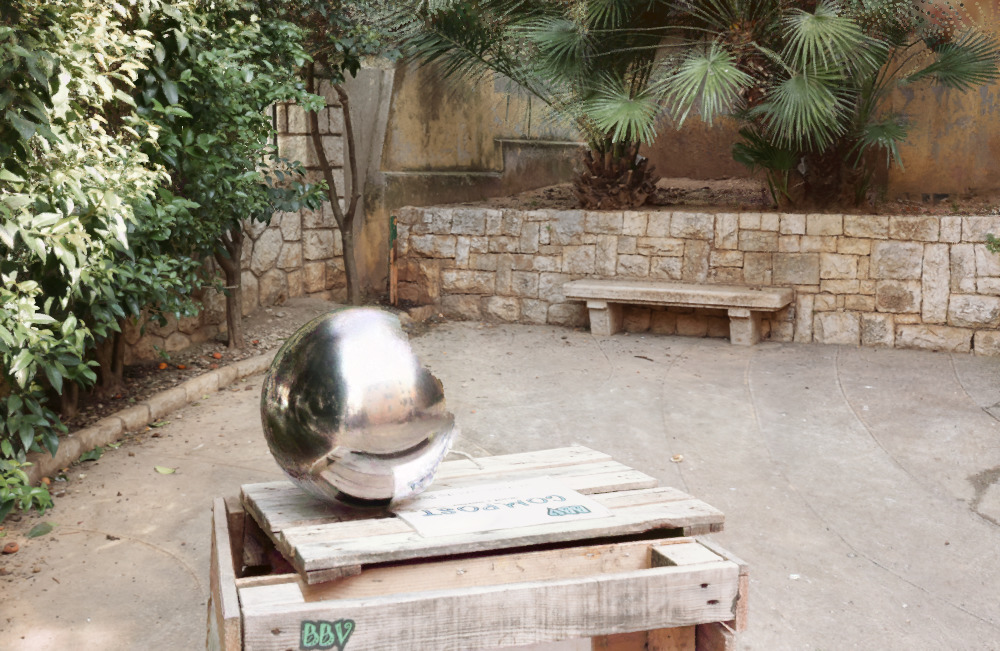}\\
	Step: 2,000 -2min-&Step: 50,000 -3h 25min-\\
	\includegraphics[trim={200 100 450 300 }, clip, width=3.96cm]{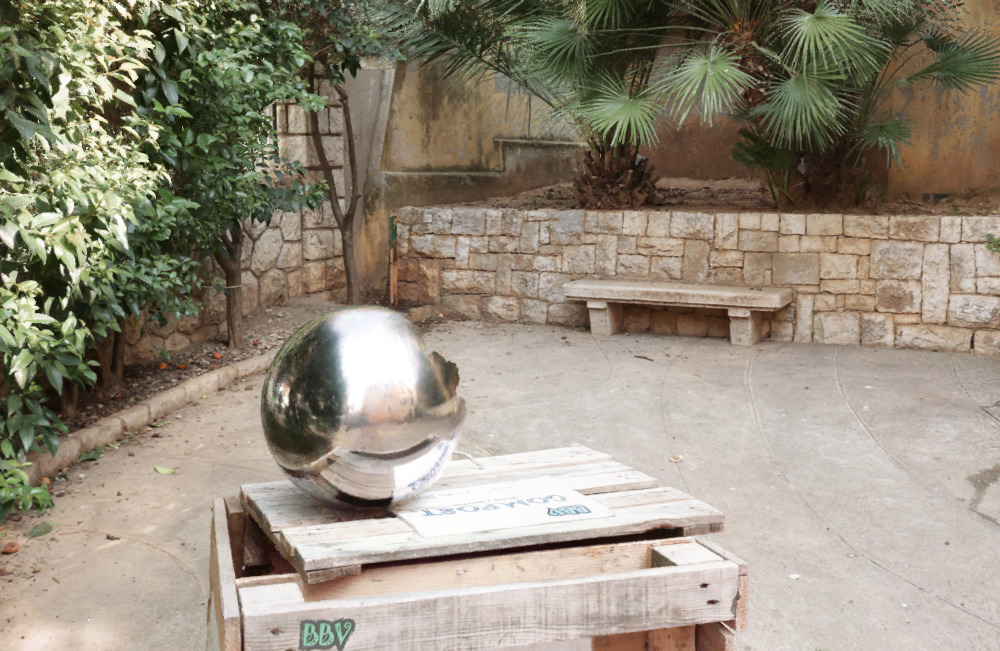}&
	\includegraphics[trim={200 100 450 300 }, clip, width=3.96cm]{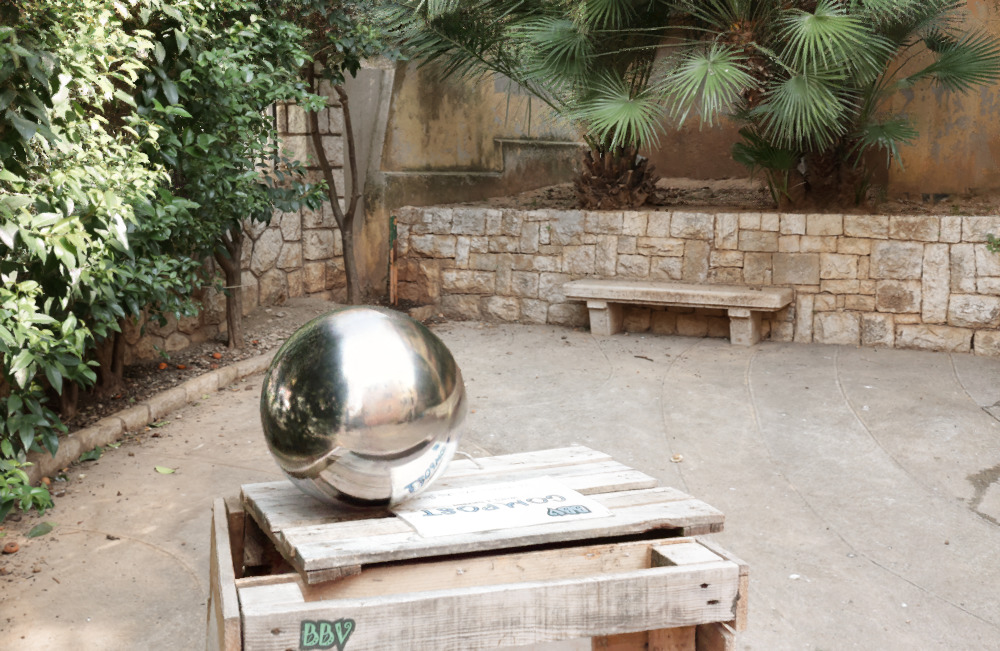}\\
	Step: 100,000 -8h 47min-&Step: 200,000 -22h 5min-\\
	\end{tabular}
	
	\caption{
	\label{fig:geometry-fix}
	\RB{The final rendering throughout the optimization process. On the top left we show the image before the warm up finishes. On the top right, we can see that the quality is greatly improved, but the full reconstruction of the specular sphere is still missing. And on the bottom right we have reconstructed the full sphere.} 
	}
\end{figure}

The multi-resolution approach starts the optimization at low resolution, i.e., with downsampled images and reduced point-clouds by taking one out of every 32 points at the outset. By operating on smaller images, and thus larger image regions, the method provides a stable initialization for the opacity and 6-channel features of the points. In a short warm-up phase we upsample twice, each time doubling resolution and retrieving progressively more of the original points. At the end of the the warm-up, we reach full image resolution and original point cloud size, and we then start optimizing all parameters.
\RB{We visualize the evolution of the optimization over time in Fig.~\ref{fig:geometry-fix}.}

MVS reconstruction can fail in some parts of the scene, creating empty regions of space. This has the consequence that points may move towards these empty regions, resulting in a sparse point cloud. To overcome this problem,  
we densify the point cloud every 2K iterations: For points with high view-space positional gradients we spawn a new point along the gradient direction. 

\section{Implementation}

We next present details of our implementation for the rasterization, neural network and our interactive renderer.
All renderings for optimization and for all results were performed at resolution 1000x666.
\RB{All source code and data are available here: \textcolor{blue}{\url{https://repo-sam.inria.fr/fungraph/neural_catacaustics/}}}

\subsection{Capture}
We take a few hundred photographs of a given scene focusing on an central reflective object.
The photos are then calibrated using off-the-sheld Structure-from-Motion (SfM) and we generate a dense point cloud using Multi-View Stereo~\cite{reality2016capture}, used to initialize the 
\diffpc. We use RealityCapture in our experiments, which takes 5-8min for SfM and 30-40min for MVS on all our scenes on our Intel Xeon 5118 Gold PC.

\subsection{Rasterization for Training}

We implement our method in Pytorch using custom CUDA kernels (inspired by ~\cite{yifan2019differentiable,kopanas2021point}). While front-to-back compositing of points is trivially differentiable, the computation of rasterized feature gradients with respect to point position and footprint 
has quadratic complexity in the number of points per pixel when implemented naively. Different from recent previous work \cite{lassner2021pulsar}, which only use a small subset of rasterized points in their backward pass, we use \emph{all} rasterized points to obtain stable gradients. 
As a remedy, we implemented a custom two-pass recursive strategy (front-to-back followed by back-to-front) to yield exact gradients while reducing time complexity to linear.

\subsection{Network Details}

The warping field is an MLP with ReLU activations of 4 layers and 256 features. We carefully normalize the camera position and the \specpc~ to the range $[-1,1]$ and we scale down the output by of this network by multiplying with 0.01 to enforce a very small magnitude of the displacements when the optimization starts.

Our environment maps have a resolution of $1000 \times 400$.
The Neural Renderer is an MLP with ReLU activations: both encoders and the decoder share the same architecture and have 32 features and 9 layers. For every 2 layers, we have a residual connection. We initialize them with FixUp. 
We opted for an MLP instead of CNNs that have been used in previous work~\cite{kopanas2021point,PMGD21}. While CNNs naturally provide high-quality renderings, they suffer from temporal artifacts due to screen-space filtering. In addition, initial tests showed that for our end-to-end optimization, the ability of the CNN to correct artifacts in screen space undermines the ability to recover an accurate 3D scene representation.

\subsection{Interactive Rendering}

We observed that the main bottleneck of our pipeline is the rasterization of the \diffpc, preventing interactive rendering. As a remedy, we have implemented an interactive OpenGL-based framework, which approximates the rasterization with hardware-accelerated EWA surface splatting \cite{ren2002object}. We use visibility splatting with depth peeling to create three semi-transparent render layers, which are composited to form the final feature tensor. Rasterization takes 100\,ms, and in conjunction with the unoptimized Python-based Pytorch implementation of the remaining parts of the pipeline, we achieve 5\,fps. We analyze the resulting image quality in Sec.~\ref{sec:eval}.

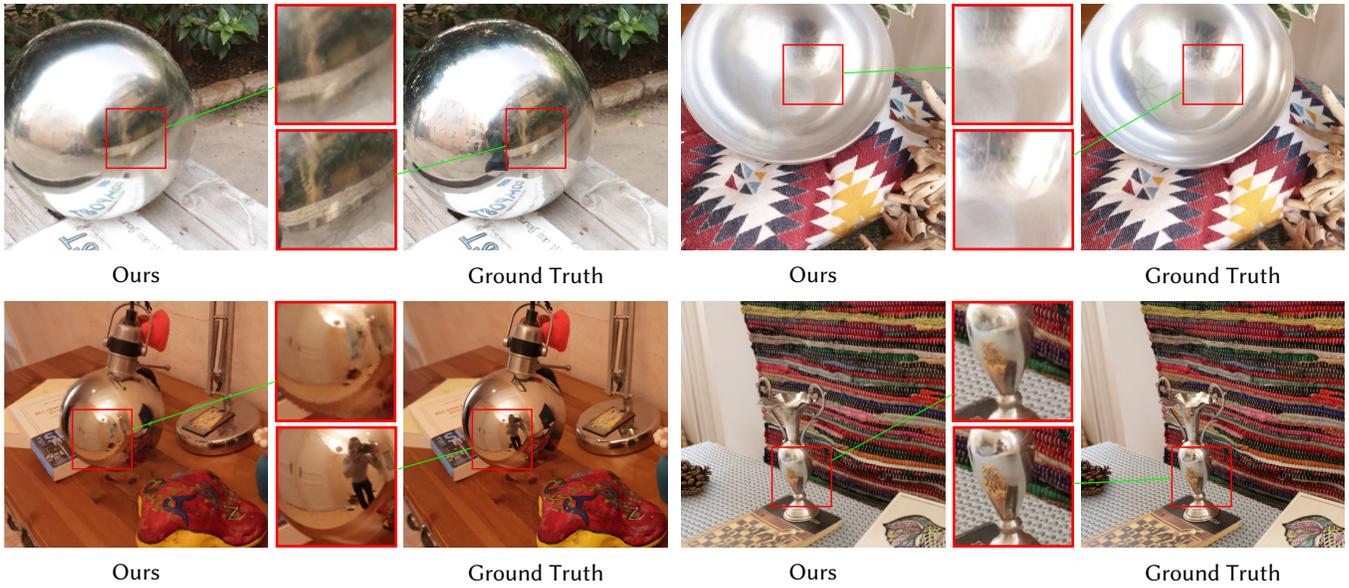
\begin{figure*}[!h]
\begin{tikzpicture}[xscale=4, spy using outlines={every spy on node/.append style={smallwindow}}]
\node[anchor=south] (FigA) at (0,0) {\includegraphics[trim={200 100 300 100 }, clip ,width=3.5cm]{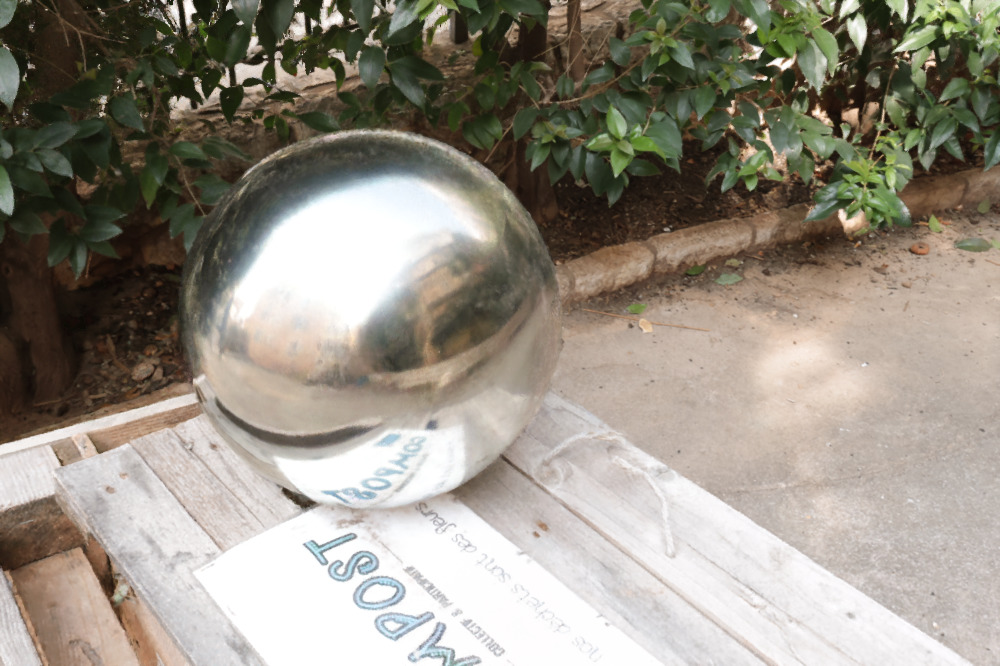}};
\spy [closeup,magnification=2] on ($(FigA)+(0.,-0.15)$) in node[largewindow,anchor=north west] at ($(FigA.north east) - (0,0.12)$);
\node [anchor=north] at ($(FigA.south)$) {\sf Ours};
\node[anchor=south] (FigB) at (1.329,0) {\includegraphics[trim={200 100 300 100 }, clip ,width=3.5cm]{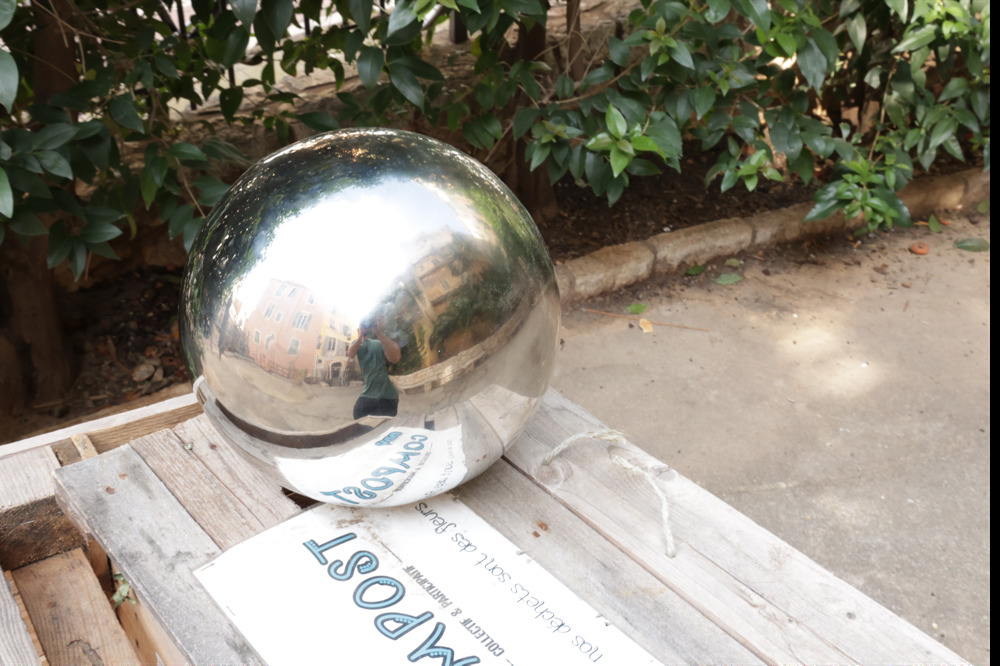}};
\spy [closeup,magnification=2] on ($(FigB)+(0.,-0.15)$) in node[largewindow,anchor=south east] at ($(FigB.south west) + (0,0.12)$);
\node [anchor=north] at ($(FigB.south)$) {\sf Ground Truth};
\node[anchor=south] (FigC) at (2.25,0) {\includegraphics[trim={200 100 300 100 }, clip ,width=3.5cm]{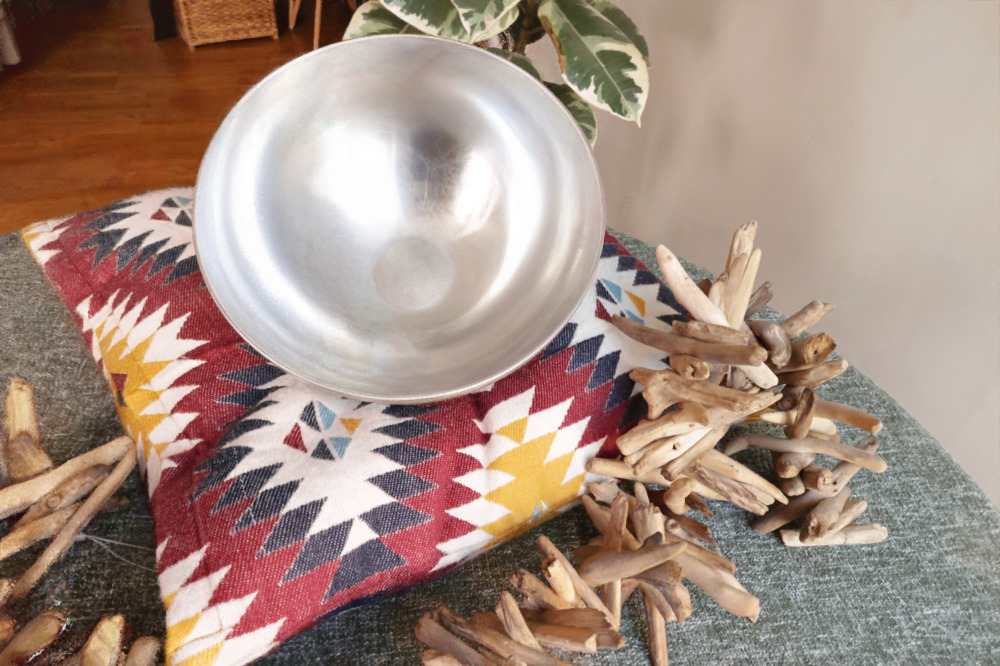}};
\spy [closeup,magnification=2] on ($(FigC)+(0.,0.7)$) in node[largewindow,anchor=north west] at ($(FigC.north east) - (0,0.12)$);
\node [anchor=north] at ($(FigC.south)$) {\sf Ours};
\node[anchor=south] (FigD) at (3.579,0) {\includegraphics[trim={200 100 300 100 }, clip ,width=3.5cm]{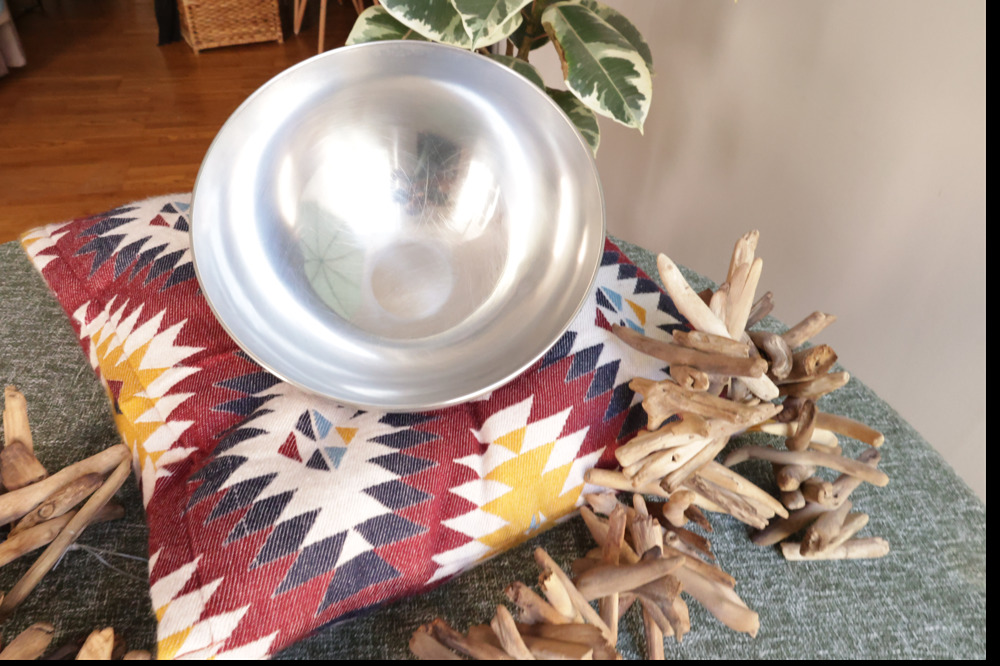}};
\spy [closeup,magnification=2] on ($(FigD)+(0.,0.7)$) in node[largewindow,anchor=south east] at ($(FigD.south west) + (0,0.12)$);
\node [anchor=north] at ($(FigD.south)$) {\sf Ground Truth};
\end{tikzpicture}

\begin{tikzpicture}[xscale=4, spy using outlines={every spy on node/.append style={smallwindow}}]
\node[anchor=south] (FigA) at (0,0) {\includegraphics[trim={100 100 400 100 }, clip ,width=3.5cm]{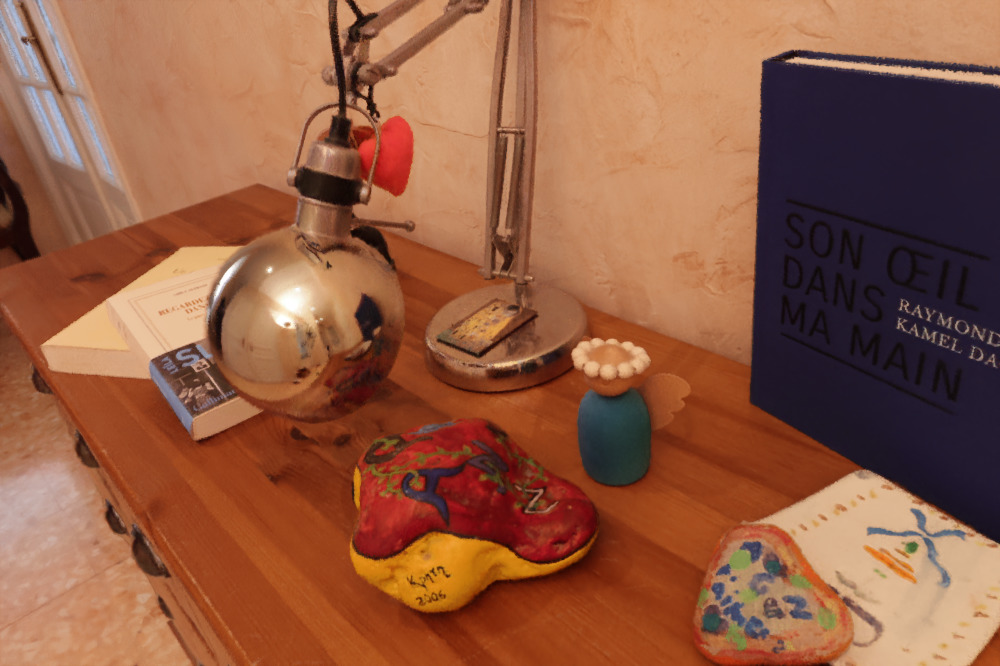}};
\spy [closeup,magnification=2] on ($(FigA)+(-0.45,-0.2)$) in node[largewindow,anchor=north west] at ($(FigA.north east) - (0,0.12)$);
\node [anchor=north] at ($(FigA.south)$) {\sf Ours};
\node[anchor=south] (FigB) at (1.329,0) {\includegraphics[trim={100 100 400 100 }, clip ,width=3.5cm]{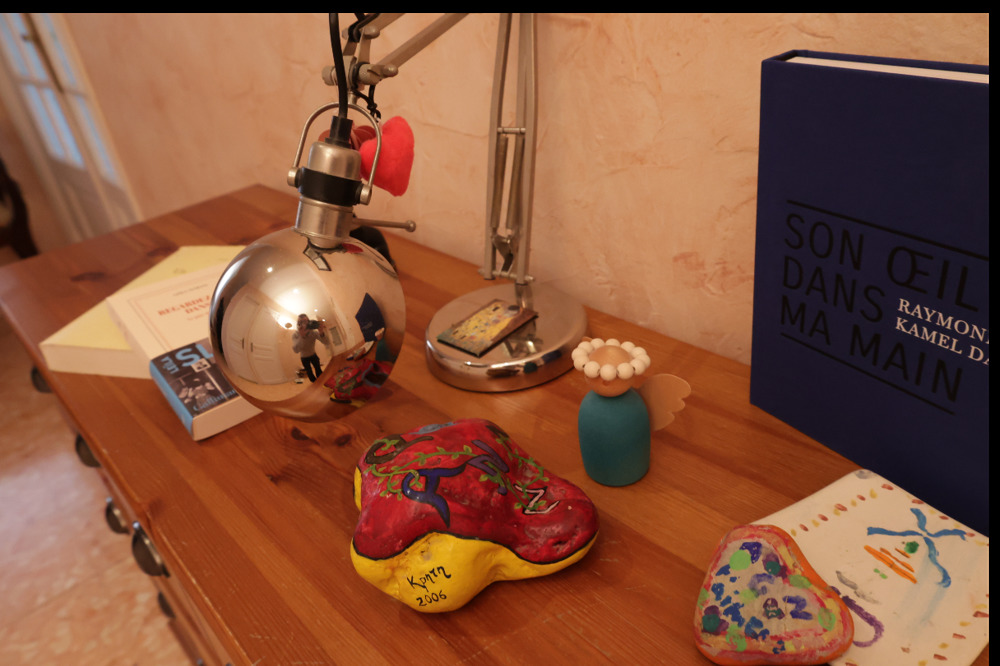}};
\spy [closeup,magnification=2] on ($(FigB)+(-0.45,-0.2)$) in node[largewindow,anchor=south east] at ($(FigB.south west) + (0,0.12)$);
\node [anchor=north] at ($(FigB.south)$) {\sf Ground Truth};
\node[anchor=south] (FigC) at (2.25,0) {\includegraphics[trim={50 100 450 100 }, clip ,width=3.5cm]{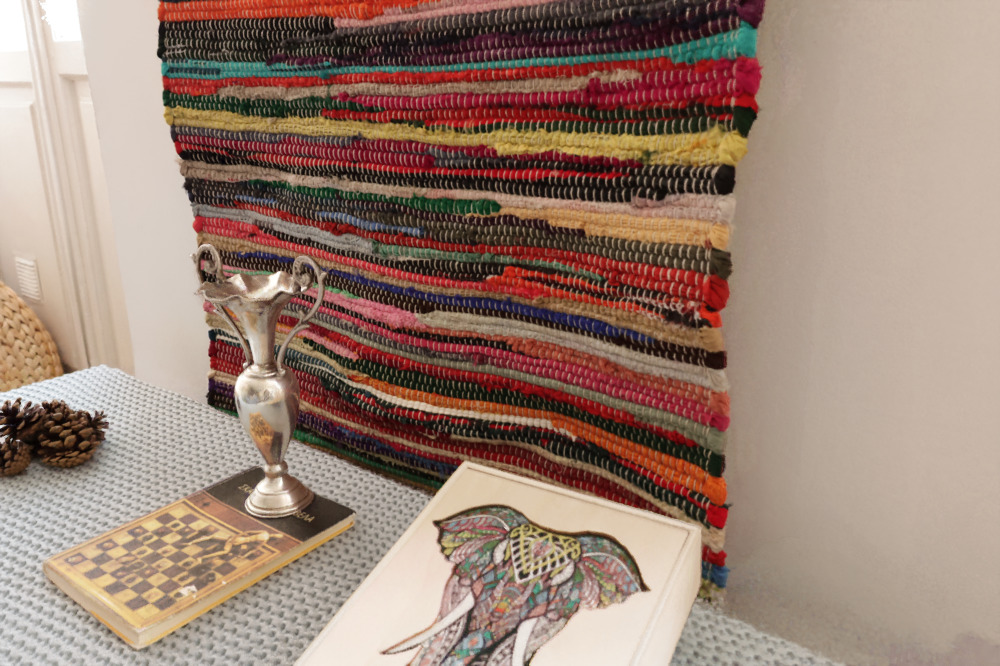}};
\spy [closeup,magnification=2] on ($(FigC)+(-0.15,-0.7)$) in node[largewindow,anchor=north west] at ($(FigC.north east) - (0,0.12)$);
\node [anchor=north] at ($(FigC.south)$) {\sf Ours};
\node[anchor=south] (FigD) at (3.579,0) {\includegraphics[trim={50 100 450 100 }, clip ,width=3.5cm]{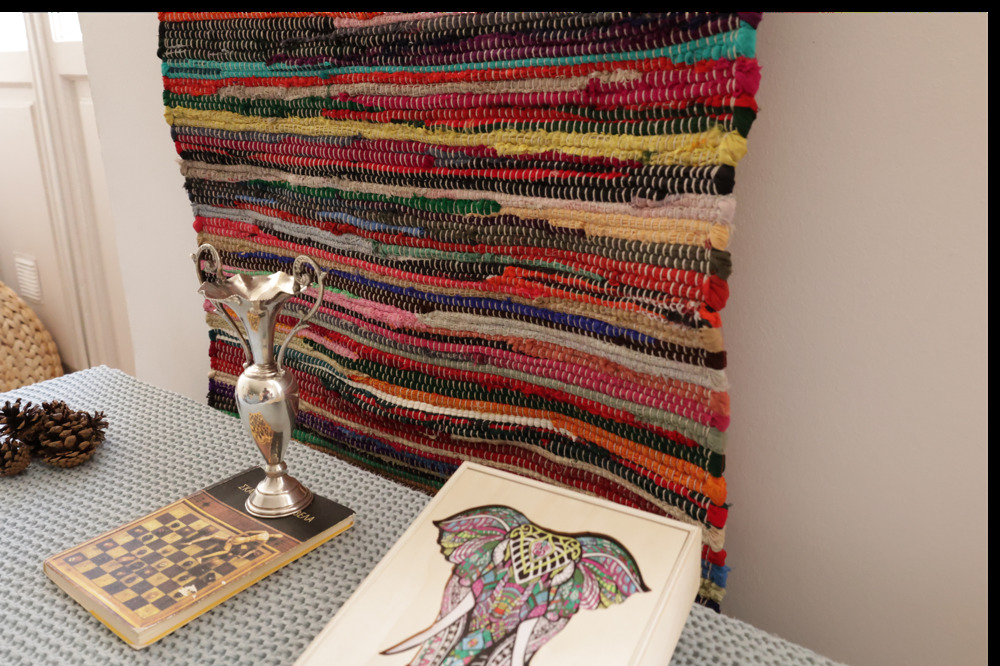}};
\spy [closeup,magnification=2] on ($(FigD)+(-0.15,-0.7)$) in node[largewindow,anchor=south east] at ($(FigD.south west) + (0,0.12)$);
\node [anchor=north] at ($(FigD.south)$) {\sf Ground Truth};
\end{tikzpicture}
\caption{
\label{fig:results}
Results of our method on \textsc{Compost, ConcaveBowl, HallwayLamp, SilverVase}: left is our rendering, right is the ground truth from images not in the input views.
Note how our renderings faithfully capture reflections.
}
\end{figure*}
\section{Results \& Evaluation}

We first present results of our method, then discuss applications to different manipulations of reflections, and finally present quantitative and qualitative evaluations.

We captured 5 scenes \RB{for evaluation}: one outdoors scene, \textsc{Compost}, and four indoors scenes, \textsc{SilverVase}, \textsc{HallwayLamp}, \textsc{ConcaveBowl}, and \textsc{CrazyBlade}. The last two contain concave reflectors.
Since we model motion of reflections, our results are best appreciated in video paths; we provide these for all scenes and comparisons in the supplemental.

\subsection{Results}

We optimize our end-to-end model for approximately 36 hours on a single RTX8000 GPU for each scene. 
We show results compared to ground truth images from \emph{paths away from the input views} in Fig.~\ref{fig:results}. 
Our method recovers and renders sharp reflections which move smoothly (please see supplemental videos), even for complex cases such as the concave reflectors in \textsc{CrazyBlade} and \textsc{ConcaveBowl}.
Overall, the shape, position and motion of the reflections is close to the ground truth.

\begin{figure}[!ht]
\includegraphics[width=0.99\linewidth]{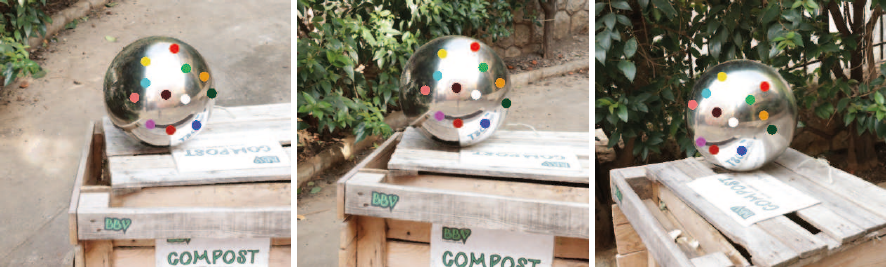}
\caption{
Our neural warpfield formulation naturally establishes correspondences of reflections across views, allowing to track reflections also in the presence of severe deformations. Example correspondences are marked as colored dots; please also refer to the supplemental video.
}
\label{fig:correspondences}
\end{figure}

\subsection{Applications}

Besides novel-view synthesis, our approach naturally enables a wider range of applications:

\subsubsection{Correspondences in reflections}

A powerful consequence of our Lagrangian design is direct support for dense correspondences in non-rigidly deforming reflections across views.
In Fig.~\ref{fig:correspondences} we show how reflection points are faithfully tracked across different views, by following the projected learned trajectories of virtual reflection points as defined by our neural catacaustic field.
We envision downstream applications like specular surface reconstruction \cite{roth2006specular, sankaranarayanan2010specular} and view-coherent image annotations \cite{caelles2017one} of reflected objects to greatly benefit from these dense correspondences.

\subsubsection{Scene and reflection editing}

\begin{figure}[!ht]
\includegraphics[width=0.99\linewidth]{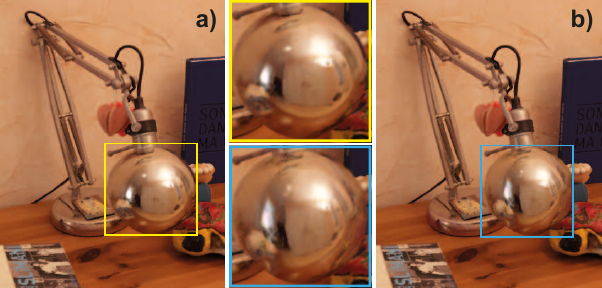}
\caption{
Reflection editing.
(\emph{a}) Original rendering.
(\emph{b}) Edited reflection, magnifying the right part of the table. The edited reflections correspond to a camera to the left of the primary view camera.
}
\label{fig:editing}
\end{figure}

In addition to physical correctness, some rendering applications also need to consider artistic goals.
Reflections are particularly amenable to artistic modification for their ability to convey spatial relations, while human observers are tolerant to deviations from physical correctness \cite{ramanarayanan2007visual}.
Tools for reflection editing have been developed for synthetic scenes \cite{ritschel2009interactive, schmidt2013path}.
Even though our pipeline is trained to faithfully reproduce input views, after training our approach allows expressive editing of reflections by decoupling the camera used for rasterization from the camera that is fed to the neural catacaustic field responsible for warping the reflections.
This way we can move reflections in a way that is coherent both spatially and across views, enabling believable edits of reflected perspective (Fig.~\ref{fig:editing}, and supplemental video).

\begin{figure}[!ht]
\includegraphics[width=0.99\linewidth]{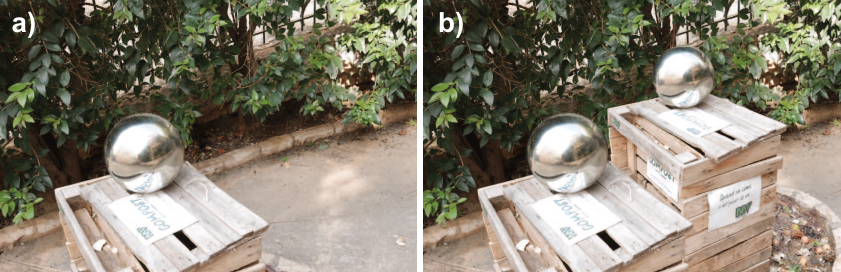}
\caption{
Our explicit scene representation allows versatile editing. 
(\emph{a}) Original scene.
(\emph{b}) Cloning of the \specpc~ as well as part of the \diffpc .
The reflections on the two spheres are different, resulting in plausible perspective across novel views (see supplemental video).
}
\label{fig:cloning}
\end{figure}

Further, our explicit scene representation supports general editing. 
In Fig.~\ref{fig:cloning} we show an object cloning example.
To achieve this result, we copy the \diffpoints~ contained in the volume we wish to clone, and transform them. We then replicate the \specpc~ and apply the inverse transformation to the camera used to warp the \specpoints. Note that we can only perform translations of the cloned objects for now; rotations would probably require specific training or parameterization.

\subsubsection{Comfortable stereo}

Stereoscopic rendering of reflections, \eg for VR applications, is challenging -- especially for curved reflectors, where excessive horizontal and vertical disparities can occur, significantly impairing viewing comfort \cite{lambooij2009visual, shibata2011zone}.
While solutions exist to enforce a comfortable disparity range for diffuse appearance \cite{jones2001controlling, lang2010nonlinear}, the corresponding problem for specular surfaces has only been addressed for \emph{synthetic} scenes with full control over the image generation process \cite{dkabala2014manipulating, templin2012highlight}.

\begin{figure}[!ht]
\includegraphics[width=0.99\linewidth]{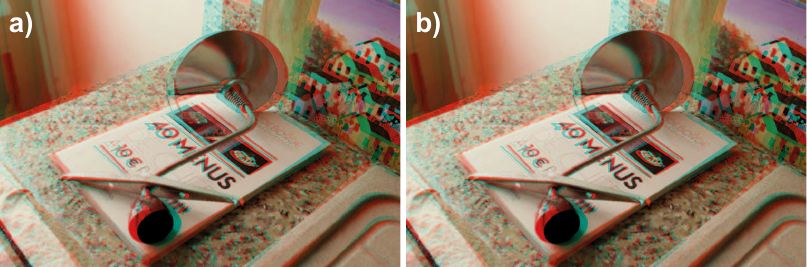}
\caption{
Our method supports comfortable stereo rendering of reflections. 
(\emph{a}) Curved reflectors frequently result in uncomfortable binocular disparities (most prominent at the top left of the blade).
(\emph{b}) Our approach allows for an explicit modification of disparities, preventing visual discomfort.
Use anaglyph glasses for stereo impression.
}
\label{fig:stereo}
\end{figure}

Our approach enables control over binocular disparity arising from curved reflectors in \emph{casually captured} scenes. 
Again, we exploit that we can decouple the stereo cameras used for rasterization from the stereo cameras steering the warpfield: 
We simply decrease the inter-ocular distance between the latter stereo cameras in case of uncomfortably large binocular disparities (Fig.~\ref{fig:stereo}).
In the limit of a single (cyclopean) camera as input to the catacaustic field, reflections for both eyes lie at the same point on the learned catacaustic, resulting in stereo characteristics of a diffuse scene element with an offset from the reflector surface \cite{templin2012highlight}, preventing visual discomfort.
Notice that the diffuse \RB{(primary)} rendering branch of our pipeline is not affected by this modification.

\subsection{Evaluation}
\label{sec:eval}

We evaluate our approach qualitatively and quantitatively in terms of rendering quality. We first show comparisons to previous work and then present ablations evaluating the tradeoffs of our different choices.

For our quantitative tests, in addition to our input views, we captured a high-speed sequence of photographs using the same settings, which gives paths of about 10-15 fps. We thus have ground truth sequences of images that are completely distinct from the input views for quantitative evaluation. 
Since we want to evaluate the quality of our renderings of \emph{reflections}, we compute all image metrics in the region of the image covered by the project of the \polytope.
All paths and renderings of all methods for these quantitative evaluations are provided in supplemental material. We have selected subsequences of these paths and provide high fps interpolation to better evaluate reflection quality and motion, also presented in the supplemental.

\begin{figure}[!h]
\includegraphics[width=0.99\linewidth]{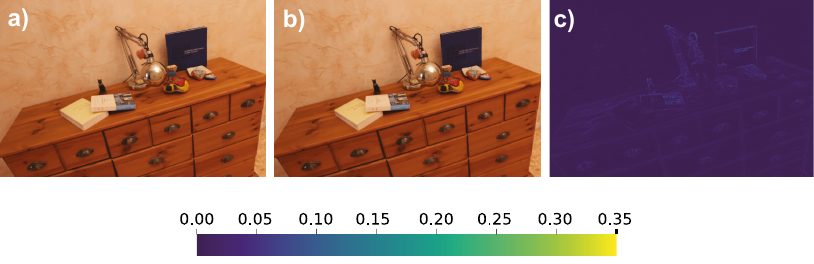}
\caption{
\label{fig:interactive-render}
(\emph{a}) Full renderer.
(\emph{b}) Interactive renderer, running at 5\,fps in our unoptimized Python implementation.
(\emph{c}) \RB{Absolute normalized} difference between \emph{a}) and \emph{b}).
}
\end{figure}

We also show the effect on quality of our unoptimized interactive renderer that runs 5 fps in Fig.~\ref{fig:interactive-render}.

\subsection{Comparisons}
\label{sec:comparisons}

We compare to MipNeRF~\cite{barron2021mipnerf} which is the current state-of-art NeRF method with code available~\footnote{At time of submission RefNeRF~\cite{verbin2021ref} code was not available.}, and to three interactive methods: Deep Blending~\cite{HPPFDB18}, the Point-Based Neural Rendering (PBNR) method of Kopanas et al.~\shortcite{kopanas2021point}, and Instant Neural Graphics Primitives~\cite{mueller2022instant}. All methods were trained with the same resolution images (1000x666). We trained MipNeRf for 1M iterations with the default batch size, for approximately 48 hours.
We compare all renderings with our full unoptimized method, currently running at 1.5s/frame.

\begin{figure*}[!h]
\setlength{\tabcolsep}{0.8pt}
\centering
\begin{tabular}{cccccc}
GT&Ours&\cite{HPPFDB18}&\cite{mueller2022instant}&\cite{barron2021mipnerf}&\cite{kopanas2021point} \\
\hline
\includegraphics[trim={380 100 200 100 }, clip ,width=2.96cm]{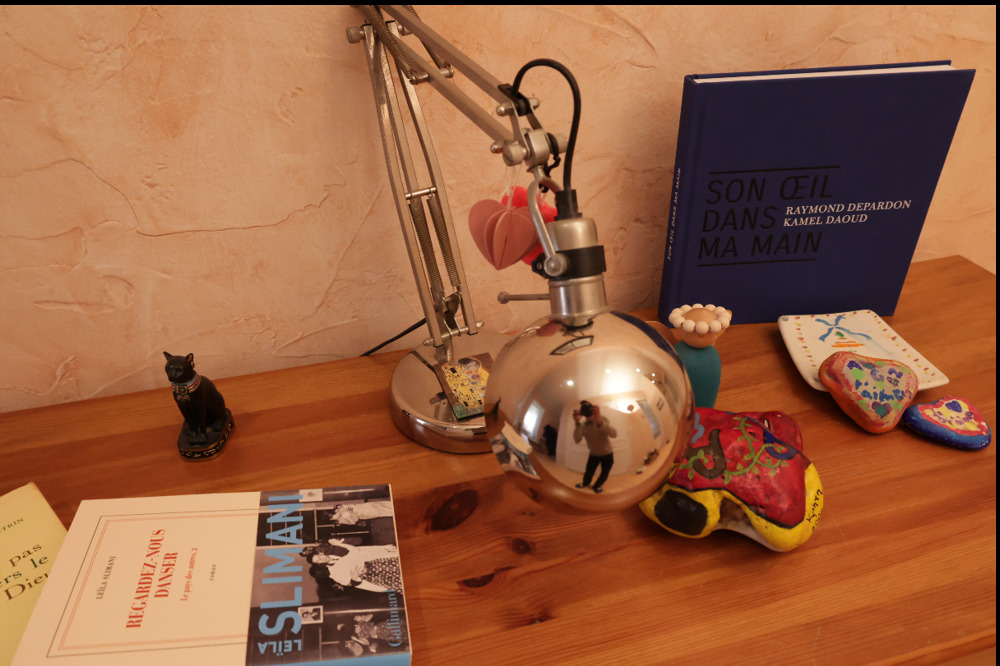}&
\includegraphics[trim={380 100 200 100 }, clip ,width=2.96cm]{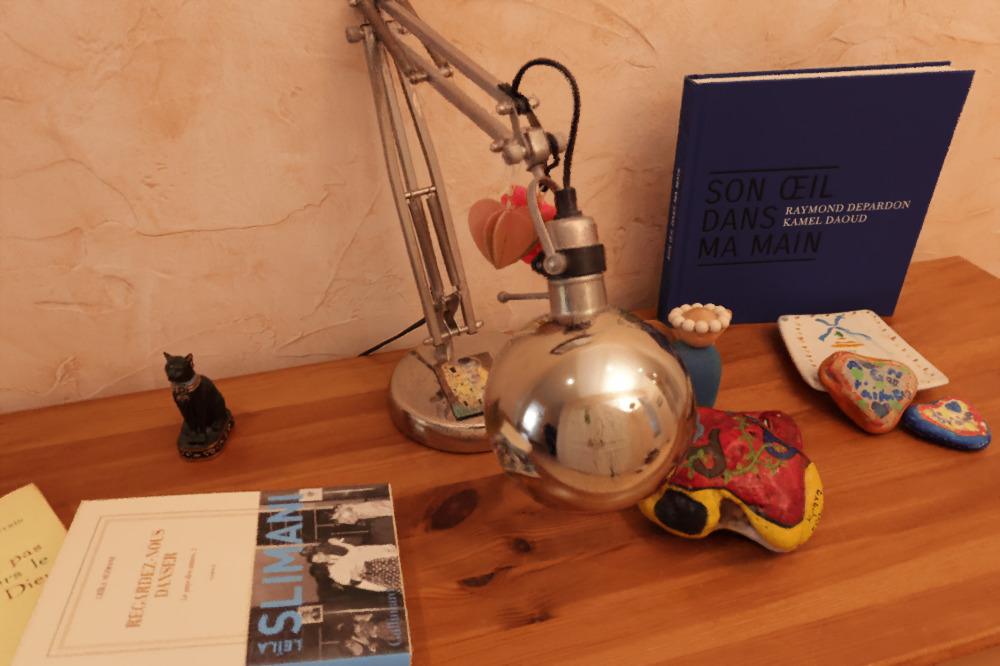}&
\includegraphics[trim={380 100 200 100 }, clip ,width=2.96cm]{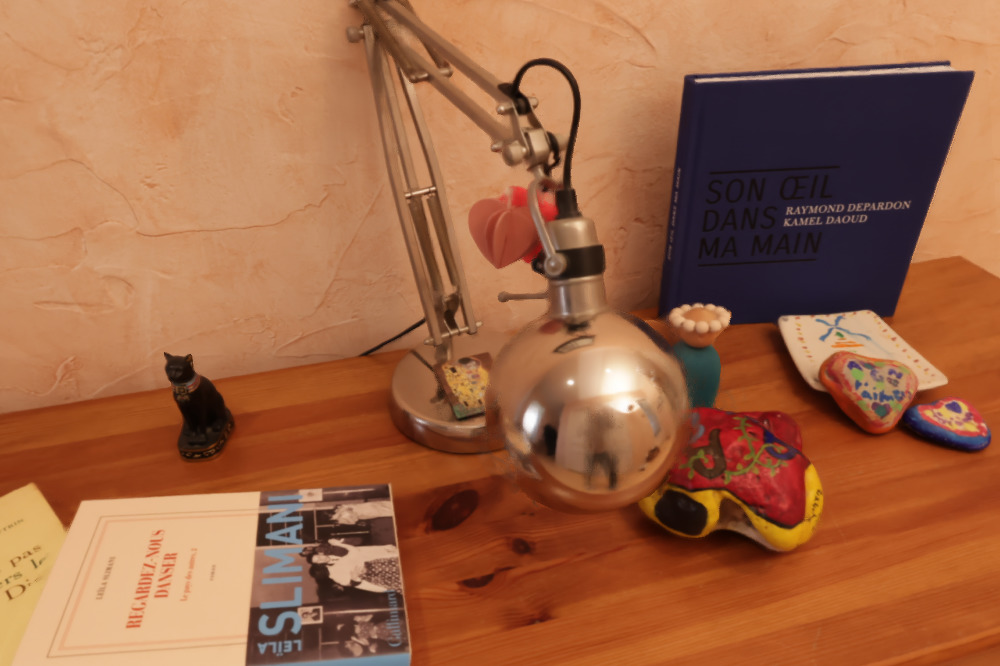}&
\includegraphics[trim={380 100 200 100 }, clip ,width=2.96cm]{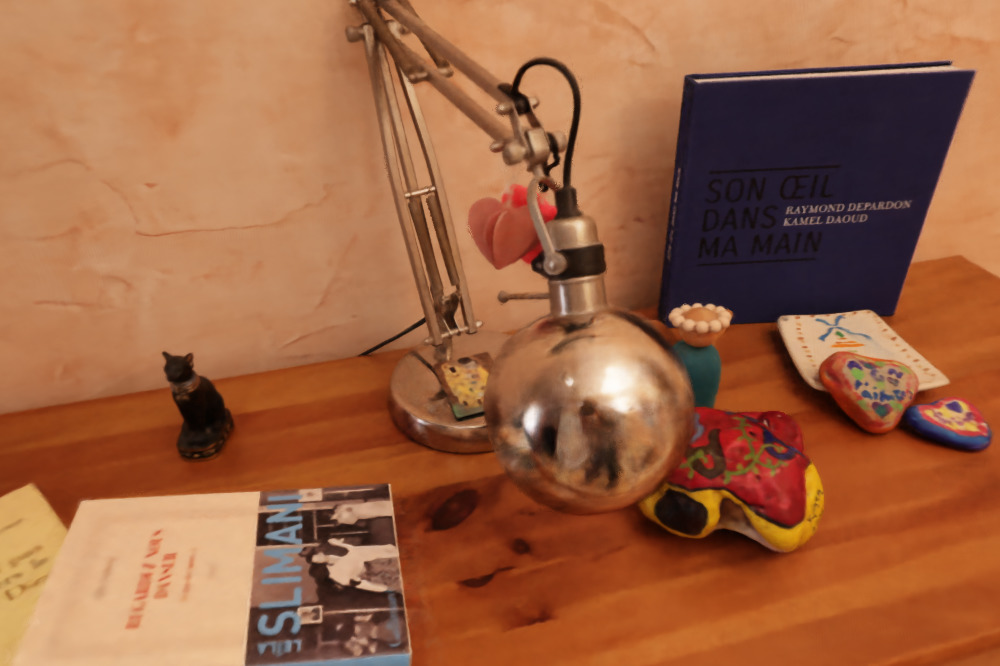}&
\includegraphics[trim={380 100 200 100 }, clip ,width=2.96cm]{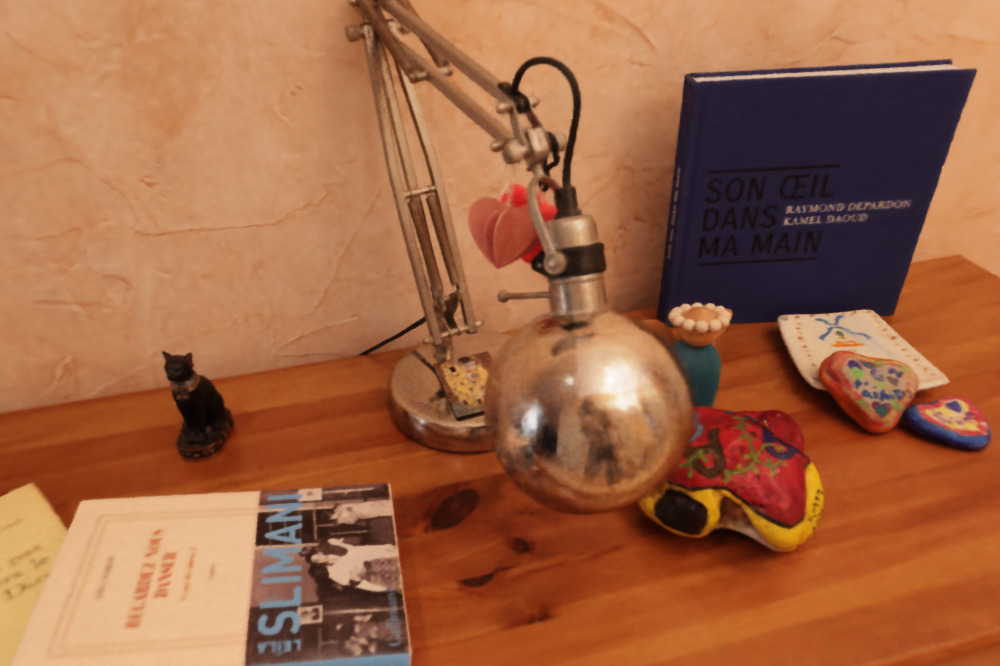}&
\includegraphics[trim={380 100 200 100 }, clip ,width=2.96cm]{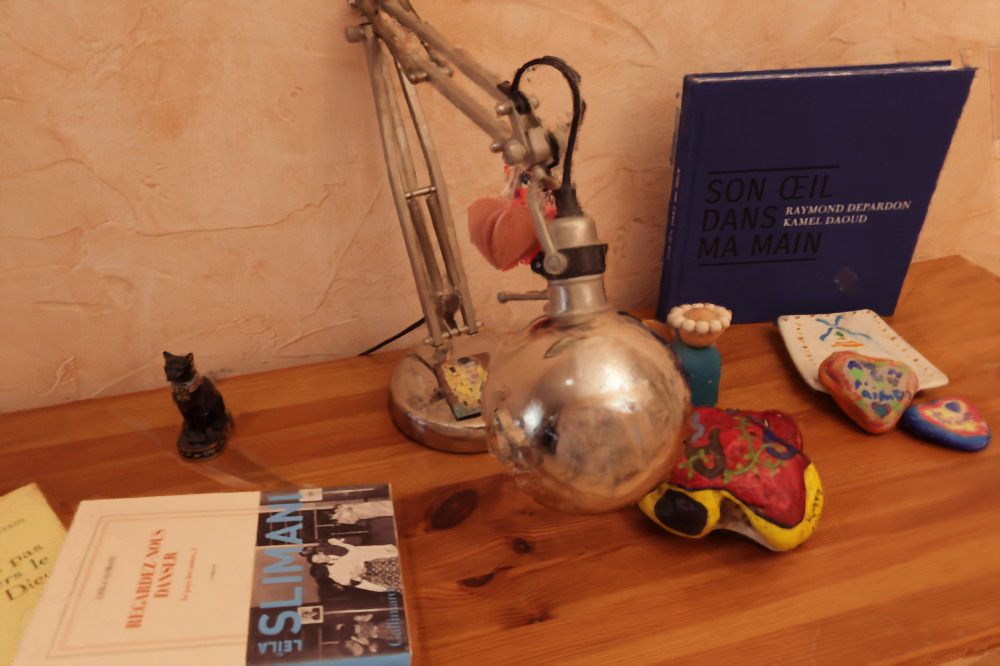}\\

\includegraphics[trim={350 200 455 200 }, clip ,width=2.96cm]{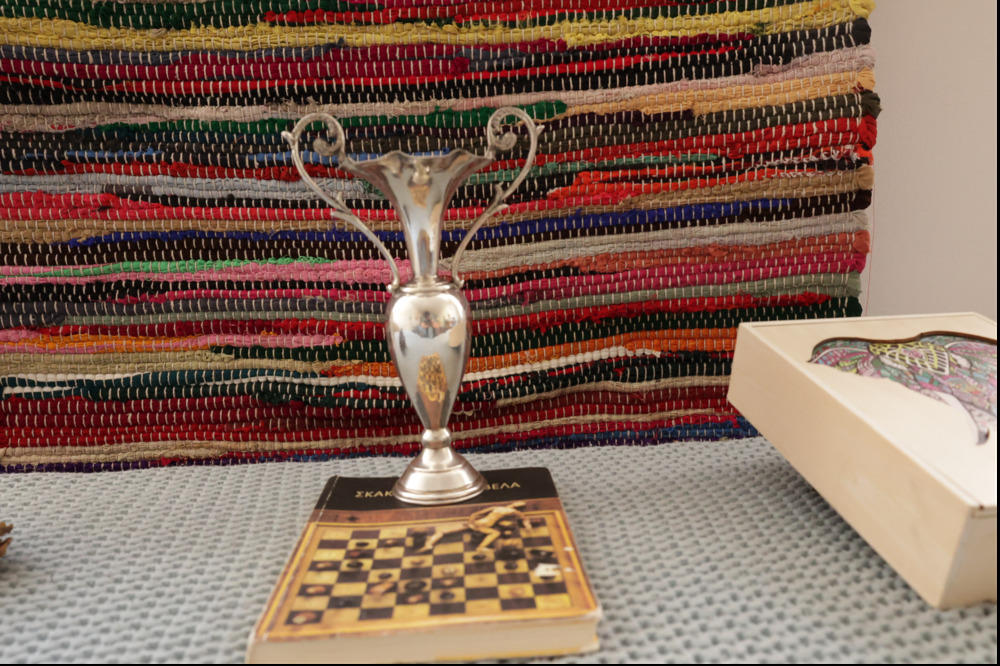}&
\includegraphics[trim={350 200 455 200 }, clip ,width=2.96cm]{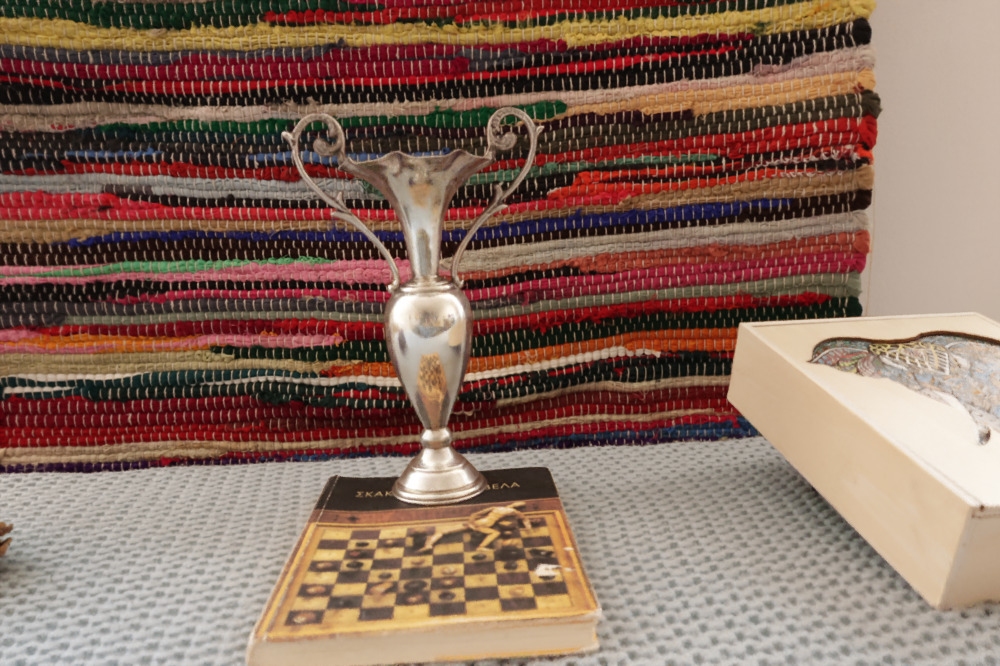}&
\includegraphics[trim={350 200 455 200 }, clip ,width=2.96cm]{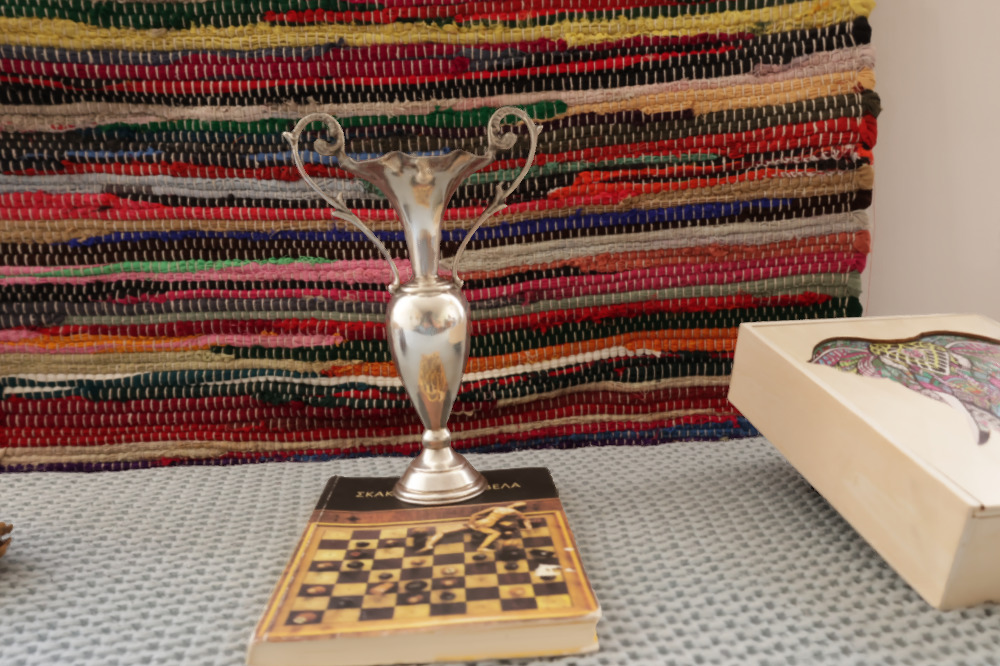}&
\includegraphics[trim={350 200 455 200 }, clip ,width=2.96cm]{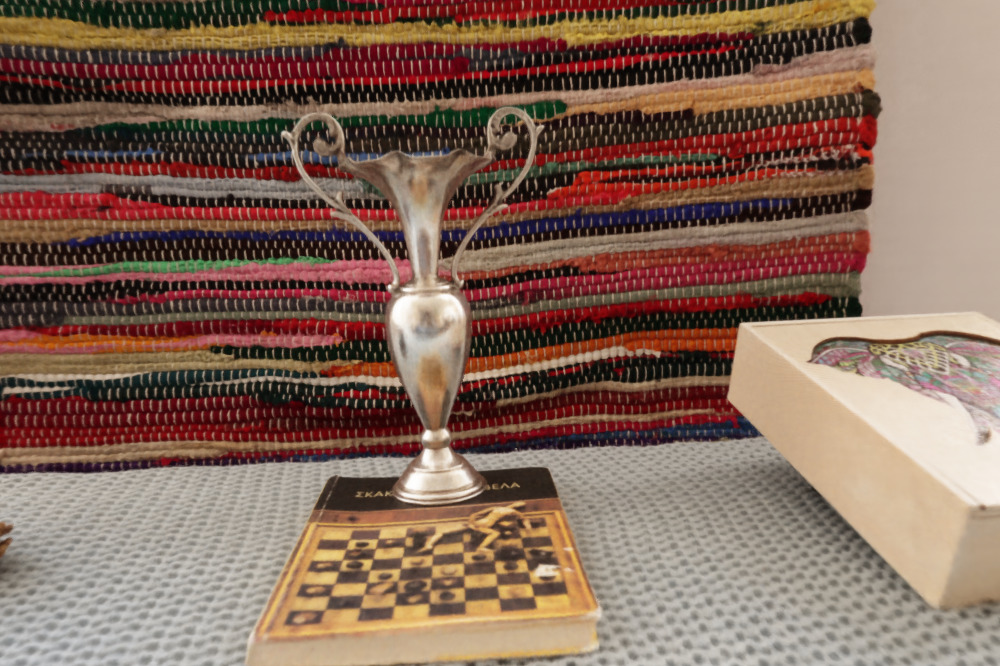}&
\includegraphics[trim={350 200 455 200 }, clip ,width=2.96cm]{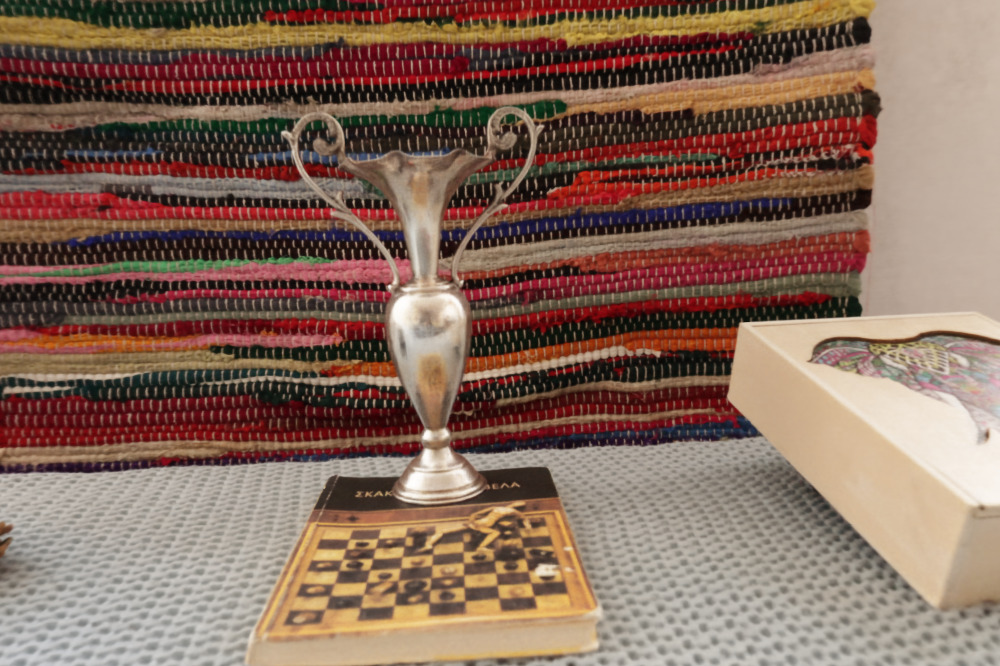}&
\includegraphics[trim={350 200 455 200 }, clip ,width=2.96cm]{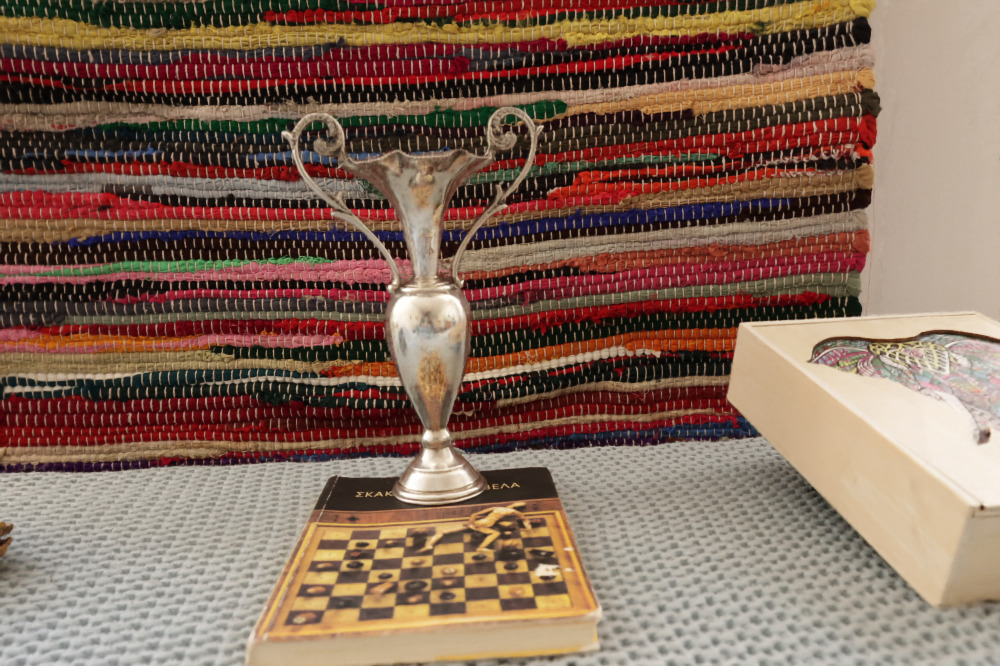}\\

\includegraphics[trim={230 300 550 150 }, clip ,width=2.96cm]{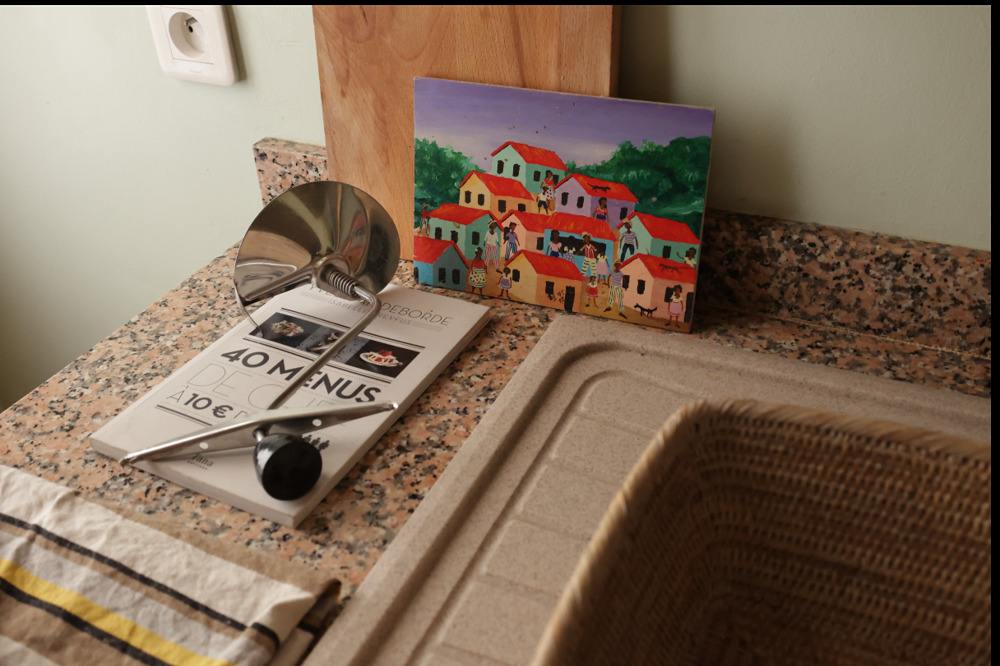}&
\includegraphics[trim={230 300 550 150 }, clip ,width=2.96cm]{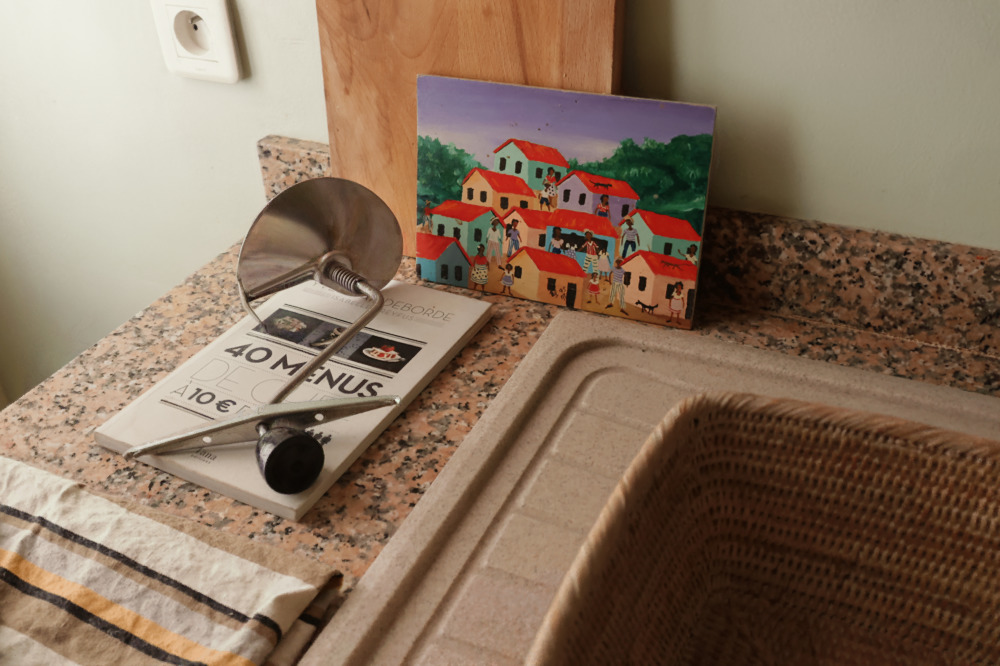}&
\includegraphics[trim={230 300 550 150 }, clip ,width=2.96cm]{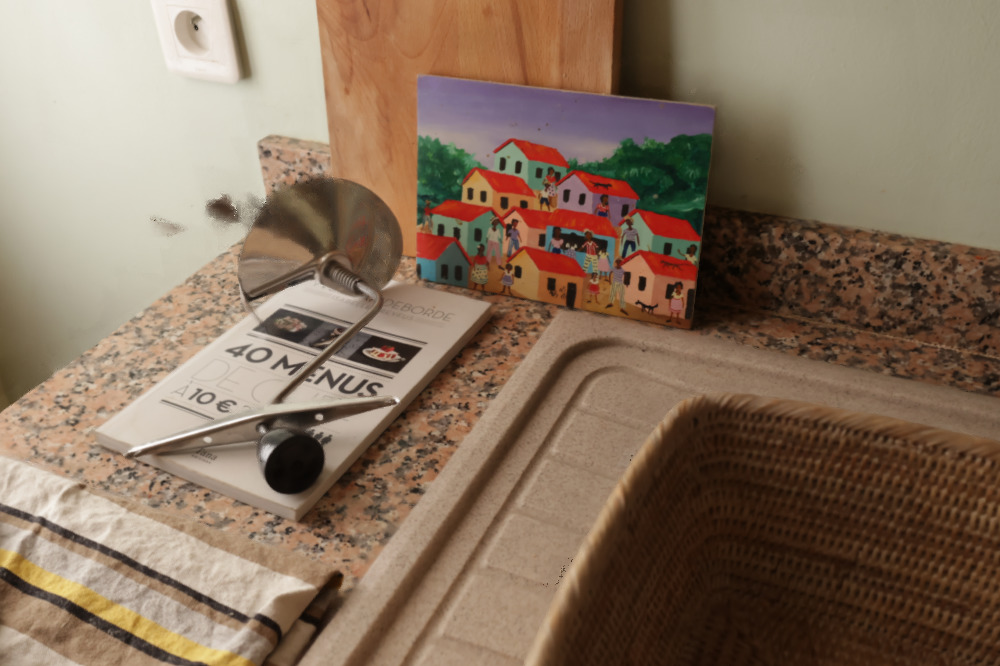}&
\includegraphics[trim={230 300 550 150 }, clip ,width=2.96cm]{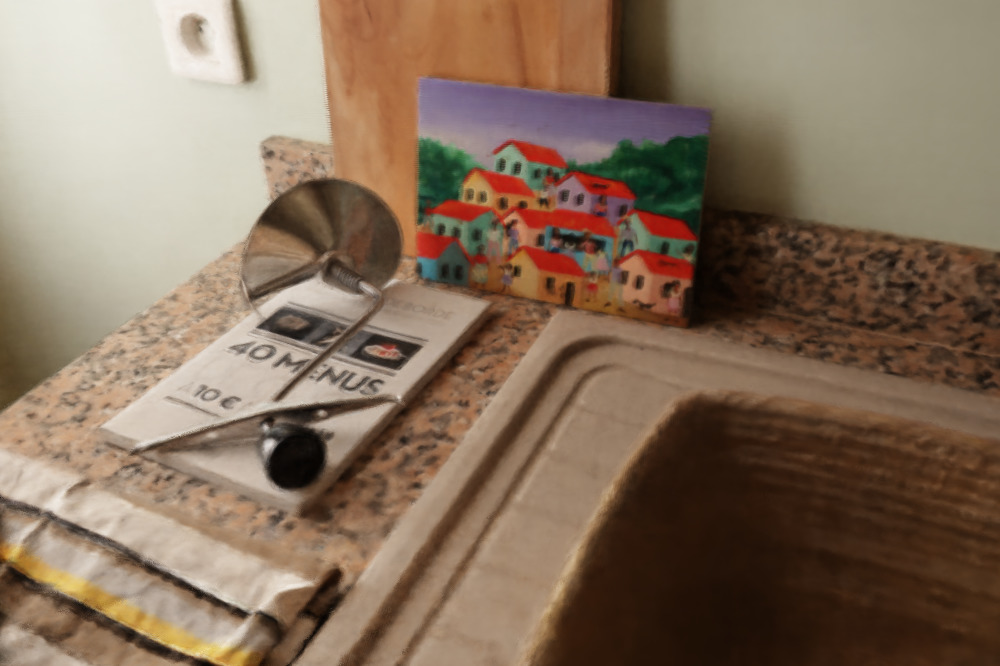}&
\includegraphics[trim={230 300 550 150 }, clip ,width=2.96cm]{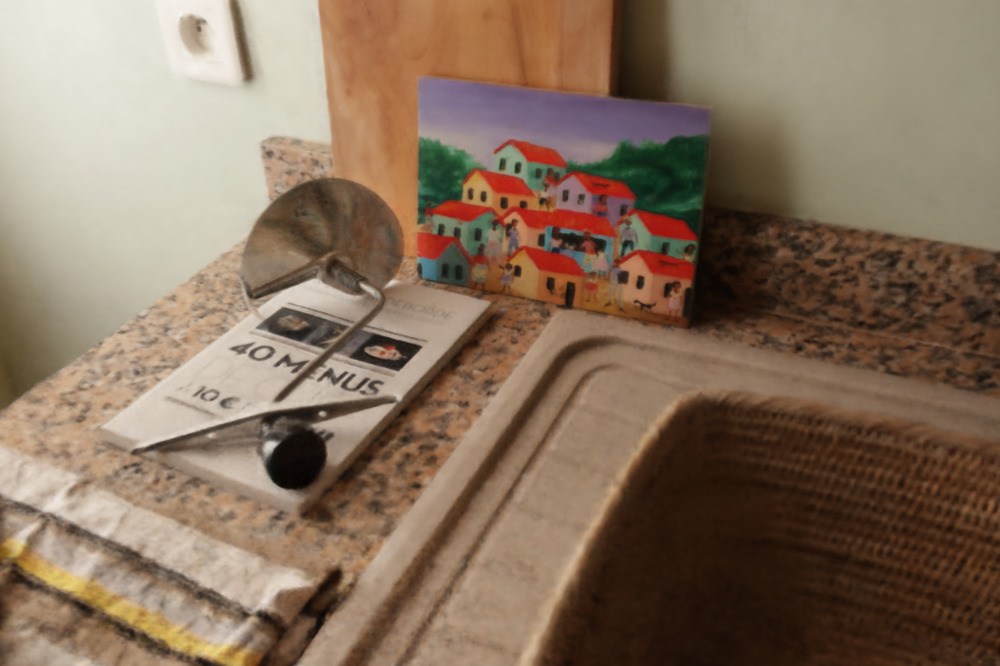}&
\includegraphics[trim={230 300 550 150 }, clip ,width=2.96cm]{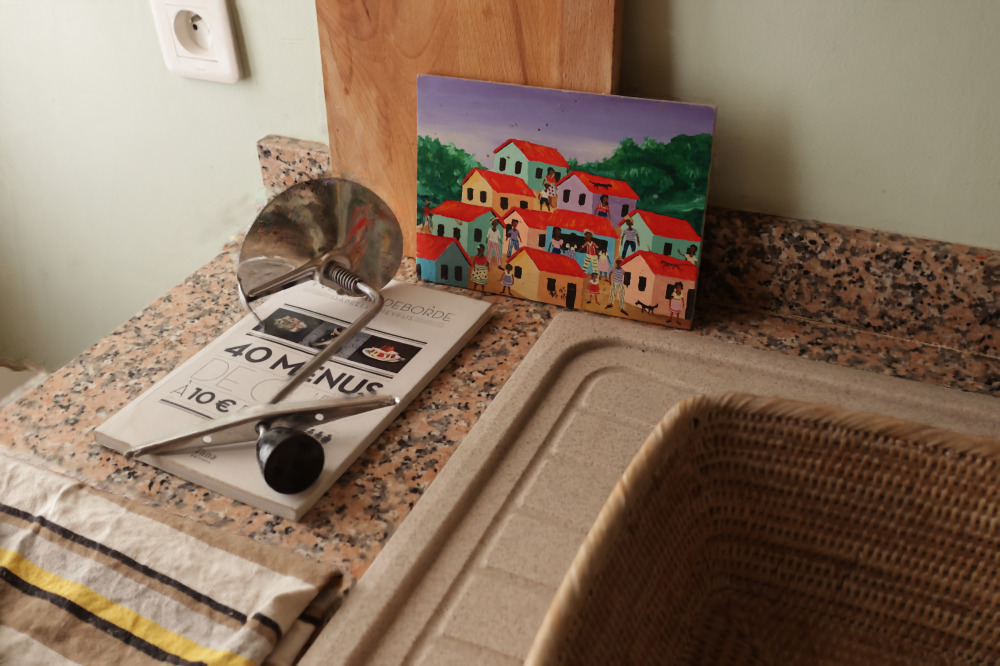}\\

\includegraphics[trim={160 50 320 50 }, clip ,width=2.96cm]{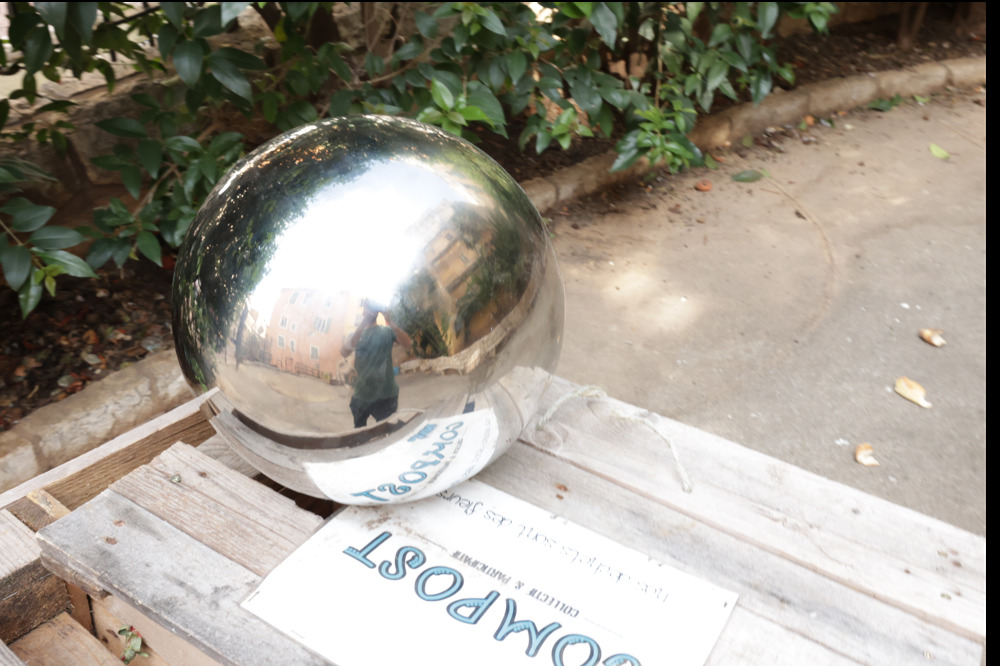}&
\includegraphics[trim={160 50 320 50 }, clip ,width=2.96cm]{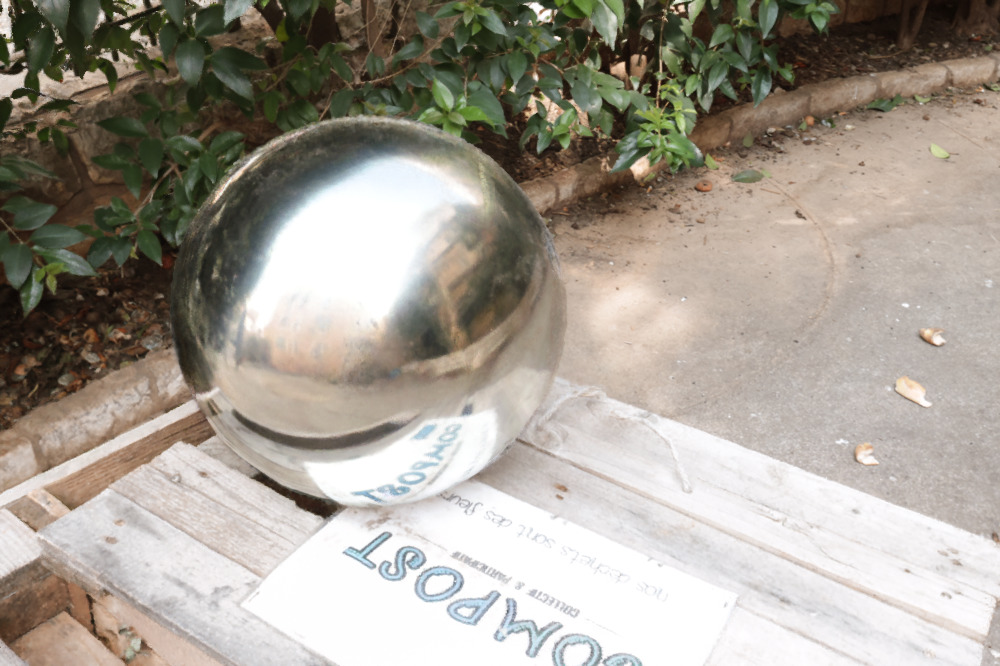}&
\includegraphics[trim={160 50 320 50 }, clip ,width=2.96cm]{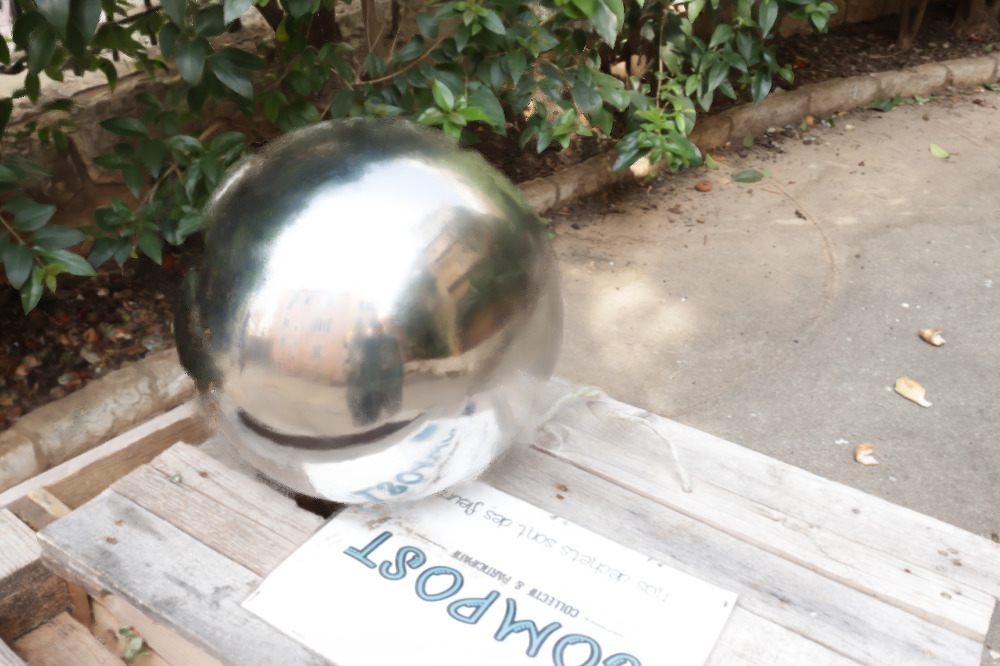}&
\includegraphics[trim={160 50 320 50 }, clip ,width=2.96cm]{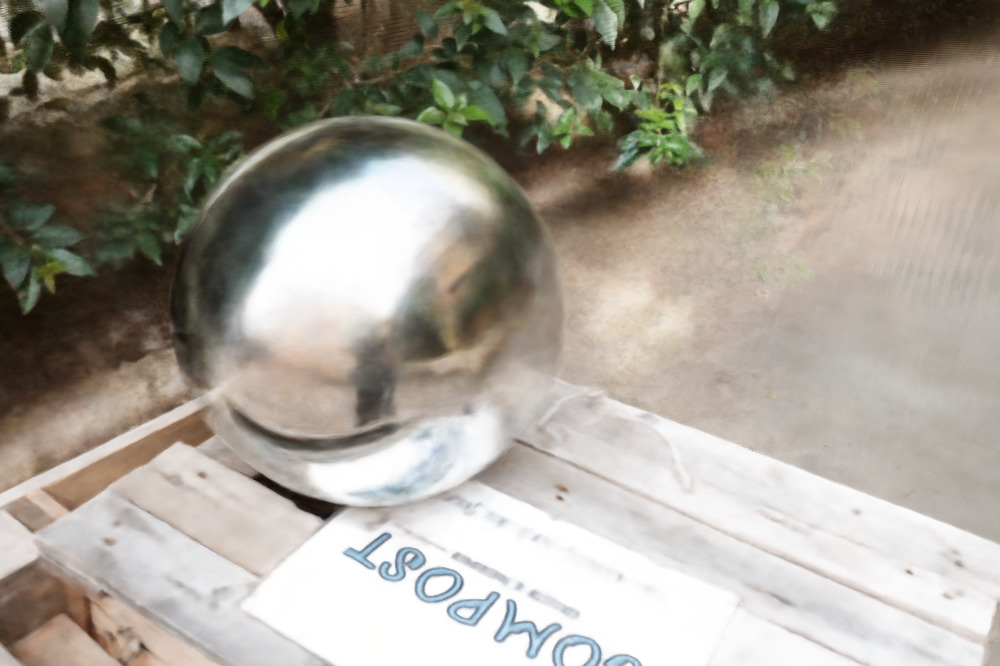}&
\includegraphics[trim={160 50 320 50 }, clip ,width=2.96cm]{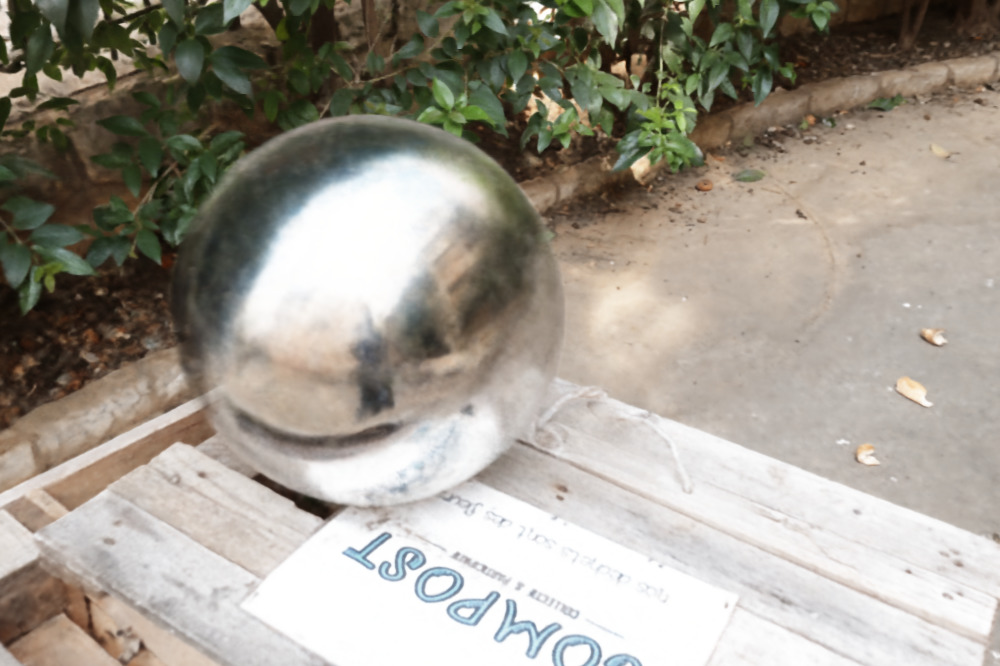}&
\includegraphics[trim={160 50 320 50 }, clip ,width=2.96cm]{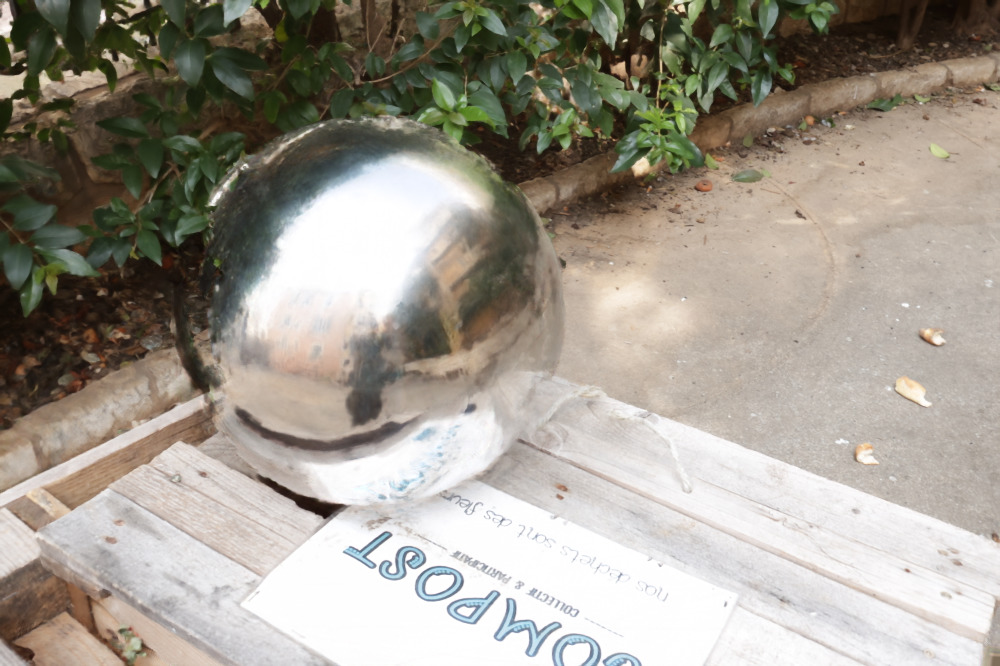}\\

\includegraphics[trim={180 200 400 0 }, clip ,width=2.96cm]{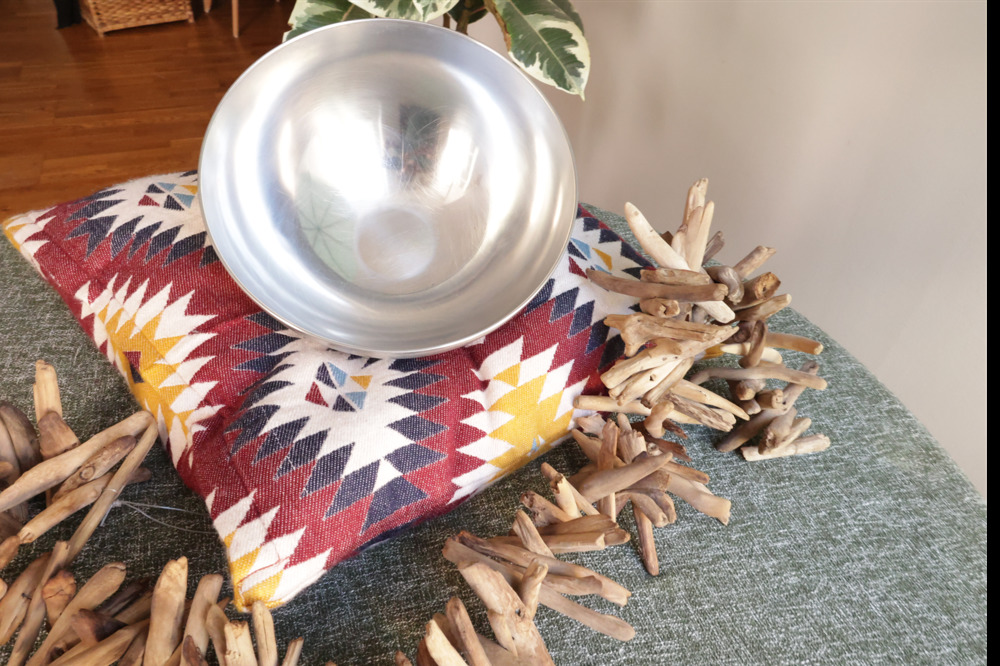}&
\includegraphics[trim={180 200 400 0 }, clip ,width=2.96cm]{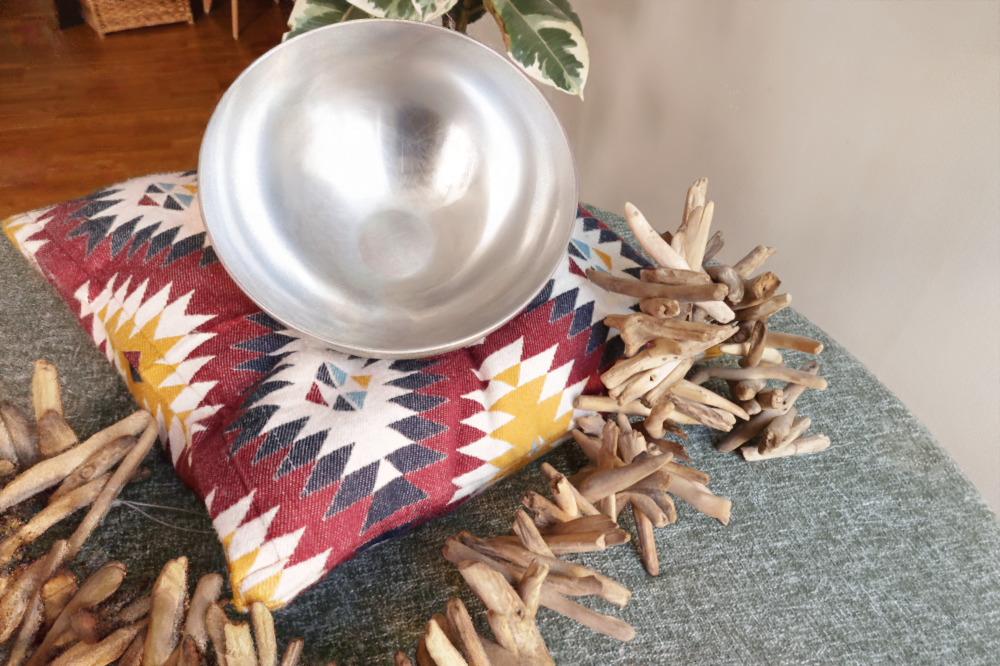}&
\includegraphics[trim={180 200 400 0 }, clip ,width=2.96cm]{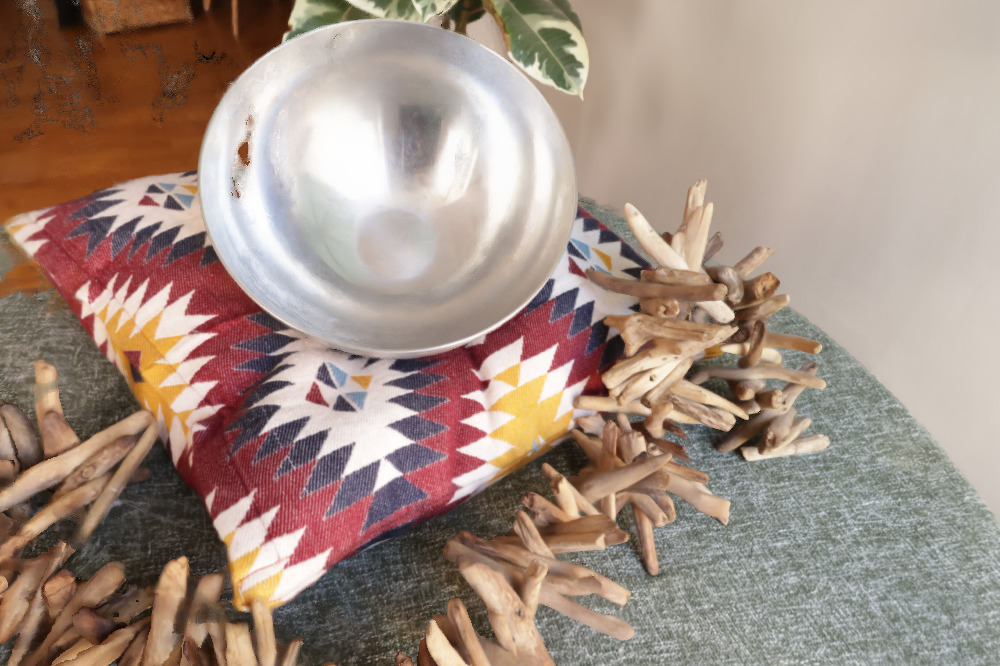}&
\includegraphics[trim={180 200 400 0 }, clip ,width=2.96cm]{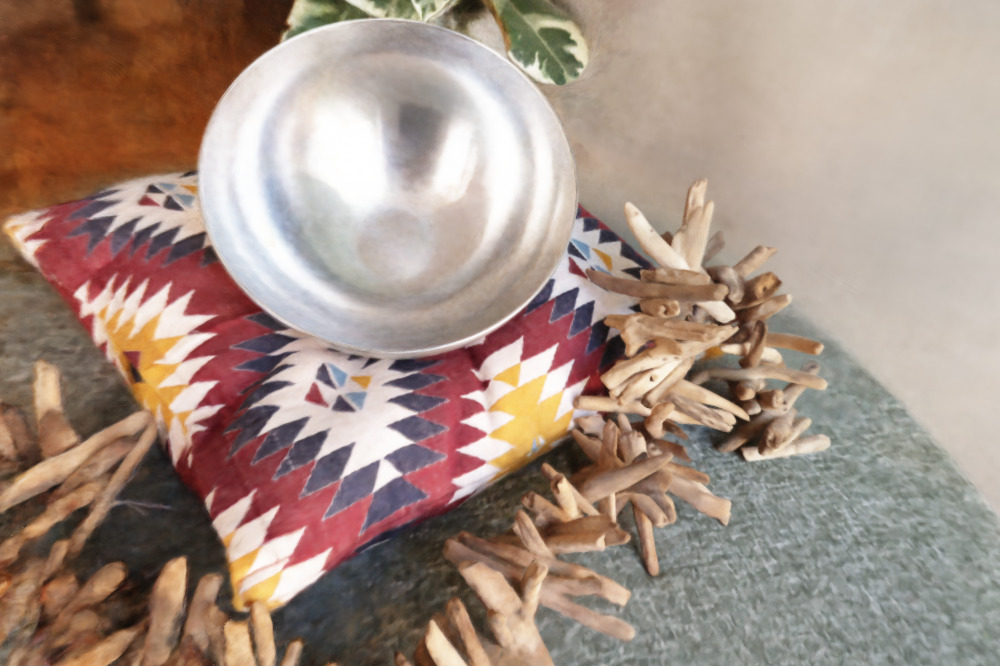}&
\includegraphics[trim={180 200 400 0 }, clip ,width=2.96cm]{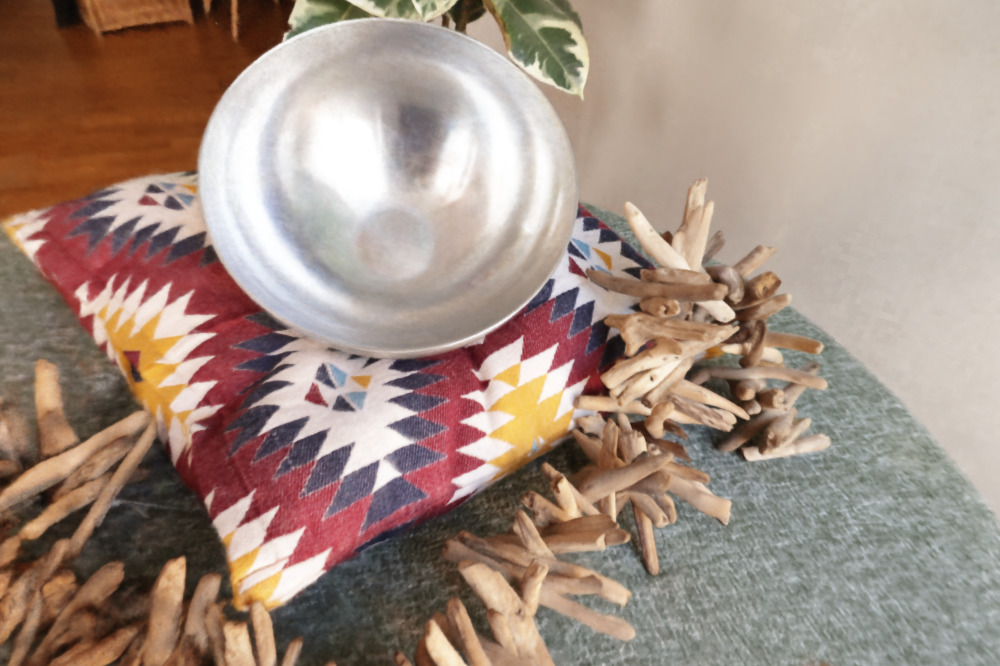}&
\includegraphics[trim={180 200 400 0 }, clip ,width=2.96cm]{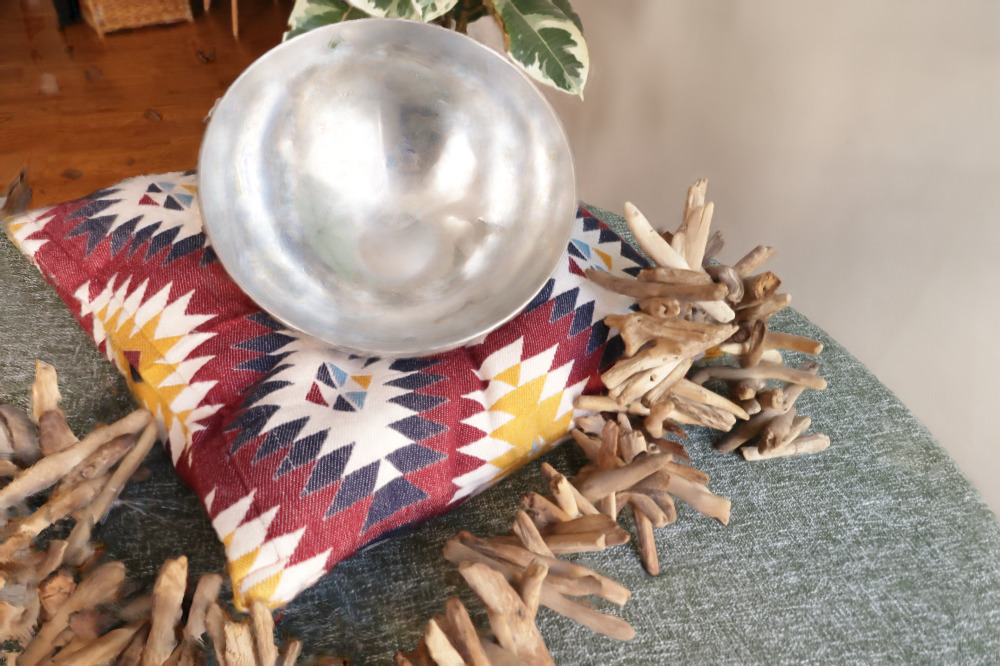}\\

\end{tabular}

\caption{
\label{fig:comparisons}
We show comparisons to previous methods and the corresponding ground truth images from paths not in the input views. The scenes are, from the top down:
\textsc{HallwayLamp}, 
\textsc{SilverVase}, 
\textsc{CrazyBlade}, 
\textsc{Compost}, 
\textsc{ConcaveBowl}. 
}
\end{figure*}

Fig.~\ref{fig:comparisons} shows qualitative comparisons of our method to previous work and the ground truth. We see that our method renders sharper and more accurate reflections in the majority of cases. Surprisingly, Deep Blending performs very well; however, it is prone to catastrophic failure for some views (see \textsc{ConcaveBowl} in Fig.~\ref{fig:comparisons}, where our method has recovered the missing geometry). Deep Blending  also does not scale with the number of images: GPU memory requirements grow with the number of images -- 300 images is about the limit of what current GPUs can handle for this method. In contrast our approach uses approximately the same amount GPU memory for any scene (around 8Gb).
Deep Blending and PBNR suffer from severe flickering and temporal artifacts in some scenes; please see \textsc{Compost} and \textsc{ConcaveBowl} in supplemental.

In Tab.~\ref{tab:comparisons} we show quantitative results for all methods averaged over all scenes. The error is computed on views not in the input views, on paths that are distinct from those used for capture. We compute D-SSIM, PSNR and L-PIPS error; our method has the best quantitative results compared to all other methods, although we are only marginally better than Deep Blending, which however has severe temporal artifacts not captured by these metrics (please see video).

\begin{table}[!h]
\caption{
	\label{tab:comparisons} \RB{Quantitative evaluation of our method compared to previous work, computed over our five scenes, on separate paths captured specifically for evalution, and separate from the input views.}
}
\small
	\begin{tabular}{|l|ccc|}
		\toprule
		 & $SSIM^\uparrow$   & $PSNR^\uparrow$    & $LPIPS^\downarrow$  \\ \midrule
	MipNeRF~\cite{barron2021mipnerf} & \cellcolor{white!40} 0.9772 & \cellcolor{yellow!40}34.6753  & \cellcolor{white!40} 0.0258 \\
	InstantNGP~\cite{mueller2022instant} & \cellcolor{white!40} 0.9769 & \cellcolor{white!40} 34.3459  & \cellcolor{white!40} 0.0253 \\ 
	Point-Based NR~\cite{kopanas2021point} & \cellcolor{yellow!40} 0.9790 &  \cellcolor{white!40} 34.3463  & \cellcolor{yellow!40} 0.0229 \\
	Deep Blending~\cite{HPPFDB18}   & \cellcolor{orange!40} 0.9832 & \cellcolor{orange!40} 35.6316 & \cellcolor{orange!40} 0.0197 \\
	Ours & \cellcolor{red!40} 0.9845 &  \cellcolor{red!40} 35.8522 & \cellcolor{red!40} 0.0179 \\
	\midrule

\end{tabular}

\end{table}

\subsection{Ablations}

We also executed an ablation study to better understand the dynamics and the influence of several components in our system. Our experiments consist of runs
on \textsc{Compost} for 200k iterations (0.5 - 1 day). The different elements of the ablation are: 
i) \emph{Primary-Only}, where we render the scene only with the \diffpc; 
ii) \emph{No-Densification}, where we do not densify the original point cloud we get from MVS.
iii) \emph{No-}$\mathcal{L}_\textrm{DSSIM}$ and \emph{No-}$\mathcal{L}_{m_{TV}}$, where we zero out these two loss terms.
iv) \emph{Half-Warp-MLP}, where we cut down the number of parameters of our warping field MLP in half.

\begin{table}[!h]
\caption{
	\label{tab:ablations}
	Quantitative effect of our ablation study for the different components of our algorithm, using the same methodology as Tab.~\ref{tab:comparisons}. 
}
		\begin{tabular}{|l|ccc|}
		\hline
		 & $SSIM^\uparrow$   & $PSNR^\uparrow$    & $LPIPS^\downarrow$  \\ \midrule
		Primary-Only   & \cellcolor{white!40} 0.9689  & \cellcolor{white!40} 32.0426  & \cellcolor{white!40} 0.0303                                                \\ 
		No-Densification     & \cellcolor{white!40} 0.9727  & \cellcolor{white!40} 32.6982  & \cellcolor{white!40} 0.0273                                                 \\ 
		No-$\mathcal{L}_\textrm{DSSIM}/\mathcal{L}_{m_{TV}}$ & \cellcolor{white!40} 0.9704  & \cellcolor{white!40} 33.2101  & \cellcolor{white!40} 0.0284                                                 \\ 
		Half-Warp-MLP  & \cellcolor{white!40} 0.9690  & \cellcolor{white!40} 32.5671  & \cellcolor{white!40} 0.0301                                                \\ 
		Full           &  \cellcolor{red!40} 0.9745 &  \cellcolor{red!40} 33.7041 & \cellcolor{red!40} 0.0254 \\ 
		\midrule 
		\end{tabular}
\end{table}

\begin{figure}[!h]
	\includegraphics[width=0.99\linewidth]{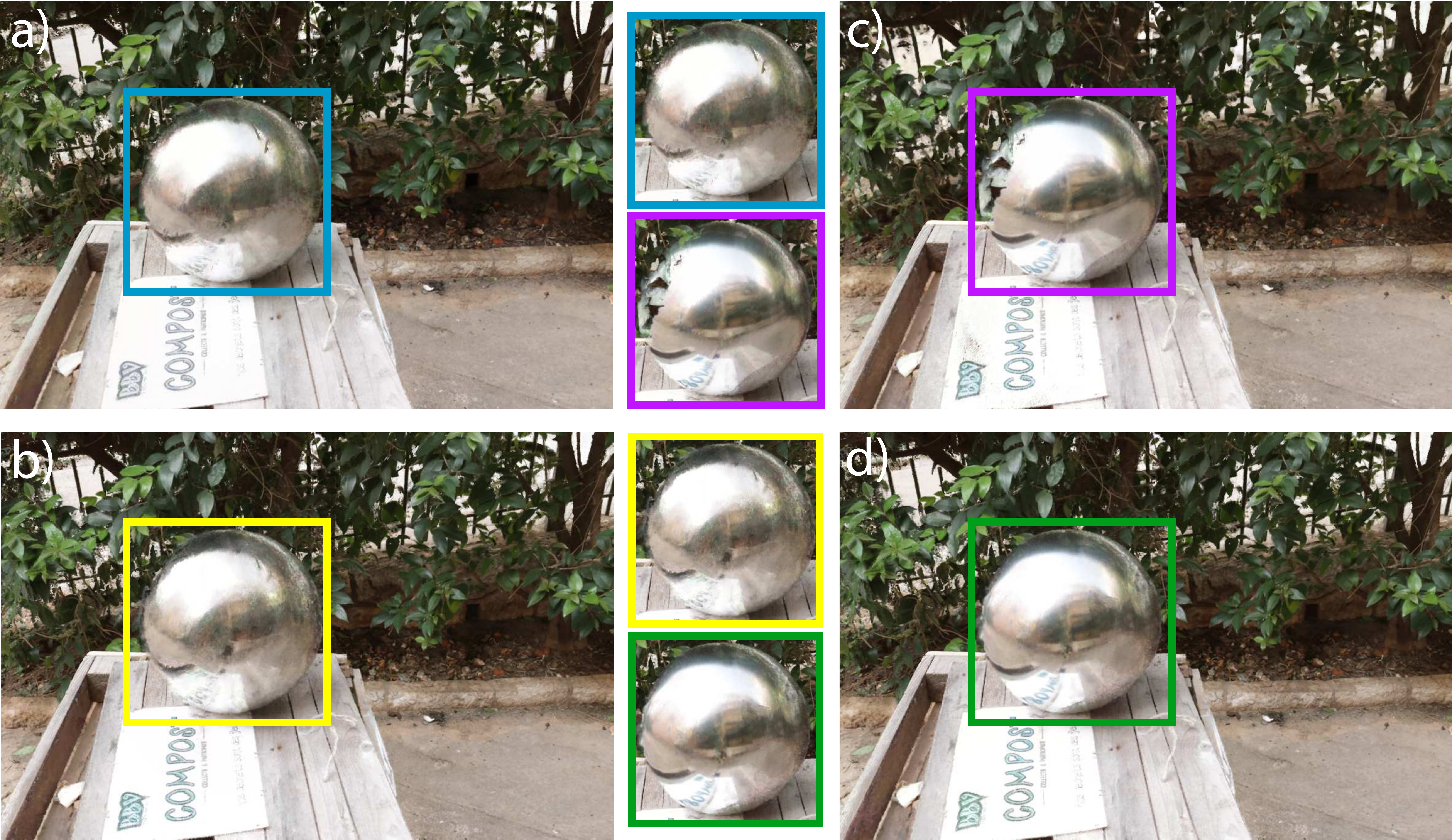}
\caption{
\label{fig:ablations}
We show the effect of the various components of our methods on the results by ablation runs. a) Primary-only, b) Half-Warp-MLP, c) No-Densification, d) Full Method.
}
\end{figure}

\begin{figure}[!h]
	\includegraphics[width=0.99\linewidth]{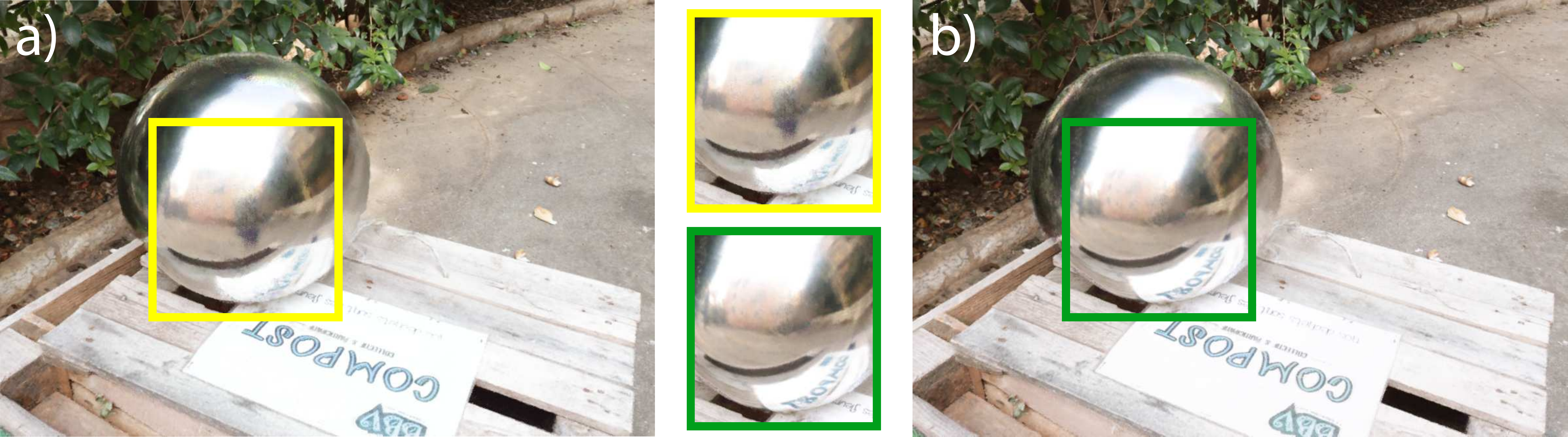}
\caption{
\label{fig:ablations-dssim}
$\mathcal{L}_\textrm{DSSIM}$ and $\mathcal{L}_{m_{TV}}$ have the surprising effect of removing the photographer from the reflections.
}
\end{figure}

As we can see in Tab.~\ref{tab:ablations} and Fig.~\ref{fig:ablations},\ref{fig:ablations-dssim}, each component of our method improves the results, in particular the importance of the \warpfield~ is clear, as well as the densification and the MLP capacity.
Primary-Only and Half-Warp-MLP cannot recover the movement of reflections and result
in blurrier renderings. The \emph{No-Densification} configuration struggles to recover from large errors in the reconstructions of the geometry during MVS.
Another interesting finding of our ablation is that the two loss terms $\mathcal{L}_\textrm{DSSIM}$ and $\mathcal{L}_{m_{TV}}$ to a large extend remove the photographer from the reflections.
This also has an effect on the numerical results in Tab.\ref{tab:ablations}, since the No-$\mathcal{L}_\textrm{DSSIM}/\mathcal{L}_{m_{TV}}$ configuration actually removes the photographer less, resulting in a bias with respect to the ground truth.
Depending on the application, this might or might not be desirable.

\begin{figure}[!h]
	\includegraphics[width=0.99\linewidth]{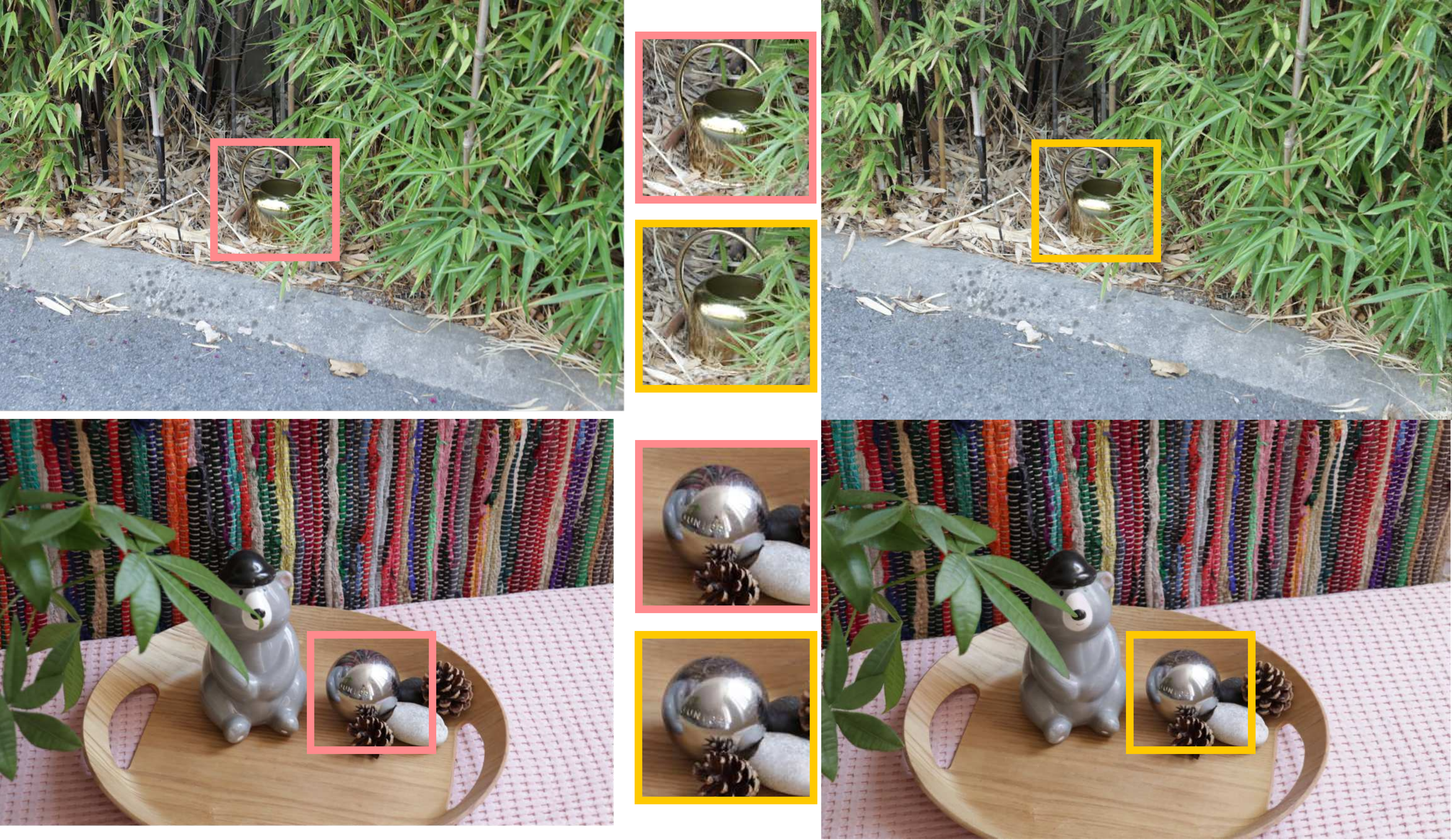}
	\caption{
		\label{fig:extra-scenes}
        Two scenes illustrating how our method handles occlusions of reflectors by the non-specular scene (vegetation in front of watering can in the first row), and handles multiple reflections (note the reflection on the bear in the reflections of the sphere in the second row). 
	}
\end{figure}

\RB{
\subsection{Occlusion, Texture and Multi-Bounce Reflections}
}

\RB{Our method handles occlusion from diffuse objects, reflectors with texture and multi-bounce reflections. To illustrate this, we provide two scenes specifically captured for this purpose, shown in Fig.~\ref{fig:extra-scenes}.}

\RB{When a diffuse object \emph{occludes} a reflector, the primary point cloud can reason about the occlusion because it contains both the reflector and the diffuse occluder. This information then propagates through the pipeline to the mask $\rho$. Specifically, in regions where the diffuse occluder is in front of the specular point cloud $\rho$ is $0$ and thus the specular point cloud will not be rendered. This can be seen in Fig.~\ref{fig:intermediate_buffers}; example renderings are shown in Fig.~\ref{fig:extra-scenes} (first row and insets). }

\RB{Specular reflectors (e.g., plastic or porcelain) can have diffuse texture. Our pipeline can model both in the separate point clouds and blends them using the value or $\rho$, correctly preserving the texture. We see this in Fig.~\ref{fig:extra-scenes} (second row), where reflections are correctly rendered despite the painted parts on the porcelain bear. However, we expect quality to degrade with more high frequency texture.}

\RB{Multiple specular objects with multi-bounce specular effects often appear together; they can be treated as one specular point cloud if they are close (e.g., for the lamp base and lampshade in \textsc{Hallway}). For multiple reflector objects far apart, multiple specular point clouds should be used. }

\RB{Since we do not explicitly model physical light transport, higher order reflections and global illumination do not become exponentially more complex as scene complexity increases. 
Such reflections are just a more complicated motion that the warp-field MLP needs to extract from the observations in the input images.  We see this in Fig.~\ref{fig:extra-scenes} (second row and insets), where the reflections on the bear are visible in the reflections of the sphere.  }

\subsection{Comparison to Physical Catacaustics}
\label{sec:gt-cata}

\begin{figure}[!h]
\includegraphics[width=0.99\linewidth]{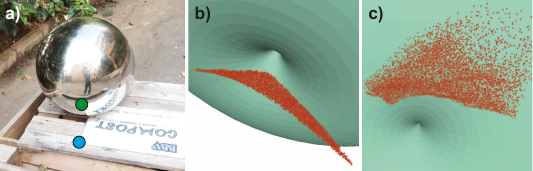}
\caption{
\label{fig:groundtruth}
Comparing neural point catacaustics to physical catacaustics.
(\emph{a}) We consider a spherical reflector, for which an analytic catacaustic surface exists.
(\emph{b},\emph{c}) Two views of the geometric configuration. The cyan mesh represents the ground truth catacaustic surface of the blue point in \emph{a}).
The red points are samples from our learned neural point catacaustics of the green point in \emph{a}). Please also refer to the main text. 
}
\end{figure}

There is no unique correct neural point catacaustic for a specific reflection due to the inherent depth ambiguity arising from our point-based representation (Sec.~\ref{sec:Background}; Fig. \ref{fig:CatacausticGeometry}d).
Our system finds \emph{a} solution that results in high-quality view-coherent renderings.
Nevertheless, we are interested in a qualitative analysis of the emerging catacaustic geometry.
To this end, we consider a scene with a known spherical reflector (Fig.~\ref{fig:groundtruth}a) and a reflected point with known position (blue point in Fig.~\ref{fig:groundtruth}a) and compute its analytic catacaustic surface (cyan mesh in Fig.~\ref{fig:groundtruth}b and c) \cite{glaeser1999reflections}.
Then, we randomly sample the volume of camera positions from which the corresponding reflection (green point in Fig.~\ref{fig:groundtruth}a) is visible, and record the output of our warp field for the corresponding specular point (red points in Fig.~\ref{fig:groundtruth}b and c).
While the two solutions roughly share some qualitative characteristics, they are quite dissimilar, despite resulting in almost identical projected reflection flow.

\section{Discussion \& Conclusion}

Our solution produces renderings of curved objects that are of higher visual quality and quantitatively better than all previous methods; our explicit representation of the \specpc~ has the additional advantage of enabling several kinds of scene manipulations, such as reflection editing, reflector cloning, comfortable stereo rendering and dense reflection tracking across views.

Nonetheless, our approach is not without limitations. We have currently tested on scenes where the reflector only covers a small part of the scene; we expect our method to work relatively well as long as SfM/MVS can produce a reasonable first approximation of the scene, but there will most probably be new issues to be dealt with. 

\RB{The focus of our work is on curved reflectors. Flat surfaces are a special case (Fig.~\ref{fig:CatacausticGeometry}(a)); we simply need to move specular points to their unique static positions. Our method would in principle be able to handle these cases, but would require adjusting our initialization to avoid excessive motion magnitudes. Depth ambiguity (Fig. ~\ref{fig:CatacausticGeometry}(d)) gives our warp field a lot of freedom in choosing point trajectories without sacrificing reflection quality. This tends to favor short and simple trajectories (Fig.~\ref{fig:groundtruth}).}

Our Lagrangian model assumes that view-dependent effects can be modeled by moving reflections. This assumption is violated for surfaces with \RB{high-frequency} spatially-varying specular materials. Even though we have demonstrated that our model handles smaller variations like surface scratches \RB{or painted porcelain}, we regard a principled solution to this exciting and challenging case as future work.

A main limitation of our method -- shared with most recent neural rendering methods -- is that we need to optimize/train our model for each scene. It is conceivable that the decoder MLPs could be adapted to be trained over a set of different scenes; training a general \warpfield~ across scenes is much more challenging.
Our training could be sped up significantly by accelerating our rasterization step. Several options could be investigated, starting with a variant of the layer-based solution we use for the interactive renderer during training, and moving on the various other approximations in the splat rasterization, although care is required for the backward pass. 

Our interactive renderer could be sped up significantly. Currently our interactive splatting-based renderer has a suboptimal implementation, in part due to memory copies between Pytorch, CUDA, and OpenGL; we expect to achieve better quality at true real-time rendering speeds with careful optimization.

In conclusion, we have presented a well-founded Lagrangian approach to render reflections from curved objects in captured multi-view scenes. We hope both our methodology, building on principles from geometric principles, and our direct point-based solution, will inspire novel solutions for other neural rendering problems.

\begin{acks}
This research was funded by the ERC Advanced grant FUNGRAPH No 788065 \textcolor{blue}{\url{http://fungraph.inria.fr}}. The authors are grateful to the OPAL infrastructure from Université Côte d’Azur  and for the HPC resources from GENCI–IDRIS (Grant 2022-AD011013409). The authors thank the anonymous reviewers for their valuable feedback, P.Hedman for proofreading earlier drafts, T.Louzi for the \textsc{SilverVase} object, S.Kousoula for help editing the video and S.Diolatzis for thoughtful discussions.
\end{acks}

\bibliographystyle{ACM-Reference-Format}
\bibliography{paper}


\begin{thebibliography}{79}


\ifx \showCODEN    \undefined \def \showCODEN     #1{\unskip}     \fi
\ifx \showDOI      \undefined \def \showDOI       #1{#1}\fi
\ifx \showISBNx    \undefined \def \showISBNx     #1{\unskip}     \fi
\ifx \showISBNxiii \undefined \def \showISBNxiii  #1{\unskip}     \fi
\ifx \showISSN     \undefined \def \showISSN      #1{\unskip}     \fi
\ifx \showLCCN     \undefined \def \showLCCN      #1{\unskip}     \fi
\ifx \shownote     \undefined \def \shownote      #1{#1}          \fi
\ifx \showarticletitle \undefined \def \showarticletitle #1{#1}   \fi
\ifx \showURL      \undefined \def \showURL       {\relax}        \fi
\providecommand\bibfield[2]{#2}
\providecommand\bibinfo[2]{#2}
\providecommand\natexlab[1]{#1}
\providecommand\showeprint[2][]{arXiv:#2}

\bibitem[\protect\citeauthoryear{Aliev, Sevastopolsky, Kolos, Ulyanov, and
  Lempitsky}{Aliev et~al\mbox{.}}{2020}]%
        {aliev2020neural}
\bibfield{author}{\bibinfo{person}{Kara-Ali Aliev}, \bibinfo{person}{Artem
  Sevastopolsky}, \bibinfo{person}{Maria Kolos}, \bibinfo{person}{Dmitry
  Ulyanov}, {and} \bibinfo{person}{Victor Lempitsky}.}
  \bibinfo{year}{2020}\natexlab{}.
\newblock \showarticletitle{Neural point-based graphics}. In
  \bibinfo{booktitle}{\emph{ECCV 2020}}. Springer, \bibinfo{pages}{696--712}.
\newblock


\bibitem[\protect\citeauthoryear{Barron, Mildenhall, Tancik, Hedman,
  Martin-Brualla, and Srinivasan}{Barron et~al\mbox{.}}{2021}]%
        {barron2021mipnerf}
\bibfield{author}{\bibinfo{person}{Jonathan~T. Barron}, \bibinfo{person}{Ben
  Mildenhall}, \bibinfo{person}{Matthew Tancik}, \bibinfo{person}{Peter
  Hedman}, \bibinfo{person}{Ricardo Martin-Brualla}, {and}
  \bibinfo{person}{Pratul~P. Srinivasan}.} \bibinfo{year}{2021}\natexlab{}.
\newblock \showarticletitle{Mip-NeRF: A Multiscale Representation for
  Anti-Aliasing Neural Radiance Fields}.
\newblock \bibinfo{journal}{\emph{ICCV}} (\bibinfo{year}{2021}).
\newblock


\bibitem[\protect\citeauthoryear{Bemana, Myszkowski, Seidel, and
  Ritschel}{Bemana et~al\mbox{.}}{2020}]%
        {Bemana2020xfields}
\bibfield{author}{\bibinfo{person}{Mojtaba Bemana}, \bibinfo{person}{Karol
  Myszkowski}, \bibinfo{person}{Hans-Peter Seidel}, {and}
  \bibinfo{person}{Tobias Ritschel}.} \bibinfo{year}{2020}\natexlab{}.
\newblock \showarticletitle{X-Fields: Implicit Neural View-, Light- and
  Time-Image Interpolation}.
\newblock \bibinfo{journal}{\emph{ACM Transactions on Graphics (Proc. SIGGRAPH
  Asia 2020)}} \bibinfo{volume}{39}, \bibinfo{number}{6}
  (\bibinfo{year}{2020}).
\newblock
\urldef\tempurl%
\url{https://doi.org/10.1145/3414685.3417827}
\showDOI{\tempurl}


\bibitem[\protect\citeauthoryear{Blinn and Newell}{Blinn and Newell}{1976}]%
        {blinn1976texture}
\bibfield{author}{\bibinfo{person}{James~F Blinn} {and}
  \bibinfo{person}{Martin~E Newell}.} \bibinfo{year}{1976}\natexlab{}.
\newblock \showarticletitle{Texture and reflection in computer generated
  images}.
\newblock \bibinfo{journal}{\emph{Commun. ACM}} \bibinfo{volume}{19},
  \bibinfo{number}{10} (\bibinfo{year}{1976}), \bibinfo{pages}{542--547}.
\newblock


\bibitem[\protect\citeauthoryear{Bruce and Giblin}{Bruce and Giblin}{1992}]%
        {bruce1992curves}
\bibfield{author}{\bibinfo{person}{James~William Bruce} {and}
  \bibinfo{person}{PJ Giblin}.} \bibinfo{year}{1992}\natexlab{}.
\newblock \bibinfo{booktitle}{\emph{Curves and Singularities: a geometrical
  introduction to singularity theory}}.
\newblock \bibinfo{publisher}{Cambridge university press}.
\newblock


\bibitem[\protect\citeauthoryear{Cabral, Olano, and Nemec}{Cabral
  et~al\mbox{.}}{1999}]%
        {cabral1999reflection}
\bibfield{author}{\bibinfo{person}{Brian Cabral}, \bibinfo{person}{Marc Olano},
  {and} \bibinfo{person}{Philip Nemec}.} \bibinfo{year}{1999}\natexlab{}.
\newblock \showarticletitle{Reflection space image based rendering}. In
  \bibinfo{booktitle}{\emph{Proceedings of the 26th annual conference on
  Computer graphics and interactive techniques}}. \bibinfo{pages}{165--170}.
\newblock


\bibitem[\protect\citeauthoryear{Caelles, Maninis, Pont-Tuset, Leal-Taix{\'e},
  Cremers, and Van~Gool}{Caelles et~al\mbox{.}}{2017}]%
        {caelles2017one}
\bibfield{author}{\bibinfo{person}{Sergi Caelles},
  \bibinfo{person}{Kevis-Kokitsi Maninis}, \bibinfo{person}{Jordi Pont-Tuset},
  \bibinfo{person}{Laura Leal-Taix{\'e}}, \bibinfo{person}{Daniel Cremers},
  {and} \bibinfo{person}{Luc Van~Gool}.} \bibinfo{year}{2017}\natexlab{}.
\newblock \showarticletitle{One-shot video object segmentation}. In
  \bibinfo{booktitle}{\emph{Proceedings of the IEEE conference on computer
  vision and pattern recognition}}. \bibinfo{pages}{221--230}.
\newblock


\bibitem[\protect\citeauthoryear{Chen and Arvo}{Chen and Arvo}{2000}]%
        {chen2000theory}
\bibfield{author}{\bibinfo{person}{Min Chen} {and} \bibinfo{person}{James
  Arvo}.} \bibinfo{year}{2000}\natexlab{}.
\newblock \showarticletitle{Theory and application of specular path
  perturbation}.
\newblock \bibinfo{journal}{\emph{ACM Transactions on Graphics (TOG)}}
  \bibinfo{volume}{19}, \bibinfo{number}{4} (\bibinfo{year}{2000}),
  \bibinfo{pages}{246--278}.
\newblock


\bibitem[\protect\citeauthoryear{D{\k{a}}ba{\l}a, Kellnhofer, Ritschel, Didyk,
  Templin, Myszkowski, Rokita, and Seidel}{D{\k{a}}ba{\l}a
  et~al\mbox{.}}{2014}]%
        {dkabala2014manipulating}
\bibfield{author}{\bibinfo{person}{{\L}ukasz D{\k{a}}ba{\l}a},
  \bibinfo{person}{Petr Kellnhofer}, \bibinfo{person}{Tobias Ritschel},
  \bibinfo{person}{Piotr Didyk}, \bibinfo{person}{Krzysztof Templin},
  \bibinfo{person}{Karol Myszkowski}, \bibinfo{person}{Przemyslaw Rokita},
  {and} \bibinfo{person}{H-P Seidel}.} \bibinfo{year}{2014}\natexlab{}.
\newblock \showarticletitle{Manipulating refractive and reflective binocular
  disparity}. In \bibinfo{booktitle}{\emph{Computer Graphics Forum}},
  Vol.~\bibinfo{volume}{33}. Wiley Online Library, \bibinfo{pages}{53--62}.
\newblock


\bibitem[\protect\citeauthoryear{Diefenbach and Badler}{Diefenbach and
  Badler}{1997}]%
        {diefenbach1997multi}
\bibfield{author}{\bibinfo{person}{Paul~J Diefenbach} {and}
  \bibinfo{person}{Norman~I Badler}.} \bibinfo{year}{1997}\natexlab{}.
\newblock \showarticletitle{Multi-pass pipeline rendering: Realism for dynamic
  environments}. In \bibinfo{booktitle}{\emph{Proceedings of the 1997 symposium
  on Interactive 3D graphics}}. \bibinfo{pages}{59--ff}.
\newblock


\bibitem[\protect\citeauthoryear{Douglas and Peucker}{Douglas and
  Peucker}{1973}]%
        {douglas1973algorithms}
\bibfield{author}{\bibinfo{person}{David~H Douglas} {and}
  \bibinfo{person}{Thomas~K Peucker}.} \bibinfo{year}{1973}\natexlab{}.
\newblock \showarticletitle{Algorithms for the reduction of the number of
  points required to represent a digitized line or its caricature}.
\newblock \bibinfo{journal}{\emph{Cartographica: the international journal for
  geographic information and geovisualization}} \bibinfo{volume}{10},
  \bibinfo{number}{2} (\bibinfo{year}{1973}), \bibinfo{pages}{112--122}.
\newblock


\bibitem[\protect\citeauthoryear{Estalella, Martin, Drettakis, Tost, Devillers,
  and Cazals}{Estalella et~al\mbox{.}}{2005}]%
        {estalella2005accurate}
\bibfield{author}{\bibinfo{person}{Pau Estalella}, \bibinfo{person}{Ignacio
  Martin}, \bibinfo{person}{George Drettakis}, \bibinfo{person}{Dani Tost},
  \bibinfo{person}{Olivier Devillers}, {and} \bibinfo{person}{Fr{\'e}d{\'e}ric
  Cazals}.} \bibinfo{year}{2005}\natexlab{}.
\newblock \showarticletitle{Accurate interactive specular reflections on curved
  objects}. In \bibinfo{booktitle}{\emph{Vision Modeling and Visualization (VMV
  2005)}}. Berlin: Akademische Verl.-Ges. Aka, 2005., \bibinfo{pages}{8}.
\newblock


\bibitem[\protect\citeauthoryear{Feng, Li, Cai, Luo, and Zhang}{Feng
  et~al\mbox{.}}{2022}]%
        {feng2022np}
\bibfield{author}{\bibinfo{person}{Wanquan Feng}, \bibinfo{person}{Jin Li},
  \bibinfo{person}{Hongrui Cai}, \bibinfo{person}{Xiaonan Luo}, {and}
  \bibinfo{person}{Juyong Zhang}.} \bibinfo{year}{2022}\natexlab{}.
\newblock \showarticletitle{Neural Points: Point Cloud Representation with
  Neural Fields for Arbitrary Upsampling}.
\newblock  (\bibinfo{year}{2022}).
\newblock


\bibitem[\protect\citeauthoryear{Foley, Van, Van~Dam, Feiner, Hughes, and
  Hughes}{Foley et~al\mbox{.}}{1996}]%
        {foley1996computer}
\bibfield{author}{\bibinfo{person}{James~D Foley}, \bibinfo{person}{Foley~Dan
  Van}, \bibinfo{person}{Andries Van~Dam}, \bibinfo{person}{Steven~K Feiner},
  \bibinfo{person}{John~F Hughes}, {and} \bibinfo{person}{J Hughes}.}
  \bibinfo{year}{1996}\natexlab{}.
\newblock \bibinfo{booktitle}{\emph{Computer graphics: principles and
  practice}}. Vol.~\bibinfo{volume}{12110}.
\newblock \bibinfo{publisher}{Addison-Wesley Professional}.
\newblock


\bibitem[\protect\citeauthoryear{Garbin, Kowalski, Johnson, Shotton, and
  Valentin}{Garbin et~al\mbox{.}}{2021}]%
        {garbin2021fastnerf}
\bibfield{author}{\bibinfo{person}{Stephan~J Garbin}, \bibinfo{person}{Marek
  Kowalski}, \bibinfo{person}{Matthew Johnson}, \bibinfo{person}{Jamie
  Shotton}, {and} \bibinfo{person}{Julien Valentin}.}
  \bibinfo{year}{2021}\natexlab{}.
\newblock \showarticletitle{Fastnerf: High-fidelity neural rendering at
  200fps}. In \bibinfo{booktitle}{\emph{Proceedings of the IEEE/CVF
  International Conference on Computer Vision}}. \bibinfo{pages}{14346--14355}.
\newblock


\bibitem[\protect\citeauthoryear{Glaeser}{Glaeser}{1999}]%
        {glaeser1999reflections}
\bibfield{author}{\bibinfo{person}{Georg Glaeser}.}
  \bibinfo{year}{1999}\natexlab{}.
\newblock \showarticletitle{Reflections on spheres and cylinders of
  revolution}.
\newblock \bibinfo{journal}{\emph{Journal for Geometry and Graphics}}
  \bibinfo{volume}{3}, \bibinfo{number}{2} (\bibinfo{year}{1999}),
  \bibinfo{pages}{121--139}.
\newblock


\bibitem[\protect\citeauthoryear{Goesele, Snavely, Curless, Hoppe, and
  Seitz}{Goesele et~al\mbox{.}}{2007}]%
        {goesele2007multi}
\bibfield{author}{\bibinfo{person}{Michael Goesele}, \bibinfo{person}{Noah
  Snavely}, \bibinfo{person}{Brian Curless}, \bibinfo{person}{Hugues Hoppe},
  {and} \bibinfo{person}{Steven~M Seitz}.} \bibinfo{year}{2007}\natexlab{}.
\newblock \showarticletitle{Multi-view stereo for community photo collections}.
  In \bibinfo{booktitle}{\emph{2007 IEEE 11th International Conference on
  Computer Vision}}. IEEE, \bibinfo{pages}{1--8}.
\newblock


\bibitem[\protect\citeauthoryear{Greene}{Greene}{1986}]%
        {greene1986environment}
\bibfield{author}{\bibinfo{person}{Ned Greene}.}
  \bibinfo{year}{1986}\natexlab{}.
\newblock \showarticletitle{Environment mapping and other applications of world
  projections}.
\newblock \bibinfo{journal}{\emph{IEEE computer graphics and Applications}}
  \bibinfo{volume}{6}, \bibinfo{number}{11} (\bibinfo{year}{1986}),
  \bibinfo{pages}{21--29}.
\newblock


\bibitem[\protect\citeauthoryear{Gross and Pfister}{Gross and Pfister}{2007}]%
        {gross2007point}
\bibfield{author}{\bibinfo{person}{Markus Gross} {and}
  \bibinfo{person}{Hanspeter Pfister}.} \bibinfo{year}{2007}\natexlab{}.
\newblock \bibinfo{booktitle}{\emph{Point-based graphics}}.
\newblock \bibinfo{publisher}{Elsevier}.
\newblock


\bibitem[\protect\citeauthoryear{Guo, Kang, Bao, He, and Zhang}{Guo
  et~al\mbox{.}}{2021}]%
        {guo2021nerfren}
\bibfield{author}{\bibinfo{person}{Yuan-Chen Guo}, \bibinfo{person}{Di Kang},
  \bibinfo{person}{Linchao Bao}, \bibinfo{person}{Yu He}, {and}
  \bibinfo{person}{Song-Hai Zhang}.} \bibinfo{year}{2021}\natexlab{}.
\newblock \showarticletitle{NeRFReN: Neural Radiance Fields with Reflections}.
\newblock \bibinfo{journal}{\emph{arXiv preprint arXiv:2111.15234}}
  (\bibinfo{year}{2021}).
\newblock


\bibitem[\protect\citeauthoryear{Hamilton}{Hamilton}{1828}]%
        {hamilton1828theory}
\bibfield{author}{\bibinfo{person}{William~Rowan Hamilton}.}
  \bibinfo{year}{1828}\natexlab{}.
\newblock \showarticletitle{Theory of systems of rays}.
\newblock \bibinfo{journal}{\emph{The Transactions of the Royal Irish Academy}}
  (\bibinfo{year}{1828}), \bibinfo{pages}{69--174}.
\newblock


\bibitem[\protect\citeauthoryear{Hedman, Philip, Price, Frahm, Drettakis, and
  Brostow}{Hedman et~al\mbox{.}}{2018}]%
        {HPPFDB18}
\bibfield{author}{\bibinfo{person}{Peter Hedman}, \bibinfo{person}{Julien
  Philip}, \bibinfo{person}{True Price}, \bibinfo{person}{Jan-Michael Frahm},
  \bibinfo{person}{George Drettakis}, {and} \bibinfo{person}{Gabriel Brostow}.}
  \bibinfo{year}{2018}\natexlab{}.
\newblock \showarticletitle{Deep Blending for Free-Viewpoint Image-Based
  Rendering}.
\newblock \bibinfo{journal}{\emph{ACM Transactions on Graphics (SIGGRAPH Asia
  Conference Proceedings)}} \bibinfo{volume}{37}, \bibinfo{number}{6}
  (\bibinfo{date}{November} \bibinfo{year}{2018}).
\newblock
\urldef\tempurl%
\url{http://www-sop.inria.fr/reves/Basilic/2018/HPPFDB18}
\showURL{%
\tempurl}


\bibitem[\protect\citeauthoryear{Hedman, Srinivasan, Mildenhall, Barron, and
  Debevec}{Hedman et~al\mbox{.}}{2021}]%
        {hedman2021snerg}
\bibfield{author}{\bibinfo{person}{Peter Hedman}, \bibinfo{person}{Pratul~P.
  Srinivasan}, \bibinfo{person}{Ben Mildenhall}, \bibinfo{person}{Jonathan~T.
  Barron}, {and} \bibinfo{person}{Paul Debevec}.}
  \bibinfo{year}{2021}\natexlab{}.
\newblock \showarticletitle{Baking Neural Radiance Fields for Real-Time View
  Synthesis}.
\newblock \bibinfo{journal}{\emph{ICCV}} (\bibinfo{year}{2021}).
\newblock


\bibitem[\protect\citeauthoryear{Jones, Lee, Holliman, and Ezra}{Jones
  et~al\mbox{.}}{2001}]%
        {jones2001controlling}
\bibfield{author}{\bibinfo{person}{Graham~R Jones}, \bibinfo{person}{Delman
  Lee}, \bibinfo{person}{Nicolas~S Holliman}, {and} \bibinfo{person}{David
  Ezra}.} \bibinfo{year}{2001}\natexlab{}.
\newblock \showarticletitle{Controlling perceived depth in stereoscopic
  images}. In \bibinfo{booktitle}{\emph{Stereoscopic Displays and Virtual
  Reality Systems VIII}}, Vol.~\bibinfo{volume}{4297}. SPIE,
  \bibinfo{pages}{42--53}.
\newblock


\bibitem[\protect\citeauthoryear{Josse and Pene}{Josse and Pene}{2014}]%
        {josse2014degree}
\bibfield{author}{\bibinfo{person}{Alfrederic Josse} {and}
  \bibinfo{person}{Fran{\c{c}}oise Pene}.} \bibinfo{year}{2014}\natexlab{}.
\newblock \showarticletitle{On the degree of caustics by reflection}.
\newblock \bibinfo{journal}{\emph{Communications in Algebra}}
  \bibinfo{volume}{42}, \bibinfo{number}{6} (\bibinfo{year}{2014}),
  \bibinfo{pages}{2442--2475}.
\newblock


\bibitem[\protect\citeauthoryear{Kopanas, Philip, Leimk{\"u}hler, and
  Drettakis}{Kopanas et~al\mbox{.}}{2021}]%
        {kopanas2021point}
\bibfield{author}{\bibinfo{person}{Georgios Kopanas}, \bibinfo{person}{Julien
  Philip}, \bibinfo{person}{Thomas Leimk{\"u}hler}, {and}
  \bibinfo{person}{George Drettakis}.} \bibinfo{year}{2021}\natexlab{}.
\newblock \showarticletitle{Point-Based Neural Rendering with Per-View
  Optimization}. In \bibinfo{booktitle}{\emph{Computer Graphics Forum}},
  Vol.~\bibinfo{volume}{40}. Wiley Online Library, \bibinfo{pages}{29--43}.
\newblock


\bibitem[\protect\citeauthoryear{Kopf, Langguth, Scharstein, Szeliski, and
  Goesele}{Kopf et~al\mbox{.}}{2013}]%
        {kopf2013image}
\bibfield{author}{\bibinfo{person}{Johannes Kopf}, \bibinfo{person}{Fabian
  Langguth}, \bibinfo{person}{Daniel Scharstein}, \bibinfo{person}{Richard
  Szeliski}, {and} \bibinfo{person}{Michael Goesele}.}
  \bibinfo{year}{2013}\natexlab{}.
\newblock \showarticletitle{Image-based rendering in the gradient domain}.
\newblock \bibinfo{journal}{\emph{ACM Transactions on Graphics (TOG)}}
  \bibinfo{volume}{32}, \bibinfo{number}{6} (\bibinfo{year}{2013}),
  \bibinfo{pages}{1--9}.
\newblock


\bibitem[\protect\citeauthoryear{Lambooij, Fortuin, Heynderickx, and
  IJsselsteijn}{Lambooij et~al\mbox{.}}{2009}]%
        {lambooij2009visual}
\bibfield{author}{\bibinfo{person}{Marc Lambooij}, \bibinfo{person}{Marten
  Fortuin}, \bibinfo{person}{Ingrid Heynderickx}, {and}
  \bibinfo{person}{Wijnand IJsselsteijn}.} \bibinfo{year}{2009}\natexlab{}.
\newblock \showarticletitle{Visual discomfort and visual fatigue of
  stereoscopic displays: A review}.
\newblock \bibinfo{journal}{\emph{Journal of imaging science and technology}}
  \bibinfo{volume}{53}, \bibinfo{number}{3} (\bibinfo{year}{2009}),
  \bibinfo{pages}{30201--1}.
\newblock


\bibitem[\protect\citeauthoryear{Lang, Hornung, Wang, Poulakos, Smolic, and
  Gross}{Lang et~al\mbox{.}}{2010}]%
        {lang2010nonlinear}
\bibfield{author}{\bibinfo{person}{Manuel Lang}, \bibinfo{person}{Alexander
  Hornung}, \bibinfo{person}{Oliver Wang}, \bibinfo{person}{Steven Poulakos},
  \bibinfo{person}{Aljoscha Smolic}, {and} \bibinfo{person}{Markus Gross}.}
  \bibinfo{year}{2010}\natexlab{}.
\newblock \showarticletitle{Nonlinear disparity mapping for stereoscopic 3D}.
\newblock \bibinfo{journal}{\emph{ACM Transactions on Graphics (TOG)}}
  \bibinfo{volume}{29}, \bibinfo{number}{4} (\bibinfo{year}{2010}),
  \bibinfo{pages}{1--10}.
\newblock


\bibitem[\protect\citeauthoryear{Lassner and Zollhofer}{Lassner and
  Zollhofer}{2021}]%
        {lassner2021pulsar}
\bibfield{author}{\bibinfo{person}{Christoph Lassner} {and}
  \bibinfo{person}{Michael Zollhofer}.} \bibinfo{year}{2021}\natexlab{}.
\newblock \showarticletitle{Pulsar: Efficient Sphere-based Neural Rendering}.
  In \bibinfo{booktitle}{\emph{Proceedings of the IEEE/CVF Conference on
  Computer Vision and Pattern Recognition}}. \bibinfo{pages}{1440--1449}.
\newblock


\bibitem[\protect\citeauthoryear{Lawrence}{Lawrence}{2013}]%
        {lawrence2013catalog}
\bibfield{author}{\bibinfo{person}{J~Dennis Lawrence}.}
  \bibinfo{year}{2013}\natexlab{}.
\newblock \bibinfo{booktitle}{\emph{A catalog of special plane curves}}.
\newblock \bibinfo{publisher}{Courier Corporation}.
\newblock


\bibitem[\protect\citeauthoryear{Levin and Weiss}{Levin and Weiss}{2007}]%
        {levin2007user}
\bibfield{author}{\bibinfo{person}{Anat Levin} {and} \bibinfo{person}{Yair
  Weiss}.} \bibinfo{year}{2007}\natexlab{}.
\newblock \showarticletitle{User assisted separation of reflections from a
  single image using a sparsity prior}.
\newblock \bibinfo{journal}{\emph{IEEE Transactions on Pattern Analysis and
  Machine Intelligence}} \bibinfo{volume}{29}, \bibinfo{number}{9}
  (\bibinfo{year}{2007}), \bibinfo{pages}{1647--1654}.
\newblock


\bibitem[\protect\citeauthoryear{Lochmann, Reinert, Ritschel, M{\"u}ller, and
  Seidel}{Lochmann et~al\mbox{.}}{2014}]%
        {lochmann2014real}
\bibfield{author}{\bibinfo{person}{Gerrit Lochmann}, \bibinfo{person}{Bernhard
  Reinert}, \bibinfo{person}{Tobias Ritschel}, \bibinfo{person}{Stefan
  M{\"u}ller}, {and} \bibinfo{person}{Hans-Peter Seidel}.}
  \bibinfo{year}{2014}\natexlab{}.
\newblock \showarticletitle{Real-time Reflective and Refractive Novel-view
  Synthesis.}. In \bibinfo{booktitle}{\emph{VMV}}. \bibinfo{pages}{9--16}.
\newblock


\bibitem[\protect\citeauthoryear{{Loza}, {Mihaylova}, {Canagarajah}, and
  {Bull}}{{Loza} et~al\mbox{.}}{2006}]%
        {loza2006structural}
\bibfield{author}{\bibinfo{person}{A. {Loza}}, \bibinfo{person}{L.
  {Mihaylova}}, \bibinfo{person}{N. {Canagarajah}}, {and} \bibinfo{person}{D.
  {Bull}}.} \bibinfo{year}{2006}\natexlab{}.
\newblock \showarticletitle{Structural Similarity-Based Object Tracking in
  Video Sequences}. In \bibinfo{booktitle}{\emph{2006 9th International
  Conference on Information Fusion}}. \bibinfo{pages}{1--6}.
\newblock
\urldef\tempurl%
\url{https://doi.org/10.1109/ICIF.2006.301574}
\showDOI{\tempurl}


\bibitem[\protect\citeauthoryear{Meshry, Goldman, Khamis, Hoppe, Pandey,
  Snavely, and Martin-Brualla}{Meshry et~al\mbox{.}}{2019}]%
        {meshry2019neural}
\bibfield{author}{\bibinfo{person}{Moustafa Meshry}, \bibinfo{person}{Dan~B
  Goldman}, \bibinfo{person}{Sameh Khamis}, \bibinfo{person}{Hugues Hoppe},
  \bibinfo{person}{Rohit Pandey}, \bibinfo{person}{Noah Snavely}, {and}
  \bibinfo{person}{Ricardo Martin-Brualla}.} \bibinfo{year}{2019}\natexlab{}.
\newblock \showarticletitle{Neural rerendering in the wild}. In
  \bibinfo{booktitle}{\emph{Proceedings of the IEEE/CVF Conference on Computer
  Vision and Pattern Recognition}}. \bibinfo{pages}{6878--6887}.
\newblock


\bibitem[\protect\citeauthoryear{Mildenhall, Srinivasan, Tancik, Barron,
  Ramamoorthi, and Ng}{Mildenhall et~al\mbox{.}}{2020}]%
        {mildenhall2020nerf}
\bibfield{author}{\bibinfo{person}{Ben Mildenhall}, \bibinfo{person}{Pratul~P
  Srinivasan}, \bibinfo{person}{Matthew Tancik}, \bibinfo{person}{Jonathan~T
  Barron}, \bibinfo{person}{Ravi Ramamoorthi}, {and} \bibinfo{person}{Ren Ng}.}
  \bibinfo{year}{2020}\natexlab{}.
\newblock \showarticletitle{Nerf: Representing scenes as neural radiance fields
  for view synthesis}. In \bibinfo{booktitle}{\emph{European conference on
  computer vision}}. Springer, \bibinfo{pages}{405--421}.
\newblock


\bibitem[\protect\citeauthoryear{Mitchell and Hanrahan}{Mitchell and
  Hanrahan}{1992}]%
        {mitchell1992illumination}
\bibfield{author}{\bibinfo{person}{Don Mitchell} {and} \bibinfo{person}{Pat
  Hanrahan}.} \bibinfo{year}{1992}\natexlab{}.
\newblock \showarticletitle{Illumination from curved reflectors}. In
  \bibinfo{booktitle}{\emph{Proceedings of the 19th annual conference on
  Computer graphics and interactive techniques}}. \bibinfo{pages}{283--291}.
\newblock


\bibitem[\protect\citeauthoryear{M\"uller, Evans, Schied, and Keller}{M\"uller
  et~al\mbox{.}}{2022}]%
        {mueller2022instant}
\bibfield{author}{\bibinfo{person}{Thomas M\"uller}, \bibinfo{person}{Alex
  Evans}, \bibinfo{person}{Christoph Schied}, {and} \bibinfo{person}{Alexander
  Keller}.} \bibinfo{year}{2022}\natexlab{}.
\newblock \showarticletitle{Instant Neural Graphics Primitives with a
  Multiresolution Hash Encoding}.
\newblock \bibinfo{journal}{\emph{ACM Trans. Graph.}} \bibinfo{volume}{41},
  \bibinfo{number}{4}, Article \bibinfo{articleno}{102} (\bibinfo{date}{July}
  \bibinfo{year}{2022}), \bibinfo{numpages}{15}~pages.
\newblock


\bibitem[\protect\citeauthoryear{Nimier-David, Vicini, Zeltner, and
  Jakob}{Nimier-David et~al\mbox{.}}{2019}]%
        {nimier2019mitsuba}
\bibfield{author}{\bibinfo{person}{Merlin Nimier-David}, \bibinfo{person}{Delio
  Vicini}, \bibinfo{person}{Tizian Zeltner}, {and} \bibinfo{person}{Wenzel
  Jakob}.} \bibinfo{year}{2019}\natexlab{}.
\newblock \showarticletitle{Mitsuba 2: A retargetable forward and inverse
  renderer}.
\newblock \bibinfo{journal}{\emph{ACM Transactions on Graphics (TOG)}}
  \bibinfo{volume}{38}, \bibinfo{number}{6} (\bibinfo{year}{2019}),
  \bibinfo{pages}{1--17}.
\newblock


\bibitem[\protect\citeauthoryear{Ofek and Rappoport}{Ofek and
  Rappoport}{1998}]%
        {ofek1998interactive}
\bibfield{author}{\bibinfo{person}{Eyal Ofek} {and} \bibinfo{person}{Ari
  Rappoport}.} \bibinfo{year}{1998}\natexlab{}.
\newblock \showarticletitle{Interactive reflections on curved objects}. In
  \bibinfo{booktitle}{\emph{Proceedings of the 25th annual conference on
  Computer graphics and interactive techniques}}. \bibinfo{pages}{333--342}.
\newblock


\bibitem[\protect\citeauthoryear{Ost, Laradji, Newell, Bahat, and Heide}{Ost
  et~al\mbox{.}}{2021}]%
        {ost2021neural}
\bibfield{author}{\bibinfo{person}{Julian Ost}, \bibinfo{person}{Issam
  Laradji}, \bibinfo{person}{Alejandro Newell}, \bibinfo{person}{Yuval Bahat},
  {and} \bibinfo{person}{Felix Heide}.} \bibinfo{year}{2021}\natexlab{}.
\newblock \showarticletitle{Neural Point Light Fields}.
\newblock \bibinfo{journal}{\emph{arXiv preprint arXiv:2112.01473}}
  (\bibinfo{year}{2021}).
\newblock


\bibitem[\protect\citeauthoryear{Park, Sinha, Barron, Bouaziz, Goldman, Seitz,
  and Martin-Brualla}{Park et~al\mbox{.}}{2021}]%
        {park2021nerfies}
\bibfield{author}{\bibinfo{person}{Keunhong Park}, \bibinfo{person}{Utkarsh
  Sinha}, \bibinfo{person}{Jonathan~T Barron}, \bibinfo{person}{Sofien
  Bouaziz}, \bibinfo{person}{Dan~B Goldman}, \bibinfo{person}{Steven~M Seitz},
  {and} \bibinfo{person}{Ricardo Martin-Brualla}.}
  \bibinfo{year}{2021}\natexlab{}.
\newblock \showarticletitle{Nerfies: Deformable neural radiance fields}. In
  \bibinfo{booktitle}{\emph{Proceedings of the IEEE/CVF International
  Conference on Computer Vision}}. \bibinfo{pages}{5865--5874}.
\newblock


\bibitem[\protect\citeauthoryear{Philip, Morgenthaler, Gharbi, and
  Drettakis}{Philip et~al\mbox{.}}{2021}]%
        {PMGD21}
\bibfield{author}{\bibinfo{person}{Julien Philip}, \bibinfo{person}{S\'ebastien
  Morgenthaler}, \bibinfo{person}{Micha{\"e}l Gharbi}, {and}
  \bibinfo{person}{George Drettakis}.} \bibinfo{year}{2021}\natexlab{}.
\newblock \showarticletitle{Free-viewpoint Indoor Neural Relighting from
  Multi-view Stereo}.
\newblock \bibinfo{journal}{\emph{ACM Transactions on Graphics}}
  (\bibinfo{year}{2021}).
\newblock
\urldef\tempurl%
\url{http://www-sop.inria.fr/reves/Basilic/2021/PMGD21}
\showURL{%
\tempurl}


\bibitem[\protect\citeauthoryear{Preparata and Shamos}{Preparata and
  Shamos}{1985}]%
        {preparata1985convex}
\bibfield{author}{\bibinfo{person}{Franco~P Preparata} {and}
  \bibinfo{person}{Michael~Ian Shamos}.} \bibinfo{year}{1985}\natexlab{}.
\newblock \showarticletitle{Convex hulls: Basic algorithms}.
\newblock In \bibinfo{booktitle}{\emph{Computational geometry}}.
  \bibinfo{publisher}{Springer}, \bibinfo{pages}{95--149}.
\newblock


\bibitem[\protect\citeauthoryear{Ramanarayanan, Ferwerda, Walter, and
  Bala}{Ramanarayanan et~al\mbox{.}}{2007}]%
        {ramanarayanan2007visual}
\bibfield{author}{\bibinfo{person}{Ganesh Ramanarayanan},
  \bibinfo{person}{James Ferwerda}, \bibinfo{person}{Bruce Walter}, {and}
  \bibinfo{person}{Kavita Bala}.} \bibinfo{year}{2007}\natexlab{}.
\newblock \showarticletitle{Visual equivalence: towards a new standard for
  image fidelity}.
\newblock \bibinfo{journal}{\emph{ACM Transactions on Graphics (TOG)}}
  \bibinfo{volume}{26}, \bibinfo{number}{3} (\bibinfo{year}{2007}),
  \bibinfo{pages}{76--es}.
\newblock


\bibitem[\protect\citeauthoryear{Reality}{Reality}{2018}]%
        {reality2016capture}
\bibfield{author}{\bibinfo{person}{Capturing Reality}.}
  \bibinfo{year}{2018}\natexlab{}.
\newblock \bibinfo{title}{RealityCapture reconstruction software.}
\newblock
  \bibinfo{howpublished}{\url{https://www.capturingreality.com/Product}}.
\newblock


\bibitem[\protect\citeauthoryear{Ren, Pfister, and Zwicker}{Ren
  et~al\mbox{.}}{2002}]%
        {ren2002object}
\bibfield{author}{\bibinfo{person}{Liu Ren}, \bibinfo{person}{Hanspeter
  Pfister}, {and} \bibinfo{person}{Matthias Zwicker}.}
  \bibinfo{year}{2002}\natexlab{}.
\newblock \showarticletitle{Object space EWA surface splatting: A hardware
  accelerated approach to high quality point rendering}. In
  \bibinfo{booktitle}{\emph{Computer Graphics Forum}},
  Vol.~\bibinfo{volume}{21}. Wiley Online Library, \bibinfo{pages}{461--470}.
\newblock


\bibitem[\protect\citeauthoryear{Riegler and Koltun}{Riegler and
  Koltun}{2021}]%
        {riegler2021stable}
\bibfield{author}{\bibinfo{person}{Gernot Riegler} {and}
  \bibinfo{person}{Vladlen Koltun}.} \bibinfo{year}{2021}\natexlab{}.
\newblock \showarticletitle{Stable view synthesis}. In
  \bibinfo{booktitle}{\emph{Proceedings of the IEEE/CVF Conference on Computer
  Vision and Pattern Recognition}}. \bibinfo{pages}{12216--12225}.
\newblock


\bibitem[\protect\citeauthoryear{Ritschel, Okabe, Thorm{\"a}hlen, and
  Seidel}{Ritschel et~al\mbox{.}}{2009}]%
        {ritschel2009interactive}
\bibfield{author}{\bibinfo{person}{Tobias Ritschel}, \bibinfo{person}{Makoto
  Okabe}, \bibinfo{person}{Thorsten Thorm{\"a}hlen}, {and}
  \bibinfo{person}{Hans-Peter Seidel}.} \bibinfo{year}{2009}\natexlab{}.
\newblock \showarticletitle{Interactive reflection editing}.
\newblock \bibinfo{journal}{\emph{ACM Transactions on Graphics (TOG)}}
  \bibinfo{volume}{28}, \bibinfo{number}{5} (\bibinfo{year}{2009}),
  \bibinfo{pages}{1--7}.
\newblock


\bibitem[\protect\citeauthoryear{Rodriguez, Leimk{\"u}hler, Prakash, Wyman,
  Shirley, and Drettakis}{Rodriguez et~al\mbox{.}}{2020a}]%
        {RLPWSD20}
\bibfield{author}{\bibinfo{person}{Simon Rodriguez}, \bibinfo{person}{Thomas
  Leimk{\"u}hler}, \bibinfo{person}{Siddhant Prakash}, \bibinfo{person}{Chris
  Wyman}, \bibinfo{person}{Peter Shirley}, {and} \bibinfo{person}{George
  Drettakis}.} \bibinfo{year}{2020}\natexlab{a}.
\newblock \showarticletitle{Glossy Probe Reprojection for Interactive Global
  Illumination}.
\newblock \bibinfo{journal}{\emph{ACM Transactions on Graphics (SIGGRAPH Asia
  Conference Proceedings)}} \bibinfo{volume}{39}, \bibinfo{number}{6}
  (\bibinfo{date}{December} \bibinfo{year}{2020}).
\newblock
\urldef\tempurl%
\url{http://www-sop.inria.fr/reves/Basilic/2020/RLPWSD20}
\showURL{%
\tempurl}


\bibitem[\protect\citeauthoryear{Rodriguez, Prakash, Hedman, and
  Drettakis}{Rodriguez et~al\mbox{.}}{2020b}]%
        {rodriguez2020image}
\bibfield{author}{\bibinfo{person}{Simon Rodriguez}, \bibinfo{person}{Siddhant
  Prakash}, \bibinfo{person}{Peter Hedman}, {and} \bibinfo{person}{George
  Drettakis}.} \bibinfo{year}{2020}\natexlab{b}.
\newblock \showarticletitle{Image-Based Rendering of Cars using Semantic Labels
  and Approximate Reflection Flow}.
\newblock \bibinfo{journal}{\emph{Proceedings of the ACM on Computer Graphics
  and Interactive Techniques}}  \bibinfo{volume}{3} (\bibinfo{year}{2020}).
\newblock


\bibitem[\protect\citeauthoryear{Roth and Black}{Roth and Black}{2006}]%
        {roth2006specular}
\bibfield{author}{\bibinfo{person}{Stefan Roth} {and}
  \bibinfo{person}{Michael~J Black}.} \bibinfo{year}{2006}\natexlab{}.
\newblock \showarticletitle{Specular flow and the recovery of surface
  structure}. In \bibinfo{booktitle}{\emph{2006 IEEE Computer Society
  Conference on Computer Vision and Pattern Recognition (CVPR'06)}},
  Vol.~\bibinfo{volume}{2}. IEEE, \bibinfo{pages}{1869--1876}.
\newblock


\bibitem[\protect\citeauthoryear{R\"uckert, Franke, and Stamminger}{R\"uckert
  et~al\mbox{.}}{2022}]%
        {adop}
\bibfield{author}{\bibinfo{person}{Darius R\"uckert}, \bibinfo{person}{Linus
  Franke}, {and} \bibinfo{person}{Marc Stamminger}.}
  \bibinfo{year}{2022}\natexlab{}.
\newblock \showarticletitle{{ADOP}: {Approximate} {Differentiable}
  {One}-{Pixel} {Point} {Rendering}}.
\newblock \bibinfo{journal}{\emph{ACM Transactions on Graphics}}
  \bibinfo{volume}{41} (\bibinfo{year}{2022}).
\newblock


\bibitem[\protect\citeauthoryear{Sankaranarayanan, Veeraraghavan, Tuzel, and
  Agrawal}{Sankaranarayanan et~al\mbox{.}}{2010}]%
        {sankaranarayanan2010specular}
\bibfield{author}{\bibinfo{person}{Aswin~C Sankaranarayanan},
  \bibinfo{person}{Ashok Veeraraghavan}, \bibinfo{person}{Oncel Tuzel}, {and}
  \bibinfo{person}{Amit Agrawal}.} \bibinfo{year}{2010}\natexlab{}.
\newblock \showarticletitle{Specular surface reconstruction from sparse
  reflection correspondences}. In \bibinfo{booktitle}{\emph{2010 IEEE Computer
  Society Conference on Computer Vision and Pattern Recognition}}. IEEE,
  \bibinfo{pages}{1245--1252}.
\newblock


\bibitem[\protect\citeauthoryear{Schmidt, Novak, Meng, Kaplanyan, Reiner,
  Nowrouzezahrai, and Dachsbacher}{Schmidt et~al\mbox{.}}{2013}]%
        {schmidt2013path}
\bibfield{author}{\bibinfo{person}{Thorsten-Walther Schmidt},
  \bibinfo{person}{Jan Novak}, \bibinfo{person}{Johannes Meng},
  \bibinfo{person}{Anton~S Kaplanyan}, \bibinfo{person}{Tim Reiner},
  \bibinfo{person}{Derek Nowrouzezahrai}, {and} \bibinfo{person}{Carsten
  Dachsbacher}.} \bibinfo{year}{2013}\natexlab{}.
\newblock \showarticletitle{Path-space manipulation of physically-based light
  transport}.
\newblock \bibinfo{journal}{\emph{ACM Transactions On Graphics (TOG)}}
  \bibinfo{volume}{32}, \bibinfo{number}{4} (\bibinfo{year}{2013}),
  \bibinfo{pages}{1--11}.
\newblock


\bibitem[\protect\citeauthoryear{Shibata, Kim, Hoffman, and Banks}{Shibata
  et~al\mbox{.}}{2011}]%
        {shibata2011zone}
\bibfield{author}{\bibinfo{person}{Takashi Shibata}, \bibinfo{person}{Joohwan
  Kim}, \bibinfo{person}{David~M Hoffman}, {and} \bibinfo{person}{Martin~S
  Banks}.} \bibinfo{year}{2011}\natexlab{}.
\newblock \showarticletitle{The zone of comfort: Predicting visual discomfort
  with stereo displays}.
\newblock \bibinfo{journal}{\emph{Journal of vision}} \bibinfo{volume}{11},
  \bibinfo{number}{8} (\bibinfo{year}{2011}), \bibinfo{pages}{11--11}.
\newblock


\bibitem[\protect\citeauthoryear{Sinha, Kopf, Goesele, Scharstein, and
  Szeliski}{Sinha et~al\mbox{.}}{2012}]%
        {sinha2012image}
\bibfield{author}{\bibinfo{person}{Sudipta~N Sinha}, \bibinfo{person}{Johannes
  Kopf}, \bibinfo{person}{Michael Goesele}, \bibinfo{person}{Daniel
  Scharstein}, {and} \bibinfo{person}{Richard Szeliski}.}
  \bibinfo{year}{2012}\natexlab{}.
\newblock \showarticletitle{Image-based rendering for scenes with reflections}.
\newblock \bibinfo{journal}{\emph{ACM Transactions on Graphics (TOG)}}
  \bibinfo{volume}{31}, \bibinfo{number}{4} (\bibinfo{year}{2012}),
  \bibinfo{pages}{1--10}.
\newblock


\bibitem[\protect\citeauthoryear{Sitzmann, Zollh{\"o}fer, and
  Wetzstein}{Sitzmann et~al\mbox{.}}{2019}]%
        {sitzmann2019srns}
\bibfield{author}{\bibinfo{person}{Vincent Sitzmann}, \bibinfo{person}{Michael
  Zollh{\"o}fer}, {and} \bibinfo{person}{Gordon Wetzstein}.}
  \bibinfo{year}{2019}\natexlab{}.
\newblock \showarticletitle{Scene Representation Networks: Continuous
  3D-Structure-Aware Neural Scene Representations}. In
  \bibinfo{booktitle}{\emph{Advances in Neural Information Processing
  Systems}}.
\newblock


\bibitem[\protect\citeauthoryear{Szeliski, Avidan, and Anandan}{Szeliski
  et~al\mbox{.}}{2000}]%
        {szeliski2000layer}
\bibfield{author}{\bibinfo{person}{Richard Szeliski}, \bibinfo{person}{Shai
  Avidan}, {and} \bibinfo{person}{Padmanabhan Anandan}.}
  \bibinfo{year}{2000}\natexlab{}.
\newblock \showarticletitle{Layer extraction from multiple images containing
  reflections and transparency}. In \bibinfo{booktitle}{\emph{Proceedings IEEE
  Conference on Computer Vision and Pattern Recognition. CVPR 2000 (Cat. No.
  PR00662)}}, Vol.~\bibinfo{volume}{1}. IEEE, \bibinfo{pages}{246--253}.
\newblock


\bibitem[\protect\citeauthoryear{Szirmay-Kalos, Umenhoffer, Patow, Sz{\'e}csi,
  and Sbert}{Szirmay-Kalos et~al\mbox{.}}{2009}]%
        {szirmay2009specular}
\bibfield{author}{\bibinfo{person}{L{\'a}szl{\'o} Szirmay-Kalos},
  \bibinfo{person}{Tam{\'a}s Umenhoffer}, \bibinfo{person}{Gustavo Patow},
  \bibinfo{person}{L{\'a}szl{\'o} Sz{\'e}csi}, {and} \bibinfo{person}{Mateu
  Sbert}.} \bibinfo{year}{2009}\natexlab{}.
\newblock \showarticletitle{Specular effects on the gpu: State of the art}. In
  \bibinfo{booktitle}{\emph{Computer Graphics Forum}},
  Vol.~\bibinfo{volume}{28}. Wiley Online Library, \bibinfo{pages}{1586--1617}.
\newblock


\bibitem[\protect\citeauthoryear{Templin, Didyk, Ritschel, Myszkowski, and
  Seidel}{Templin et~al\mbox{.}}{2012}]%
        {templin2012highlight}
\bibfield{author}{\bibinfo{person}{Krzysztof Templin}, \bibinfo{person}{Piotr
  Didyk}, \bibinfo{person}{Tobias Ritschel}, \bibinfo{person}{Karol
  Myszkowski}, {and} \bibinfo{person}{Hans-Peter Seidel}.}
  \bibinfo{year}{2012}\natexlab{}.
\newblock \showarticletitle{Highlight microdisparity for improved gloss
  depiction}.
\newblock \bibinfo{journal}{\emph{ACM Transactions on Graphics (TOG)}}
  \bibinfo{volume}{31}, \bibinfo{number}{4} (\bibinfo{year}{2012}),
  \bibinfo{pages}{1--5}.
\newblock


\bibitem[\protect\citeauthoryear{Tewari, Fried, Thies, Sitzmann, Lombardi,
  Sunkavalli, Martin-Brualla, Simon, Saragih, Nie{\ss}ner,
  et~al\mbox{.}}{Tewari et~al\mbox{.}}{2020}]%
        {tewari2020state}
\bibfield{author}{\bibinfo{person}{Ayush Tewari}, \bibinfo{person}{Ohad Fried},
  \bibinfo{person}{Justus Thies}, \bibinfo{person}{Vincent Sitzmann},
  \bibinfo{person}{Stephen Lombardi}, \bibinfo{person}{Kalyan Sunkavalli},
  \bibinfo{person}{Ricardo Martin-Brualla}, \bibinfo{person}{Tomas Simon},
  \bibinfo{person}{Jason Saragih}, \bibinfo{person}{Matthias Nie{\ss}ner},
  {et~al\mbox{.}}} \bibinfo{year}{2020}\natexlab{}.
\newblock \showarticletitle{State of the art on neural rendering}. In
  \bibinfo{booktitle}{\emph{Computer Graphics Forum}},
  Vol.~\bibinfo{volume}{39}. Wiley Online Library, \bibinfo{pages}{701--727}.
\newblock


\bibitem[\protect\citeauthoryear{Tewari, Thies, Mildenhall, Srinivasan,
  Tretschk, Wang, Lassner, Sitzmann, Martin-Brualla, Lombardi,
  et~al\mbox{.}}{Tewari et~al\mbox{.}}{2021}]%
        {tewari2021advances}
\bibfield{author}{\bibinfo{person}{Ayush Tewari}, \bibinfo{person}{Justus
  Thies}, \bibinfo{person}{Ben Mildenhall}, \bibinfo{person}{Pratul
  Srinivasan}, \bibinfo{person}{Edgar Tretschk}, \bibinfo{person}{Yifan Wang},
  \bibinfo{person}{Christoph Lassner}, \bibinfo{person}{Vincent Sitzmann},
  \bibinfo{person}{Ricardo Martin-Brualla}, \bibinfo{person}{Stephen Lombardi},
  {et~al\mbox{.}}} \bibinfo{year}{2021}\natexlab{}.
\newblock \showarticletitle{Advances in neural rendering}.
\newblock \bibinfo{journal}{\emph{arXiv preprint arXiv:2111.05849}}
  (\bibinfo{year}{2021}).
\newblock


\bibitem[\protect\citeauthoryear{Thies, Zollh{\"o}fer, and Nie{\ss}ner}{Thies
  et~al\mbox{.}}{2019}]%
        {thies2019deferred}
\bibfield{author}{\bibinfo{person}{Justus Thies}, \bibinfo{person}{Michael
  Zollh{\"o}fer}, {and} \bibinfo{person}{Matthias Nie{\ss}ner}.}
  \bibinfo{year}{2019}\natexlab{}.
\newblock \showarticletitle{Deferred neural rendering: Image synthesis using
  neural textures}.
\newblock \bibinfo{journal}{\emph{ACM Transactions on Graphics (TOG)}}
  \bibinfo{volume}{38}, \bibinfo{number}{4} (\bibinfo{year}{2019}),
  \bibinfo{pages}{1--12}.
\newblock


\bibitem[\protect\citeauthoryear{Tretschk, Tewari, Golyanik, Zollh\"{o}fer,
  Lassner, and Theobalt}{Tretschk et~al\mbox{.}}{2021}]%
        {tretschk2021nonrigid}
\bibfield{author}{\bibinfo{person}{Edgar Tretschk}, \bibinfo{person}{Ayush
  Tewari}, \bibinfo{person}{Vladislav Golyanik}, \bibinfo{person}{Michael
  Zollh\"{o}fer}, \bibinfo{person}{Christoph Lassner}, {and}
  \bibinfo{person}{Christian Theobalt}.} \bibinfo{year}{2021}\natexlab{}.
\newblock \showarticletitle{Non-Rigid Neural Radiance Fields: Reconstruction
  and Novel View Synthesis of a Dynamic Scene From Monocular Video}. In
  \bibinfo{booktitle}{\emph{{IEEE} International Conference on Computer Vision
  ({ICCV})}}. {IEEE}.
\newblock


\bibitem[\protect\citeauthoryear{Verbin, Hedman, Mildenhall, Zickler, Barron,
  and Srinivasan}{Verbin et~al\mbox{.}}{2021}]%
        {verbin2021ref}
\bibfield{author}{\bibinfo{person}{Dor Verbin}, \bibinfo{person}{Peter Hedman},
  \bibinfo{person}{Ben Mildenhall}, \bibinfo{person}{Todd Zickler},
  \bibinfo{person}{Jonathan~T Barron}, {and} \bibinfo{person}{Pratul~P
  Srinivasan}.} \bibinfo{year}{2021}\natexlab{}.
\newblock \showarticletitle{Ref-NeRF: Structured View-Dependent Appearance for
  Neural Radiance Fields}.
\newblock \bibinfo{journal}{\emph{arXiv preprint arXiv:2112.03907}}
  (\bibinfo{year}{2021}).
\newblock


\bibitem[\protect\citeauthoryear{Wang, Wang, Zhao, Wu, Xu, and Yu}{Wang
  et~al\mbox{.}}{2021}]%
        {MirrorNeRF}
\bibfield{author}{\bibinfo{person}{Ziyu Wang}, \bibinfo{person}{Liao Wang},
  \bibinfo{person}{Fuqiang Zhao}, \bibinfo{person}{Minye Wu},
  \bibinfo{person}{Lan Xu}, {and} \bibinfo{person}{Jingyi Yu}.}
  \bibinfo{year}{2021}\natexlab{}.
\newblock \showarticletitle{MirrorNeRF: One-shot Neural Portrait Radiance Field
  from Multi-mirror Catadioptric Imaging}. In \bibinfo{booktitle}{\emph{2021
  IEEE International Conference on Computational Photography (ICCP)}}.
  \bibinfo{pages}{1--12}.
\newblock
\urldef\tempurl%
\url{https://doi.org/10.1109/ICCP51581.2021.9466270}
\showDOI{\tempurl}


\bibitem[\protect\citeauthoryear{Whelan, Goesele, Lovegrove, Straub, Green,
  Szeliski, Butterfield, Verma, Newcombe, Goesele, et~al\mbox{.}}{Whelan
  et~al\mbox{.}}{2018}]%
        {whelan2018reconstructing}
\bibfield{author}{\bibinfo{person}{Thomas Whelan}, \bibinfo{person}{Michael
  Goesele}, \bibinfo{person}{Steven~J Lovegrove}, \bibinfo{person}{Julian
  Straub}, \bibinfo{person}{Simon Green}, \bibinfo{person}{Richard Szeliski},
  \bibinfo{person}{Steven Butterfield}, \bibinfo{person}{Shobhit Verma},
  \bibinfo{person}{Richard~A Newcombe}, \bibinfo{person}{M Goesele},
  {et~al\mbox{.}}} \bibinfo{year}{2018}\natexlab{}.
\newblock \showarticletitle{Reconstructing scenes with mirror and glass
  surfaces.}
\newblock \bibinfo{journal}{\emph{ACM Trans. Graph.}} \bibinfo{volume}{37},
  \bibinfo{number}{4} (\bibinfo{year}{2018}), \bibinfo{pages}{102--1}.
\newblock


\bibitem[\protect\citeauthoryear{Whitted}{Whitted}{1979}]%
        {whitted1979improved}
\bibfield{author}{\bibinfo{person}{Turner Whitted}.}
  \bibinfo{year}{1979}\natexlab{}.
\newblock \showarticletitle{An improved illumination model for shaded display}.
  In \bibinfo{booktitle}{\emph{Proceedings of the 6th annual conference on
  Computer graphics and interactive techniques}}. \bibinfo{pages}{14}.
\newblock


\bibitem[\protect\citeauthoryear{Wieschollek, Gallo, Gu, and Kautz}{Wieschollek
  et~al\mbox{.}}{2018}]%
        {wieschollek2018separating}
\bibfield{author}{\bibinfo{person}{Patrick Wieschollek},
  \bibinfo{person}{Orazio Gallo}, \bibinfo{person}{Jinwei Gu}, {and}
  \bibinfo{person}{Jan Kautz}.} \bibinfo{year}{2018}\natexlab{}.
\newblock \showarticletitle{Separating reflection and transmission images in
  the wild}. In \bibinfo{booktitle}{\emph{Proceedings of the European
  Conference on Computer Vision (ECCV)}}. \bibinfo{pages}{89--104}.
\newblock


\bibitem[\protect\citeauthoryear{Wiles, Gkioxari, Szeliski, and Johnson}{Wiles
  et~al\mbox{.}}{2020}]%
        {wiles2020synsin}
\bibfield{author}{\bibinfo{person}{Olivia Wiles}, \bibinfo{person}{Georgia
  Gkioxari}, \bibinfo{person}{Richard Szeliski}, {and} \bibinfo{person}{Justin
  Johnson}.} \bibinfo{year}{2020}\natexlab{}.
\newblock \showarticletitle{Synsin: End-to-end view synthesis from a single
  image}. In \bibinfo{booktitle}{\emph{Proceedings of the IEEE/CVF Conference
  on Computer Vision and Pattern Recognition}}. \bibinfo{pages}{7467--7477}.
\newblock


\bibitem[\protect\citeauthoryear{Wizadwongsa, Phongthawee, Yenphraphai, and
  Suwajanakorn}{Wizadwongsa et~al\mbox{.}}{2021}]%
        {Wizadwongsa2021NeX}
\bibfield{author}{\bibinfo{person}{Suttisak Wizadwongsa},
  \bibinfo{person}{Pakkapon Phongthawee}, \bibinfo{person}{Jiraphon
  Yenphraphai}, {and} \bibinfo{person}{Supasorn Suwajanakorn}.}
  \bibinfo{year}{2021}\natexlab{}.
\newblock \showarticletitle{NeX: Real-time View Synthesis with Neural Basis
  Expansion}. In \bibinfo{booktitle}{\emph{IEEE Conference on Computer Vision
  and Pattern Recognition (CVPR)}}.
\newblock


\bibitem[\protect\citeauthoryear{Wu, Rubinstein, Shih, Guttag, Durand, and
  Freeman}{Wu et~al\mbox{.}}{2012}]%
        {wu2012eulerian}
\bibfield{author}{\bibinfo{person}{Hao-Yu Wu}, \bibinfo{person}{Michael
  Rubinstein}, \bibinfo{person}{Eugene Shih}, \bibinfo{person}{John Guttag},
  \bibinfo{person}{Fr{\'e}do Durand}, {and} \bibinfo{person}{William Freeman}.}
  \bibinfo{year}{2012}\natexlab{}.
\newblock \showarticletitle{Eulerian video magnification for revealing subtle
  changes in the world}.
\newblock \bibinfo{journal}{\emph{ACM transactions on graphics (TOG)}}
  \bibinfo{volume}{31}, \bibinfo{number}{4} (\bibinfo{year}{2012}),
  \bibinfo{pages}{1--8}.
\newblock


\bibitem[\protect\citeauthoryear{Xie, Takikawa, Saito, Litany, Yan, Khan,
  Tombari, Tompkin, Sitzmann, and Sridhar}{Xie et~al\mbox{.}}{2021}]%
        {xie2021neural}
\bibfield{author}{\bibinfo{person}{Yiheng Xie}, \bibinfo{person}{Towaki
  Takikawa}, \bibinfo{person}{Shunsuke Saito}, \bibinfo{person}{Or Litany},
  \bibinfo{person}{Shiqin Yan}, \bibinfo{person}{Numair Khan},
  \bibinfo{person}{Federico Tombari}, \bibinfo{person}{James Tompkin},
  \bibinfo{person}{Vincent Sitzmann}, {and} \bibinfo{person}{Srinath Sridhar}.}
  \bibinfo{year}{2021}\natexlab{}.
\newblock \showarticletitle{Neural Fields in Visual Computing and Beyond}.
\newblock \bibinfo{journal}{\emph{arXiv preprint arXiv:2111.11426}}
  (\bibinfo{year}{2021}).
\newblock


\bibitem[\protect\citeauthoryear{Xu, Wu, Zhu, Huang, Yang, Bao, and Xu}{Xu
  et~al\mbox{.}}{2021}]%
        {xu2021scalable}
\bibfield{author}{\bibinfo{person}{Jiamin Xu}, \bibinfo{person}{Xiuchao Wu},
  \bibinfo{person}{Zihan Zhu}, \bibinfo{person}{Qixing Huang},
  \bibinfo{person}{Yin Yang}, \bibinfo{person}{Hujun Bao}, {and}
  \bibinfo{person}{Weiwei Xu}.} \bibinfo{year}{2021}\natexlab{}.
\newblock \showarticletitle{Scalable image-based indoor scene rendering with
  reflections}.
\newblock \bibinfo{journal}{\emph{ACM Transactions on Graphics (TOG)}}
  \bibinfo{volume}{40}, \bibinfo{number}{4} (\bibinfo{year}{2021}),
  \bibinfo{pages}{1--14}.
\newblock


\bibitem[\protect\citeauthoryear{Xu, Xu, Philip, Bi, Shu, Sunkavalli, and
  Neumann}{Xu et~al\mbox{.}}{2022}]%
        {pointnerf}
\bibfield{author}{\bibinfo{person}{Qiangeng Xu}, \bibinfo{person}{Zexiang Xu},
  \bibinfo{person}{Julien Philip}, \bibinfo{person}{Sai Bi},
  \bibinfo{person}{Zhixin Shu}, \bibinfo{person}{Kalyan Sunkavalli}, {and}
  \bibinfo{person}{Ulrich Neumann}.} \bibinfo{year}{2022}\natexlab{}.
\newblock \showarticletitle{Point-NeRF: Point-based Neural Radiance Fields}.
\newblock  (\bibinfo{year}{2022}).
\newblock


\bibitem[\protect\citeauthoryear{Yifan, Serena, Wu, {\"O}ztireli, and
  Sorkine-Hornung}{Yifan et~al\mbox{.}}{2019}]%
        {yifan2019differentiable}
\bibfield{author}{\bibinfo{person}{Wang Yifan}, \bibinfo{person}{Felice
  Serena}, \bibinfo{person}{Shihao Wu}, \bibinfo{person}{Cengiz {\"O}ztireli},
  {and} \bibinfo{person}{Olga Sorkine-Hornung}.}
  \bibinfo{year}{2019}\natexlab{}.
\newblock \showarticletitle{Differentiable surface splatting for point-based
  geometry processing}.
\newblock \bibinfo{journal}{\emph{ACM Transactions on Graphics (TOG)}}
  \bibinfo{volume}{38}, \bibinfo{number}{6} (\bibinfo{year}{2019}),
  \bibinfo{pages}{1--14}.
\newblock


\bibitem[\protect\citeauthoryear{Yu, Li, Tancik, Li, Ng, and Kanazawa}{Yu
  et~al\mbox{.}}{2021}]%
        {yu2021plenoctrees}
\bibfield{author}{\bibinfo{person}{Alex Yu}, \bibinfo{person}{Ruilong Li},
  \bibinfo{person}{Matthew Tancik}, \bibinfo{person}{Hao Li},
  \bibinfo{person}{Ren Ng}, {and} \bibinfo{person}{Angjoo Kanazawa}.}
  \bibinfo{year}{2021}\natexlab{}.
\newblock \showarticletitle{{PlenOctrees} for Real-time Rendering of Neural
  Radiance Fields}. In \bibinfo{booktitle}{\emph{ICCV}}.
\newblock


\bibitem[\protect\citeauthoryear{Zwicker, Pfister, Van~Baar, and Gross}{Zwicker
  et~al\mbox{.}}{2001}]%
        {zwicker2001surface}
\bibfield{author}{\bibinfo{person}{Matthias Zwicker},
  \bibinfo{person}{Hanspeter Pfister}, \bibinfo{person}{Jeroen Van~Baar}, {and}
  \bibinfo{person}{Markus Gross}.} \bibinfo{year}{2001}\natexlab{}.
\newblock \showarticletitle{Surface splatting}. In
  \bibinfo{booktitle}{\emph{Proceedings of the 28th annual conference on
  Computer graphics and interactive techniques}}. \bibinfo{pages}{371--378}.
\newblock


\end{thebibliography}

\end{document}